\documentclass[conference]{IEEEtran}
\IEEEoverridecommandlockouts
\usepackage{cite}
\usepackage{amsmath,amssymb,amsfonts}
\usepackage{algorithm}
\usepackage{algorithmicx}
\usepackage[noend]{algpseudocode}
\usepackage{float}
\usepackage[inline]{enumitem}
\usepackage{graphicx}
\usepackage{textcomp}
\usepackage{xcolor}
\usepackage[export]{adjustbox}

\usepackage[utf8]{inputenc}
\usepackage{url}
\usepackage{subfig}

\MakeRobust{\Call}
\newcommand\fixme[1]{}
\algdef{SE}[DOWHILE]{Do}{doWhile}{\algorithmicdo}[1]{\algorithmicwhile\ #1}%

\begin{document}

\def\arxivpreprint{}

\title{Scheduling Methods to Reduce Response Latency of Function as a Service
\ifdefined\arxivpreprint
\thanks{Preprint of the paper accepted at IEEE 32nd International Symposium on Computer Architecture and High Performance Computing,
Porto, Portugal, 2020}
\fi
}
\author{
\IEEEauthorblockN{Pawel Zuk}
\IEEEauthorblockA{\textit{Institute of Informatics} \\
\textit{University of Warsaw}\\
Warsaw, Poland \\
p.zuk@mimuw.edu.pl}
\and
\IEEEauthorblockN{Krzysztof Rzadca,~\IEEEmembership{Member,~IEEE,}}
\IEEEauthorblockA{\textit{Institute of Informatics} \\
\textit{University of Warsaw}\\
Warsaw, Poland \\
krzadca@mimuw.edu.pl}
}
\maketitle

\begin{abstract}
  Function as a Service (FaaS) permits cloud customers to deploy to cloud individual functions,
  in contrast to complete virtual machines or Linux containers. 
  All major cloud providers offer FaaS products (Amazon Lambda, Google Cloud Functions, Azure Serverless);
  there are also popular open-source implementations (Apache OpenWhisk) with commercial offerings (Adobe I/O Runtime, IBM Cloud Functions).
  A new feature of FaaS is function composition: a function may (sequentially) call another function, which, in turn, may call yet another function -- forming a chain of invocations.
  From the perspective of the infrastructure, a composed FaaS is less opaque than a virtual machine or a container.
  We show that this additional information enables the infrastructure to reduce the response latency.
  In particular, knowing the sequence of future invocations,
  the infrastructure can schedule these invocations along with environment preparation.
  We model resource management in FaaS as a scheduling problem combining
  (1) sequencing of invocations,
  (2) deploying execution environments on machines, and
  (3) allocating invocations to deployed environments.
  For each aspect, we propose heuristics. 
  We explore their performance by simulation on a range of synthetic workloads.
  Our results show that if the setup times are long compared to invocation times,
  algorithms that use information about the composition of functions consistently outperform greedy, myopic algorithms,
  leading to significant decrease in response latency.

\end{abstract}

\begin{IEEEkeywords}
scheduling, workflow, setup time, function-as-a-service, serverless
\end{IEEEkeywords}

\section{Introduction}

Serverless computing allows a cloud customer to run their code in production without configuring and allocating the software and the infrastructure stack~\cite{Castro:2019:RSC:3372896.3368454}. A cloud customer can thus focus on their application, rather than on managing the production environment. Major cloud providers offer serverless products (Amazon Lambda, 
Google Cloud Functions, 
Microsoft Azure Serverless). 
We focus on a variant of serverless computing called \emph{Function as a Service (FaaS)}~\cite{fox2017status}. In FaaS, a cloud customer uploads the source code of a (stateless) function to the provider.
When an end-user issues a request, this code is executed on the infrastructure provided and managed by the FaaS system. The FaaS system isolates requests by providing a prepared execution environment (e.g., a Linux container) for each invocation.
%

We focus on a relatively new element of FaaS, \emph{composition} of functions~\cite{baldini2017serverless}. During an invocation of a composed FaaS initiated by a single incoming event (e.g., an HTTP request), a function calls another function, that, in turn, may call yet another function and so on. 
If these invocations are all synchronous, the call structure is a chain; if some are asynchronous, it is a DAG.
In this paper, we narrow our focus to chains, as they are natively supported in OpenWhisk; and chains are sufficient to show the benefits resulting from better scheduling.
However, our algorithms and our conclusions generalize to DAGs
\ifdefined\arxivpreprint
(we refer to the Appendix).
\else
(we refer to the Appendix in~\cite{preprint}).
\fi
%
%
%

The existing open-source FaaS systems (OpenWhisk, Fission Workflows) do not use the information about the structure of the function compositions. Each invocation in a composition chain is treated independently. 
However, once the first function is invoked, the scheduler knows that the functions that follow in a chain will be eventually called too --- thus, the scheduler can prepare their execution environments in advance.

The contributions of this paper are as follows:
\begin{itemize}
\item We model scheduling in FaaS as a combination of the multiple knapsack problem, scheduling with dependencies and with setup times (Section~\ref{sec:new-model}). 
\item We propose a number of heuristics for each aspect (Section~\ref{sec:algorithms}). These heuristics derive from classic approaches, but we adjust them to the FaaS specificity.
\item By simulations, we show that heuristics examining the composition structure lead to lower response latencies (Section~\ref{sec:experiments}).
\end{itemize}



\section{Modeling FaaS Resource Management}
\label{sec:new-model}
\subsection{Resource Management in OpenWhisk}
\label{sec:implementations}
In this section, we describe from the resource management perspective a representative implementation of a serverless cloud platform, the open-source Apache OpenWhisk~\cite{org2017apache}.
OpenWhisk is mature, actively-developed software also offered commercially (IBM Cloud Functions, Adobe I/O Runtime).
OpenWhisk alternatives include OpenLambda~\cite{hendrickson2016serverless} and Fission~\cite{kritikos2018review}. OpenLambda uses containers to provide runtime environment for functions. Fission is designed for Kubernetes~\cite{kubernetes}; it can be deployed on existing cluster among other applications, which makes its adoption significantly easier. 
This section forms a background for our scheduling model that follows in Section~\ref{sec:model}.

OpenWhisk allows a cloud \emph{customer} to upload \emph{functions} (essentially, code snippets). A function is executed when \emph{end-users} issue requests.
A function executes in an \emph{environment} --- an initialized Docker container. 
Different Docker images are used for each of supported languages;  a customer can also provide a custom image (with, e.g., additional libraries).
Before the first execution of a function, the container must be initialized (e.g., setting up the container or compiling a Go function). This initialization can take a considerable amount of time (called later the \emph{setup time}) --- \cite{shahrad2019architectural} reports at least 500ms.
An environment is specific to a function (it is not reused between different functions).
Subsequent invocations may reuse the same environment (no further setup times are necessary).
By default, in OpenWhisk an environment executes at most a single invocation at any given moment.
%

OpenWhisk also allows to compose several functions into a chain (a sequence). After one function finishes, its result are passed to the next function; the last function responds to the end-user.
While sequences are natively supported, in order to spawn two or more functions in parallel (resulting in a DAG), the developer may use an additional OpenWhisk Composer module or call the OpenWhisk API from the function code.

Architecture of OpenWhisk is complex. However, from our perspective the key components are the \emph{controller} and the \emph{invoker}. The controller communicates with the invokers by message passing (via Apache Kafka).

The invoker is an agent program running on a worker node. The invoker is responsible for executing actions scheduled on a particular node. Each invoker has a unique identifier; it announces itself to the controller while starting.

The controller acts as a scheduler handling incoming events and routing function invocations to invokers.
The controller monitors the status of workers and the currently executing invocations.

The controller attempts to balance load across nodes.
The algorithm selects the initial worker node for each function based on a hash of the workspace name and the function name. 
Similarly, the algorithm picks for each function another number, called the \emph{step size} (a number co-prime with the count of worker nodes).
Each time a function is invoked, the controller attempts to schedule the invocation on its initial worker. If a worker doesn't have sufficient resources immediately available, the controller tries to schedule the invocation on the next node (increased by the \emph{step size}).
If the invocation cannot be immediately scheduled on any node, it is queued on a randomly chosen node.


\subsection{A Scheduling Model for FaaS}
\label{sec:model}
In this section we define the optimization model for the FaaS resource management problem. The aim of this model is to have the simplest possible (yet still realistic) approximation of a FaaS system that enables us to show that considering FaaS compositions allow optimizations. We thus deliberately do not take into account some factors that we argue are orthogonal for this work. 
%
%

We use the standard notation from~\cite{brucker2007scheduling}.
A single end-user request corresponds to a \emph{job} $J_i$. A job is composed of one or more \emph{tasks} $O_{i,k}$, each corresponding to a single FaaS invocation. The request is responded to (the job completes) at time $C_i$ when the last task completes, $C_i = \max_j C_{i,j}$. Tasks have dependencies resulting from, e.g., before-after relationships in the code. While in general such dependencies can be modeled by a DAG, in this work we concentrate on chains of tasks, i.e., task $O_{i,k+1}$ starts (at time $\sigma_{i, k+1}$) only after $O_{i,k}$ completes, $\sigma_{i,k+1} \geq C_{i,k} $
\ifdefined\arxivpreprint
(we show additional results for DAGs the Appendix).
\else
(we show additional results for DAGs in~\cite{preprint}).
\fi

We assume that individual functions are repeatedly executed (modeling similar requests from many end-users but also shared modules like authorization). We model such grouping by mapping each task $O_{i,k}$ to exactly one family $f(O_{i,k})$ (obviously, two tasks $O_{i,k}$ and $O_{i,l}$ from a job $J_i$ might belong to different families). All tasks from a family $f$ require the same environment $E_f$, have the same execution time (duration) $p_f$ and require the same amount of resources $q_f$.

A task $O_{i,k}$ from a family $f(O_{i,k})$ is executed on exactly one machine in an \emph{environment} (OpenWhisk container) $E_f$. $E_f$ requires set-up time $s_f$ (initialization of the environment) before executing the first task (subsequent tasks do not require set-up times).  Typically, $s_f$ is non-negligible and longer than the task's duration, $s_f > p_f$ (but we don't assume this).

A machine commonly hosts many environments (thus supporting parallel execution of tasks). 
Since the moment the environment's preparation starts -- and until it is removed -- each environment $e_f$ uses $q_f$ of the machine's resources (e.g., bytes of memory) whether a task executes or not.  The number of hosted environments is limited by the capacity of the machine $Q$ ($\sum q_f \leq Q$).
We consider only a single dimension of the resource requests as OpenWhisk assumes a linear relation between memory and CPU limits of the underlying containers. Similarly, Google Cloud Functions allow customers to specify only a single dimension (memory requirement). However, it should be relatively easy to extend our model to vector packing~\cite{chekuri1999multi}. 

We do not consider the additional latency caused by communication between tasks because we assume that a high-throughput, low-latency network of a modern datacenter is less of a limit than the link between the datacenter and the Internet.
We assume that the machines are homogeneous (machine resources $Q$ and execution times $p_f$ are the same). If a FaaS system is deployed on VMs rented from an IaaS cloud, it is natural to use a Managed Instance Group (MIG) that requires all VMs to have the same instance type. If FaaS is deployed on a bare-metal data-center, the amount of machines having the same hardware configuration should be higher that other scalability limits (e.g. at a Google data-center, 98\% of machines from a 10,000-machine cluster belong to one of just 4 hardware configurations~\cite{verma2015large}).

We assume that \emph{jobs} have no release times, i.e., the first tasks of all the jobs are ready to be scheduled at time 0. This assumption approximates a system under peak load --- there is a queue of requests to be scheduled now.
Note that in contrast to \emph{jobs}, individual tasks (in particular, the tasks that follow the first task of a job) do have non-zero release times, resulting from inter-task dependencies.

Our model is \emph{clairvoyant}.
A FaaS system repeatedly (thousands of times) executes individual functions.
Thus, once a particular family is known for some time,
$q_f$, $p_f$ and the function structure should be easy to estimate using standard statistical methods --- and before that, the system can use conservative upper bounds (e.g., defaults used by OpenWhisk). \cite{rzadca2020autopilot} shows that even simple methods estimate precisely memory and CPU for long-running containers (which, in principle, is harder than estimating FaaS, as FaaS are shorter, thus repeated much more frequently than a container).
%
%

The system optimizes the average response latency. As all $N$ jobs are ready at time 0, this metric corresponds to $\frac{1}{N}\sum_{i=1}^N C_i$.

To summarize, the scheduling problem consists of finding for each task $O_{i,k}$ a machine and a start time $\sigma_{i,k}$ so that:
\begin{enumerate}
	\item at $\sigma_{i,k}$, there is a prepared environment for $f(O_{i,k})$ on that machine that does not execute any other task during $[\sigma_{i,k}, \sigma_{i,k}+p_f]$ (a scheduling constraint);
	\item dependencies are fulfilled: if $k>1$, $\sigma_{i,k} \geq C_{i,k-1}$ (a dependency constraint);
	\item at any time, for each machine, the sum of requirements of the installed environments is smaller than the machine capacity (a multiple knapsack constraint).
\end{enumerate}
This problem is NP-hard, as generalizing several NP-hard problems (knapsack~\cite{garey1979computers}, $P2 | chains | \sum C_i$~\cite{brucker2007scheduling}).

\section{Algorithms}
\label{sec:algorithms}


\begin{algorithm}[tb]
  \begin{algorithmic}
	\Function{schedulingStep}{t, queue, wait, policy}
	\State \Comment{$policy \in \left\{default, start \right\}$, $wait \in \left\{ true, false \right\}$}
	\For{$task \in \Call{finishedTasks}{t}$}
		\If {$policy == default$}
			\State{$\Call{queueDependentTasks}{task, t}$}
		\EndIf
	\EndFor
	
	\For{$task \in \Call{order}{queue}$}
		\State {$e \gets \Call{FindUnusedEnvironment}{task}$}
		\If {$e$ is nil and $wait$}
			\State{$e \gets \Call{FindEnvironmentToWait}{task}$}
		\EndIf
		\If {$e$ is nil}
			\State{$e \gets \Call{PlaceNewEnvironment}{task}$}
		\EndIf
		\If {$e$ is nil}
			\State{$e \gets \Call{RemoveAndPlaceEnvironment}{task}$}
		\EndIf
		\If {$e$ is not nil}
			\State{$\Call{assignTask}{c, task, \Call{releaseTime}{task}}$}
			\State{$\Call{removeFromQueue}{task}$}
			\If {$policy == start$}
				\State{$p \gets \Call{duration}{task}$}
				\State{$\Call{queueDependentTasks}{task, t + p}$}
			\EndIf
		\EndIf
	\EndFor
	\EndFunction
  \end{algorithmic}
  \caption{Framework scheduling algorithm.}\label{alg:schedule}
\end{algorithm}

In this section we describe heuristics to schedule FaaS invocations.
We decompose the FaaS scheduling problem into three aspects:
\emph{sequencing} of invocations; \emph{deployment} of execution environments on machines; and \emph{allocation} of invocations to deployed environments.
We start with a framework algorithm (Algorithm~\ref{alg:schedule}) to show how these aspects are combined to build a schedule; we then describe specific heuristics for each of the aspects.
\emph{Sequencing} corresponds to the ordering policy (Section~\ref{sec:ordering-policy}) and the awareness of task dependencies (Section~\ref{sec:awareness-dependencies}). \emph{Deployment} corresponds to the removal policy (Section~\ref{sec:removal-policy}). \emph{Allocation} corresponds to the waiting/non-waiting variants (Section~\ref{sec:waiting}).

The framework algorithm is a standard scheduling loop executing \emph{schedulingStep} at time $t$ when at least one task completes.
The algorithm maintains a queue of tasks $[O_{i,k}]$ to schedule.
\begin{enumerate}
\item Queue the successors $O_{i,k+1}$ of tasks completed at $t$ ($\{O_{i,k}: \sigma_{i,k}+p_f=t\}$) (\emph{queueDependentTasks}). 
\item Apply a scheduling policy to the queued tasks (\emph{Order}).
\item Try to find an environment $e$ for each queued task:
	\begin{enumerate}
		\item Try to claim an initialized environment of the required type (\emph{FindUnusedEnvironment}, and -- if $wait$ -- \emph{FindEnvironmentToWait}).
			In this step we iterate over all machines and take the first matching environment.
			(Section~\ref{sec:waiting} describes the $wait$ variant).
		\item If (a) fails, try to create a new environment without removing any existing one (\emph{PlaceNewEnvironment}).
			As above, we use the first fitting machine.
		\item If (b) fails, try to find a machine with sufficient capacity for $e$ that is currently claimed by environments that do not execute any task; remove these environments, and install $e$ (\emph{RemoveAndPlaceEnvironment}).
		\item If (c) fails, the task remains in the queue.
	\end{enumerate}
\item If an environment $e$ is found, assign the task (\emph{AssignTask}); otherwise (3.a-c all fail) the task remains in the queue.
\end{enumerate}
\emph{AssignTask} starts a task on an environment as follows. Each environment has a queue of assigned task. Immediately after creating an environment, it is initialized (which takes time $s_f$). Then, the environment starts to execute tasks sequentially from its queue. If the head task is not ready (waiting for dependencies), the environment waits (no backfilling).
This may happen in the \emph{start} policy (see Section~\ref{sec:awareness-dependencies}).

In the following, we propose concrete variants for these functions. 
We denote the full scheduling policy by a tuple $(A, B, C, D)$, e.g., , $(FIFO, LRU, wait, start)$, where $A$ denotes the ordering policy, $B$ denotes the removal policy, $C$ indicates if variant is \emph{waiting} and $D$ describes whether the variant is dependency-aware.

\subsection{Ordering policy (Order)}
\label{sec:ordering-policy}
We compare the standard FIFO and SJF with three orderings taking into account the dependencies:

\begin{itemize}
	\item \emph{FIFO (First Come First Served)} -- use the order in which the tasks were added.
	\item \emph{EF (Existing First)} -- 
      partition the tasks into two groups: (1) there is at least one idle, initialized environment $e$ of matching type $E_{f(O_{i,k})}$; (2) the rest.
	  Schedule the first group before the second group. The relative order of the tasks in both groups remains stable (FIFO).
	  For example, if queue contains five tasks $[O_{i_1,k_1}, O_{i_2,k_2}, O_{i_3,k_3}, O_{i_4,k_4}, O_{i_5,k_5}]$, there is only one environment $e$ that is idle and only tasks $O_{i_1,k_1}$, $O_{i_3,k_3}$, $O_{i_4,k_4}$ require environment with type matching $e$, the resulting order is $[O_{i_1,k_1}, O_{i_3,k_3}, O_{i_4,k_4}, O_{i_2,k_2}, O_{i_5,k_5}]$.
	\item \emph{SJF (Shortest Jobs First)} -- order by increasing durations $p_f$;
	\item \emph{SW (Smallest Work)} -- order by increasing remaining work \emph{in a job}, i.e. for a task $O_{i,k}$, order by $\sum_{k' \geq k} p_{f(O_{i,k'})}$. 
	\item \emph{RT (Release Time)} -- ordered by the time the task's predecessors are completed. 
\end{itemize}

\subsection{Removal policy}
\label{sec:removal-policy}
\emph{RemoveAndPlaceEnvironment} removes environments according to either a standard LRU, or one of policies considering either initialization time $s_f$ or environment popularity:
\begin{itemize}
	\item LRU -- remove the LRU (Least Recently Used) environment(s) from first fitting machine (i.e. having enough space to be freed).
	\item min time removal -- remove the environment(s) with the smallest setup time $s_f$ (if more than one, select a single machine having environments with the smallest total $s_f$).
	\item min family removal -- remove the environment(s) from the family with the highest number of currently initialized environments.
      As it may be needed to remove more than one environment, choose a machine to minimize resulting number of families without any environment.
\end{itemize}

\subsection{Greedy environment creation}
\label{sec:waiting}

If there is no unused environment of the required type $E_f$, a greedy algorithm (i.e. when \emph{wait} is false) just attempts to create a new one.
However, when setup times $s_f$ are longer than task's duration $p_f$, it might be faster just to wait until one of currently initialized environments completes its assigned task.
We implement this policy by setting \emph{wait} to \emph{true} in Algorithm~\ref{alg:schedule}.
When no idle environment is available,
function \emph{FindEnvironmentToWait} computes for each initialized environment $e$ of type $E_f$ the time $C_e$ the last task currently assigned to this environment completes.
If an environment $e^*$ is available sooner than the time needed to set up a new environment ($\min C_e \leq t + s_f$), the task is assigned to $e^*$.
This variant use the (limited) clairvoyance of the scheduler by taking into account the knowledge of tasks' durations and setup times of their execution environments.

The \emph{waiting} variant is analogous to scheduling tasks in Heterogeneous Earliest Finish Time (HEFT~\cite{zhao2003experimental,bittencourt2010dag}) that places a task on a processor that will finish the task as the earliest. 

\subsection{Awareness of task dependencies}
\label{sec:awareness-dependencies}
A myopic (\emph{default}) scheduler queues just the tasks that are currently ready to execute: $O_{i,0}$ (the first tasks in the jobs), or the tasks for which the predecessors completed $\{ O_{i,k}: C_{i,k-1} \leq t\}$. However, when a task's $O_{i,k}$ predecessors complete, it might happen that there is no idle environment $e_{f(O_{i,k})}$, and thus $O_{i,k}$ must still wait $s_f$ until a new environment is initialized.

We propose two policies, \emph{start} and \emph{start with break (stbr)}, that use the structure of the job to prepare environments in advance. 
Both policies put the successor $O_{i,k+1}$ to the end of the queue when scheduling $O_{i,k}$; the successor has the release time $t+p_{f(O_{i,k})}$ (the time when $O_{i,k}$ completes). The notion of the release time allows us to block $O_{i,k+1}$'s execution until it is ready (as described in \emph{AssignTask}).
Note that \emph{start} and \emph{stbr}  may result in an environment that is (temporarily) blocked: e.g., if an empty system schedules a chain of two tasks, the second task from the chain is added to the queue immediately after scheduling the first task; this second task will be assigned to its environment, but cannot be started until the first task is completed. 
In \emph{start} variant, after \emph{schedulingStep} completes and new tasks were added to queue, scheduler tries placing them following the same procedure.
Compared with \emph{start}, \emph{stbr} immediately after adding $O_{i,k+1}$ reorders tasks in the queue according to the scheduling policy and restarts the placement (for clarity, \emph{stbr} is not presented in Algorithm~\ref{alg:schedule}).



\section{Evaluation} 
\label{sec:experiments}
We evaluate our algorithms with a calibrated simulator.
We use a simulator rather than modify the OpenWhisk scheduler for the following reasons.
First, a discrete-time simulator enables us to execute much more test scenarios and on a considerably larger scale (we perform tests on $1440 \cdot 15$ problem instances).
Second,
as our results will show, to schedule tasks more efficiently, the OpenWhisk controller (the central scheduler) should take over some of the decisions currently made by the invokers (agents residing on machines).
For example, \emph{min family removal} needs to know which family has the highest number of installed environments in the whole cluster ---  thus, the state of the whole cluster (note that this policy can be implemented in a distributed way: the cluster state can be broadcasted to the invokers).
%
To ensure that our simulator's results can be generalized to an OpenWhisk installation, we compare the performance of an actual OpenWhisk system with its simulation; the Pearson correlation between these results is very high (Section~\ref{sec:ow_comparison}).

\subsection{Method}
\label{sec:setup}


To test the performance of our algorithms, we generated synthetic instances with a wide range of parameter values. We are not aware of any publicly-available workloads for FaaS or related systems (having dependencies, function families and setup times).
Nevertheless, we also attempted to create instances resembling real scenarios by using Google Cluster Trace~\cite{clusterdata:Wilkes2020a} and generating only missing data.
\ifdefined\arxivpreprint
We present results of this approach in the Appendix.
\else
We present results of this approach in the Appendix~\cite{preprint}.
\fi

Many parameters of instances have a relative, rather than absolute, effect on the result. For example, multiplying by a constant both $Q$, the machine capacity, and $q_f$, the size of the task, results in an instance that has very similar scheduling properties. 
There is a similar relationship between setup times $s_f$ and durations $p_f$; and between the total number of tasks $n$ and the number of tasks in a chain $l$.
We thus fix one parameter from each pair to a constant (or a small range); and vary the other.
We have $n=1000$ tasks; $p_f$ is generated by the uniform distribution over integers $p_f \sim U[1, 10]$; similarly $q_f \sim U[1, 10]$.
The remaining parameters have ranges:
\begin{itemize}
	\item family count $n_f$: 10, 20, 50, 100, 200, 500;
	\item setup times $s_f$: $[0, 0]$, $[10, 20]$, $[100, 200]$, $[1000, 2000]$;
	\item chain lengths $l$: $[2, 10]$, $[10, 20]$, $[50, 100]$;
	\item machine count $m$: 2, 5, 10, 20, 50;
	\item machine sizes $Q$: 10, 20, 50.
\end{itemize}
For each combination of the parameters (or ranges) $n_f$, $s_f$, $l$, we generate 20 random instances, resulting in  $1440$ instances. We evaluate each instance on each of the 15 machine environments.

These ranges of parameters are wide. As we experiment on synthetic data, one of our goals is to explore trends -- characterize instances for which our proposed method works better (or worse) than the current baseline. In particular, chains longer than 10 ($l>10$) are longer than what we suspect is the current FaaS usage. On the other hand, it is not a lot compared with a call graph depth on any non-trivial software. At this point of FaaS evolution it is difficult to foresee the degree of compartmentalization future FaaS software will have -- and chains longer than 10 invocations represent fine-grained decomposition (similar to modern non-FaaS software). 

Given $n_f$, $[s_{\min}, s_{\max}], [l_{\min}, l_{\max}]$ we generate an instance as follows. For each of $n_f$, we set $s_f \sim U[s_{\min}, s_{\max}]$ and $p_f \sim U[1,10]$. For each of $n=1000$ tasks, we set its family $f$ to $U[1, n_f]$.
We then chain tasks to jobs. Until all tasks are assigned, we are creating jobs by, first, setting the number of tasks in a job to $l \sim U[l_{\min}, l_{\max}]$ (the last created job could be smaller, taking the remaining tasks); and then choosing $l$ unassigned tasks and putting them in a random sequence.

For each experiment, our simulator computes the average response latency, $(1/n) \sum C_i$.
Due to space constraints, we omit results on tail, 95\%-ile latency --  the 95\%-ile results also support our conclusions (unsurprisingly, the ranges are larger than for the averages).

In addition to testing variants of Algorithm~\ref{alg:schedule}, we simulate the current, round-robin behavior of the OpenWhisk scheduler (Section~\ref{sec:implementations}) with an algorithm \emph{OW}.
\emph{OW} randomly selects for each family $f$ the initial machine $m_f$ and the \emph{step size} $k_f$, an integer co-prime with the number of machines $m$. When scheduling a task $O_{i,k}$ in family $f$, \emph{OW} checks machines $m_f$, $m_f + k_f$, $m_f + 2k_f$, \dots (all additions modulo $m$), stopping at the first machine that has either the environment $E_f$  ready to process, or $q_f$ free resources (including unused environments that could be removed) to install a new environment $E_f$. If there is no such machine, $O_{i,k}$ is queued on a randomly-chosen machine.

\subsection{Validation of the simulator against OpenWhisk}
\label{sec:ow_comparison}

\begin{figure*}[t]
	\centering
	\subfloat[{by chain length ($n_f=50$, $s_f\in[10,20]$)}]{{\includegraphics[width=0.33\textwidth,trim=-10 12 -10 13,clip]{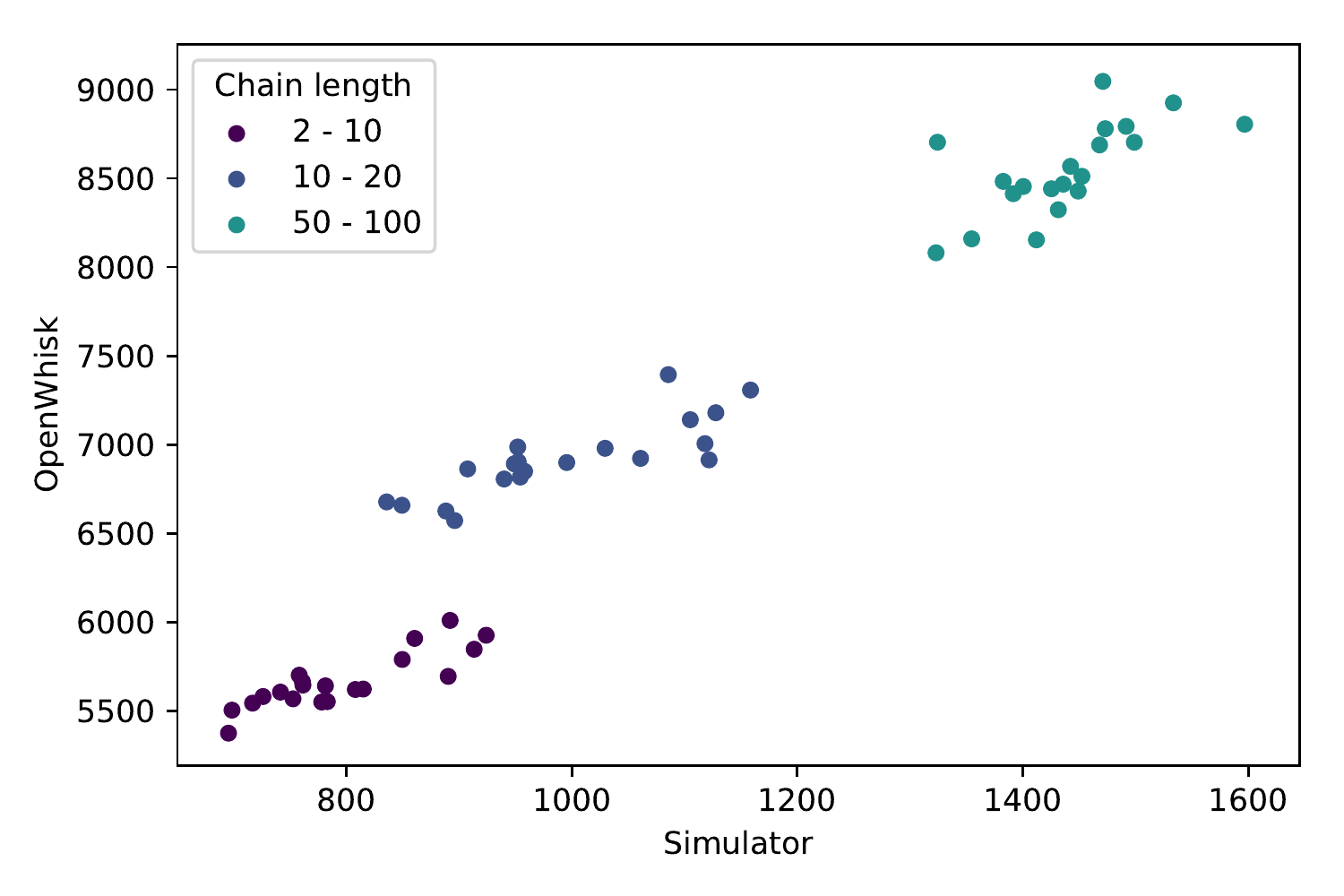}}}%
	\subfloat[{by family count ($l\in[10, 20]$, $s_f \in [10, 20]$)}]{{\includegraphics[width=0.33\textwidth,trim=-10 12 -10 13,clip]{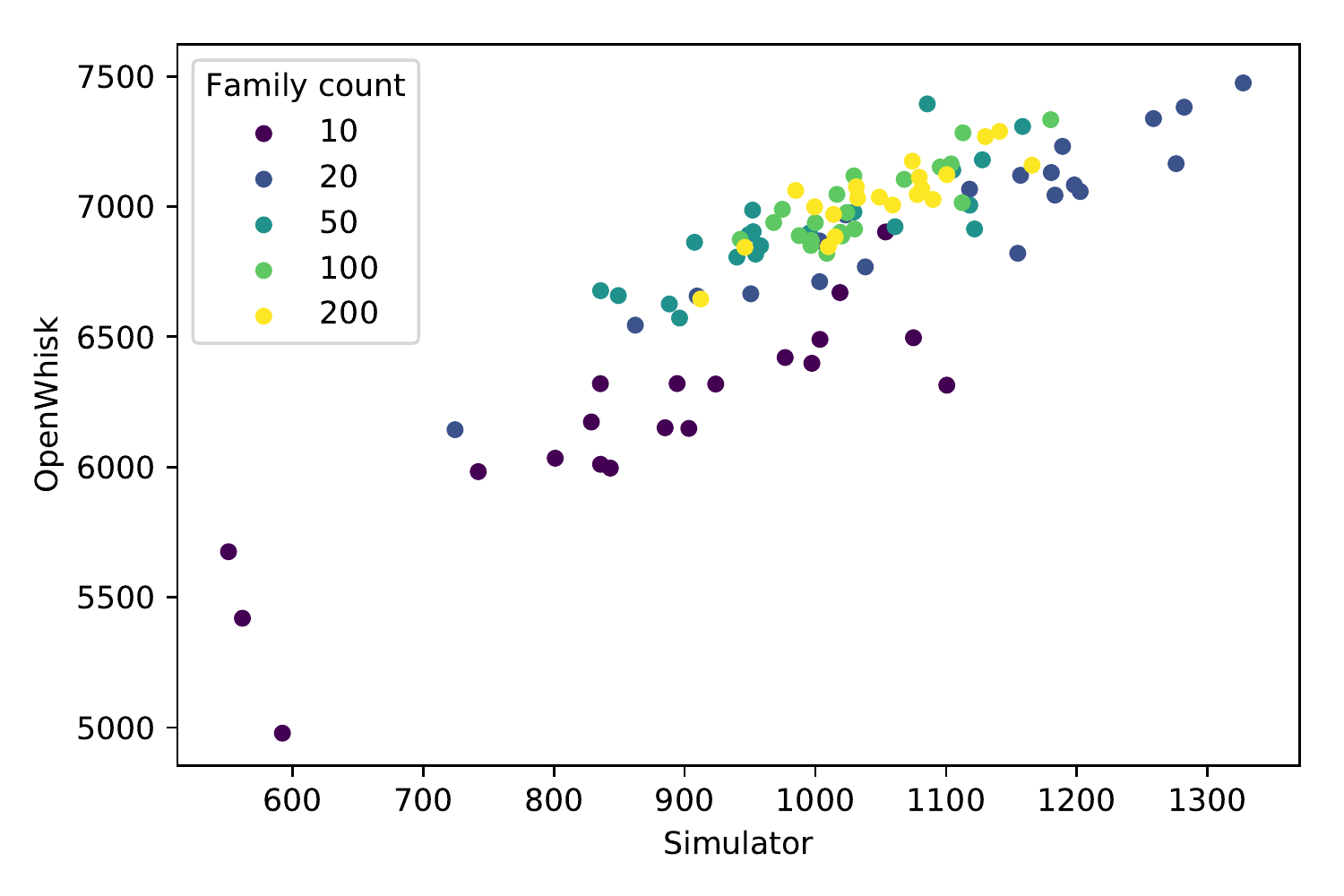}}}%
	\subfloat[{by setup time ($n_f=50$, $l\in[10, 20]$)}]{{\includegraphics[width=0.33\textwidth,trim=-10 12 -10 13,clip]{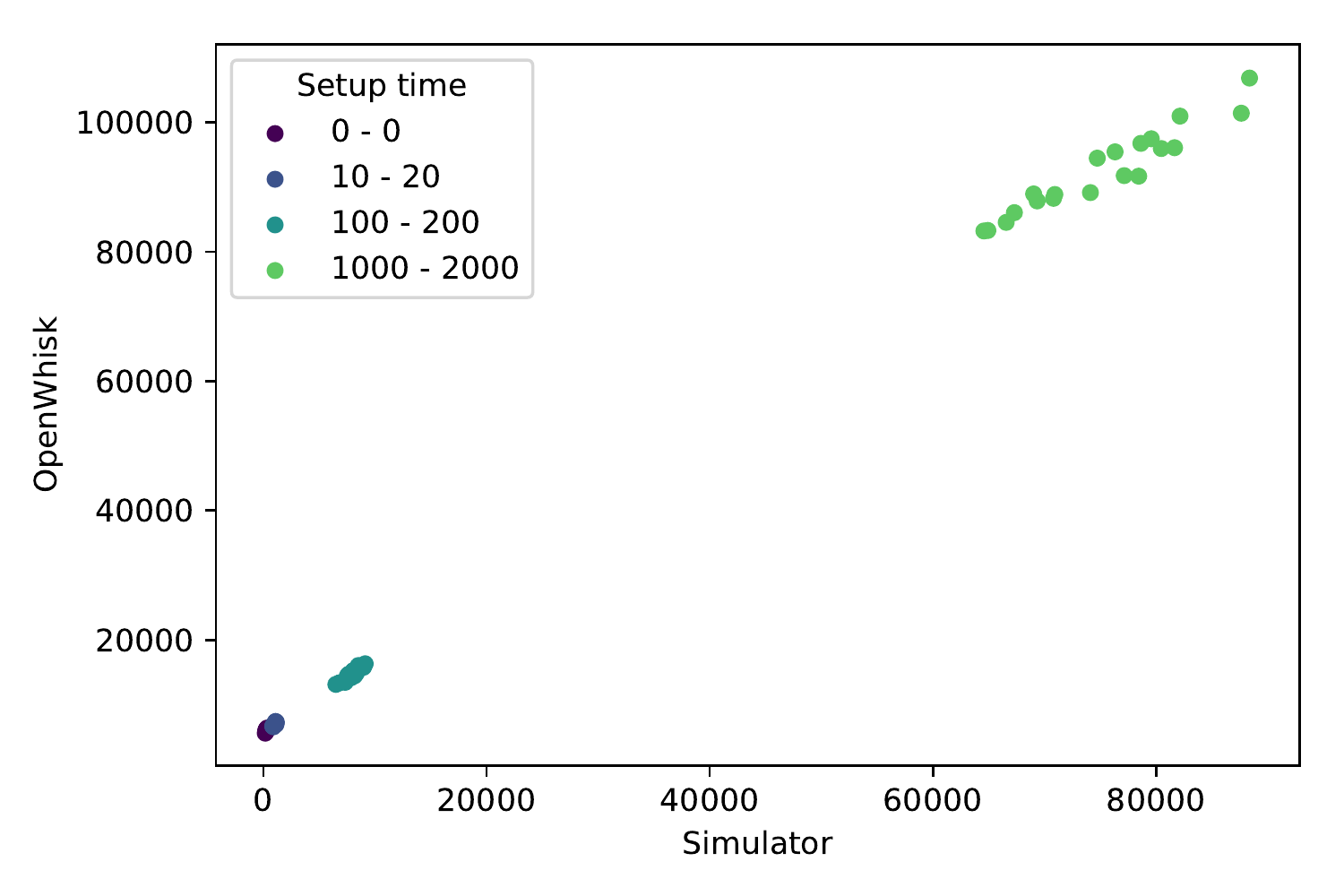}}}%
	\caption{Average latency on OpenWhisk system (Y axis) and simulation (X axis).
      1 unit is 10ms.
	Each point corresponds to a single instance executed on both OpenWhisk and simulator.
	}
	\label{fig:ow_sim_cmp}
\end{figure*}

To compare the results of our simulator with OpenWhisk, we developed a customized OpenWhisk execution environment that emulates a function with a certain setup time $s_f$, execution time $p_f$ and resource requirement $q_f$.
This environment emulates initialization by sleeping for $s_f * 10ms$; and it emulates execution by sleeping for $p_f * 10ms$. While sleeping does not use the requested memory ($q_f * 128MB$), the memory is blocked (through Linux cgroup limits) and therefore cannot be simultaneously used by other environments. We chose 10ms as the time unit to reduce impact of possible fluctuations of VM or network parameters in the datacenter (we performed some early experiments with 1ms and this noise was significant; and  with a longer time unit tests take unreasonable time).
%
We emulate a single instance from our simulator by creating, for each job $J_i$, an equivalent sequence of invocations in OpenWhisk.
To avoid caching of results in OpenWhisk, we ensure that each invocation is executed with a distinct set of parameters.
We deployed an OpenWhisk cluster (1 controller and $m=10$ invokers) on 11 VMs in GCE.
All machines have 2 vCPU and 16GB RAM. 
We further restrict the memory OpenWhisk can use on machines to 1280MB (equivalent to $Q=10$).
In order to reduce impact of cloud storage on system performance, we used a ramdisk to store OpenWhisk accounting database.
We also extended limits (maximum duration and sequence length) and changed the default log level to WARN.
To reduce the impact of brief performance changes, we executed each test instance thrice and reported the median.

In Figure~\ref{fig:ow_sim_cmp} we compare the average response latency in OpenWhisk and in our simulator varying
chain lengths, the number of families and the ranges of setup times.
For consistency, OpenWhisk results are rescaled to the simulator time unit (divided by 10) 
The Pearson correlation between OpenWhisk and simulator is very high (between $0.86$ when varying family count, Fig.~\ref{fig:ow_sim_cmp}.b, and $0.999$ when varying the setup time, Fig.~\ref{fig:ow_sim_cmp}.c).
There is, however, an additive factor in OpenWhisk noticeable especially in smaller instances in Fig.,~\ref{fig:ow_sim_cmp}.(a) and Fig.~\ref{fig:ow_sim_cmp}.(b): the range of OpenWhisk results in $[5000, 9000]$, while the range of simulated results is in $[550, 1600]$; on larger instances, as in Fig.~\ref{fig:ow_sim_cmp}.(c), this constant factor is less noticeable.
This additive factor is caused by an additional system overhead added to every function execution: 
each invocation stores data in a database and requires internal communication.
We conclude that the high correlation between the simulator and the OpenWhisk results validates our simulator -- that the differences between algorithms observed in the simulator are transferable to the results in OpenWhisk.

\subsection{Relative Performance of Policies}\label{sec:policy}

\begin{figure*}[t]
	\centering
	\subfloat[]{{\includegraphics[width=0.33\textwidth,trim=0 12 0 13,clip]{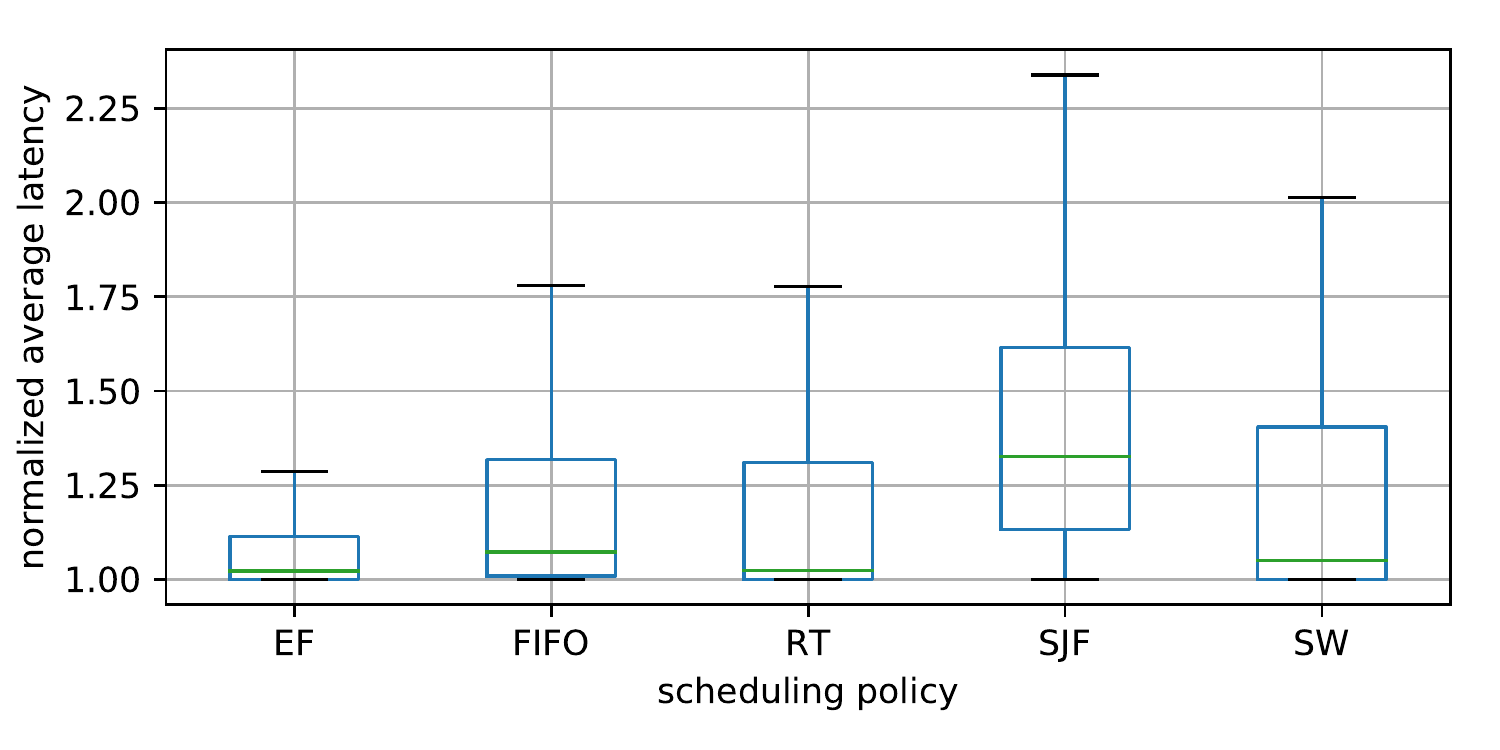}}}%
	\subfloat[]{{\includegraphics[width=0.33\textwidth,trim=0 12 0 13,clip]{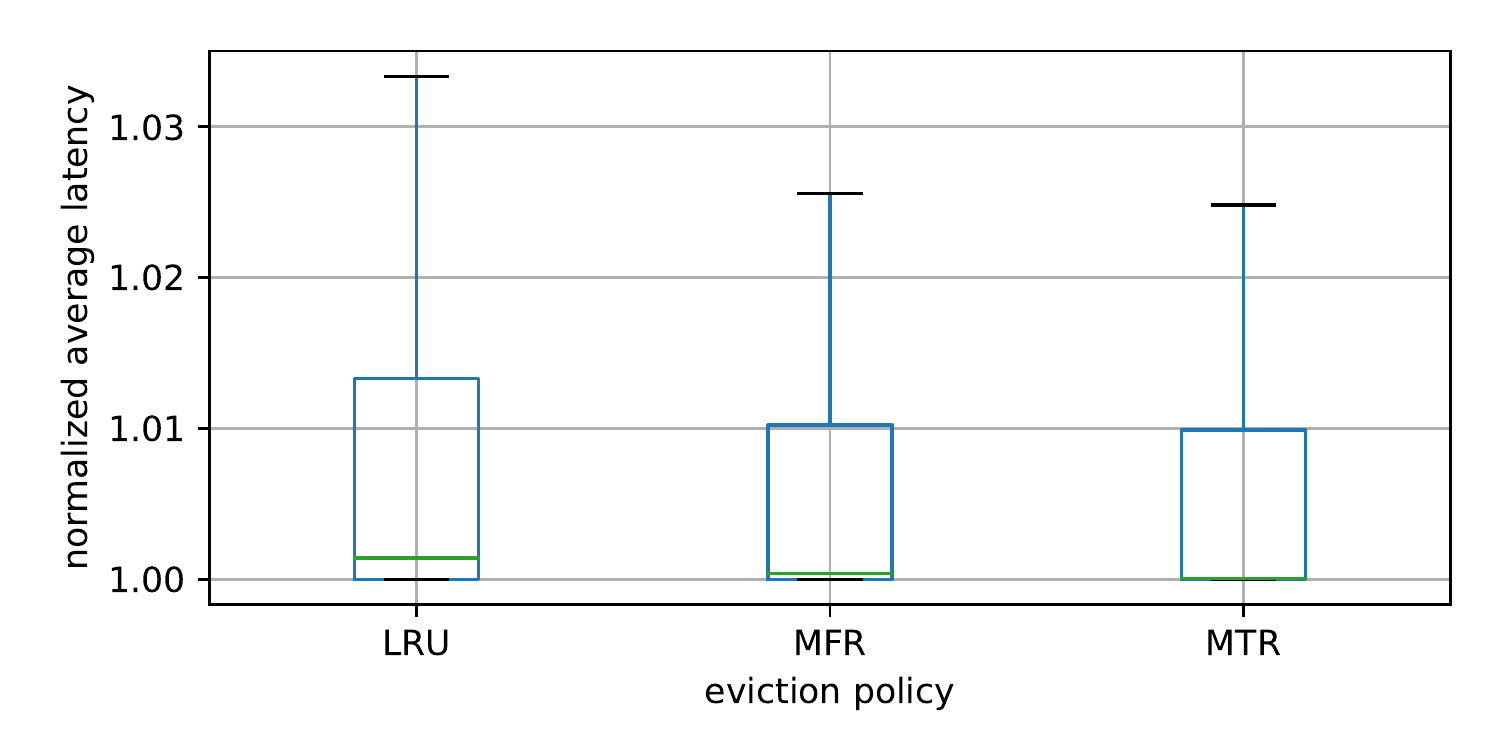}}}%
	\subfloat[]{{\includegraphics[width=0.33\textwidth,trim=0 12 0 13,clip]{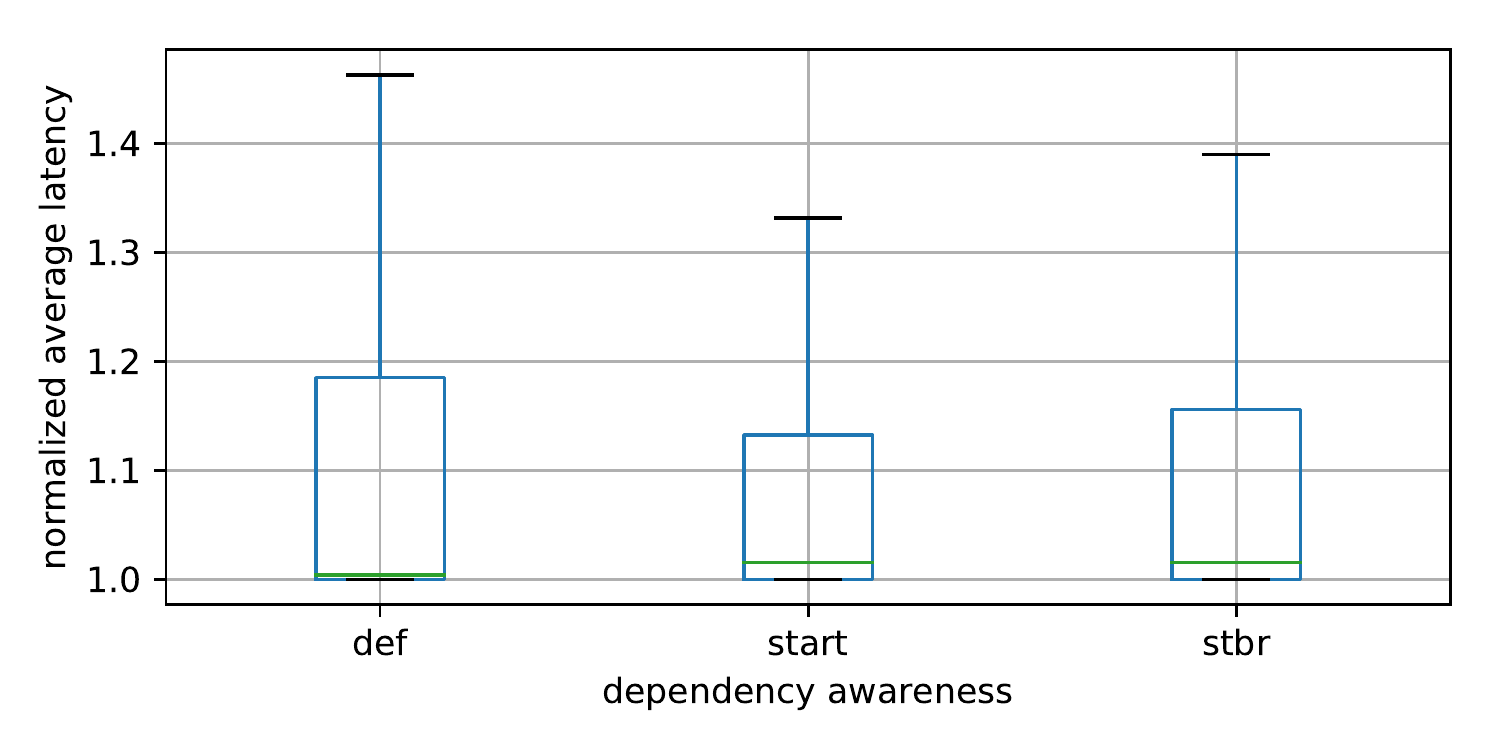}}}%
	\caption{Comparison of resulting average latency under: different scheduling policies (a), removal policies (b) and variants of dependency--awareness (c).
	In (a), for each instance (the same tasks, machine capacities and machine count), and having other variants of the algorithm set (removal policy, \emph{waiting} and dependency-awareness), we find the minimal average latency among the 5 scheduling policies; we then normalize the results from all 5 scheduling policies by this minimal average latency.
	Each box corresponds to a statistics over experiments with all the removal policies (both in \emph{waiting} non-\emph{waiting} variant) and all dependency-awareness variants (def, start, stbr), performed on all instances and all possible machine environments (over 300k individual data points).
	For (b) and (c) results are normalized as in a), but for different removal policies (b) and for different dependency-aware variants (c), rather than scheduling policies.
	Here and in all following box plots, the box height indicates the first and the third quartile, the line inside the box indicates the median, and the whiskers extend to the most extreme data point within 1.5 $\times$ IQR.
	}
	\label{fig:compare_avg}
\end{figure*}

We first analyze the impact of each policy 
by analyzing their relative performance. For each variant (A, B, C, D), on each instance, we compute the relative performance of the policy we measure by finding the minimal average latency across all variants of the measured policy while keeping the rest of the variants the same.
For example, when measuring the effect of the scheduling policy (A),
on an instance, we find the minimum average latency from the 5 variants of the scheduling policy: (EF, b, c, d), (FIFO, b, c, d), (RT, b, c, d), (SFJ, b, c, d), (SW, b, c, d) (keeping b, c, d the same); and then we divide all 5 by this value.
The goal of this analysis is to narrow down our focus to the aspects of the problem that are crucial for the performance. Using this method, we show that, e.g., all removal policies result in very similar outcomes.
Figure~\ref{fig:compare_avg} shows the results. 

\emph{Ordering}: \emph{EF} policy
dominates other ordering policies, confirming that it is better to avoid environment setup by reusing existing environments.
Its median is similar to RT (and lower than other algorithms), and the range of values (including the third quartile) is the lowest.
%
\emph{Removal}: Unlike scheduling policies, all the removal policies result in virtually the same schedule length: the range of Y axis is 1.035; thus outliers are only 3.5\% worse than the minimal schedule found in the alternative methods.

\emph{Dependency awareness}: Both \emph{start} and \emph{stbr} result in similar performance. We confirmed this result by looking at individual instances: the performance of \emph{start} and \emph{stbr} were similar.

To improve the readability in the remainder, given that the removal policies have little effect on the schedule length (Figure~\ref{fig:compare_avg}), we show only the results for LRU. Similarly, we skip results for SJF and RT orderings: RT is close to FIFO and SJF is clearly dominated by other variants. Finally, as the difference between \emph{start} and \emph{stbr} variants is small, we show results only for \emph{start}.

\subsection{Impact of the length of the chain}
\label{sec:chain_len_cmp}

\begin{figure*}[!h]
	\centering
	\subfloat[length 2-10]{{\includegraphics[width=0.33\textwidth,trim=0 11 0 10,clip]{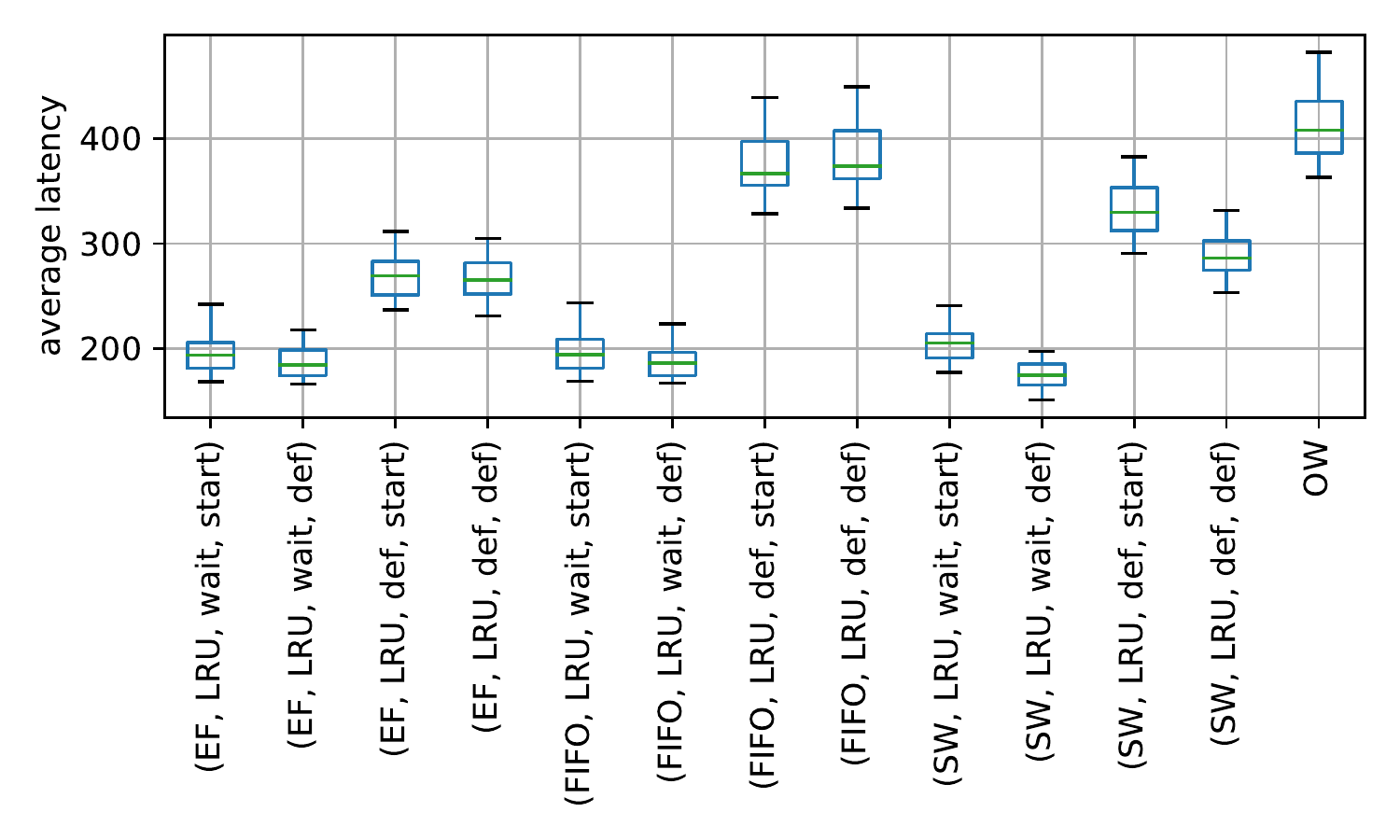}}}%
	\subfloat[length 10-20]{{\includegraphics[width=0.33\textwidth,trim=0 11 0 10,clip]{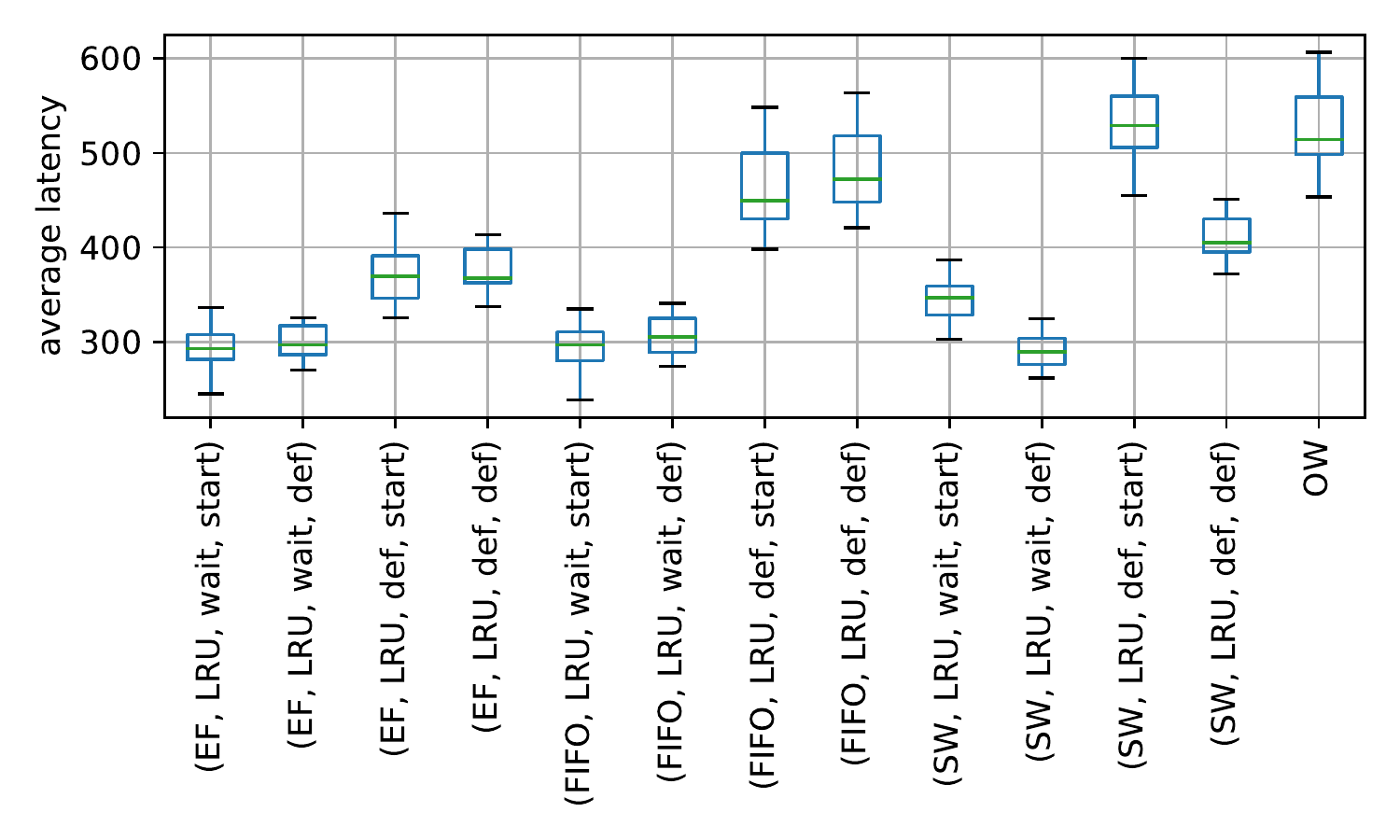}}}%
	\subfloat[length 50-100]{{\includegraphics[width=0.33\textwidth,trim=0 11 0 10,clip]{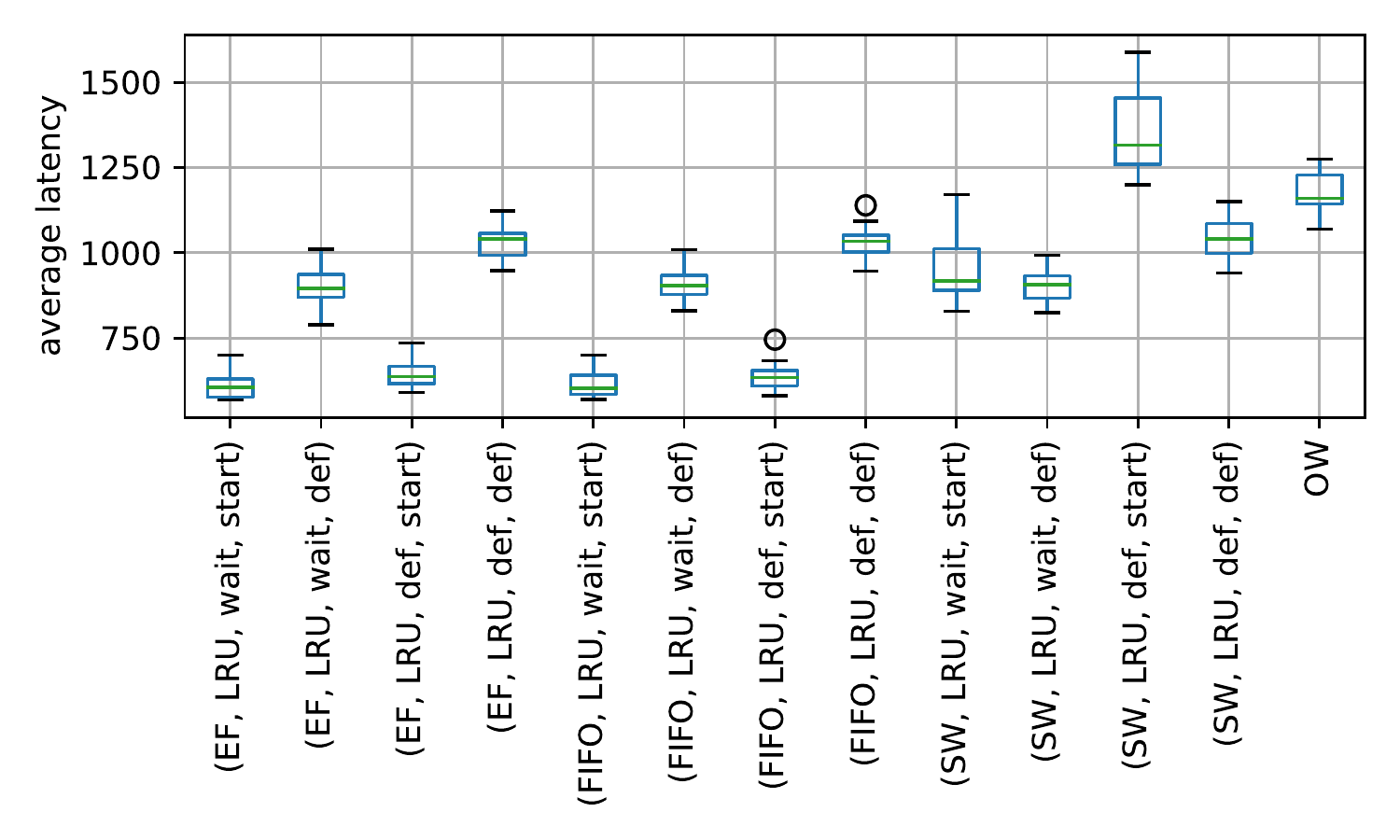}}}%
	\caption{Influence of the length of the chain.
	For all instances $n_f=50$, $m=20$, $Q=10$ with setup times 10-20.}
	\label{fig:compare_chain_len}
\end{figure*}
In the rest of the experimental section, we analyze the sensitivity of the policies to various parameters of the instance, starting with the average length of the chain. In Figure~\ref{fig:compare_chain_len}, in all instances $n_f=50$, $s_f \in [10, 20]$, $m=20$, $Q=10$ (results for larger $n_f$, $s_f$ $m$ and $Q$ are similar; we omit them due to space constraints).
All scheduling algorithms using EF as the ordering policy significantly reduce latency compared to the baseline OW (1.3-2.4x), with larger reductions for shorter chains.
The \emph{start} dependency-aware variant further reduces latency, especially for longer chains ($[50-100]$), and also for other scheduling methods (FIFO).
Therefore, for deployments with long (50 tasks and above) chains, at least 100 families, setup times 100 (and larger) with at least 20 machines of size 10 (or more), implementing dependency-aware scheduler can provide measurable benefits.

\subsection{Impact of the number of families}
\label{sec:family_count_cmp}

\begin{figure*}[t]
	\centering
	\subfloat[10 families]{{\includegraphics[width=0.33\textwidth,trim=0 11 0 10,clip]{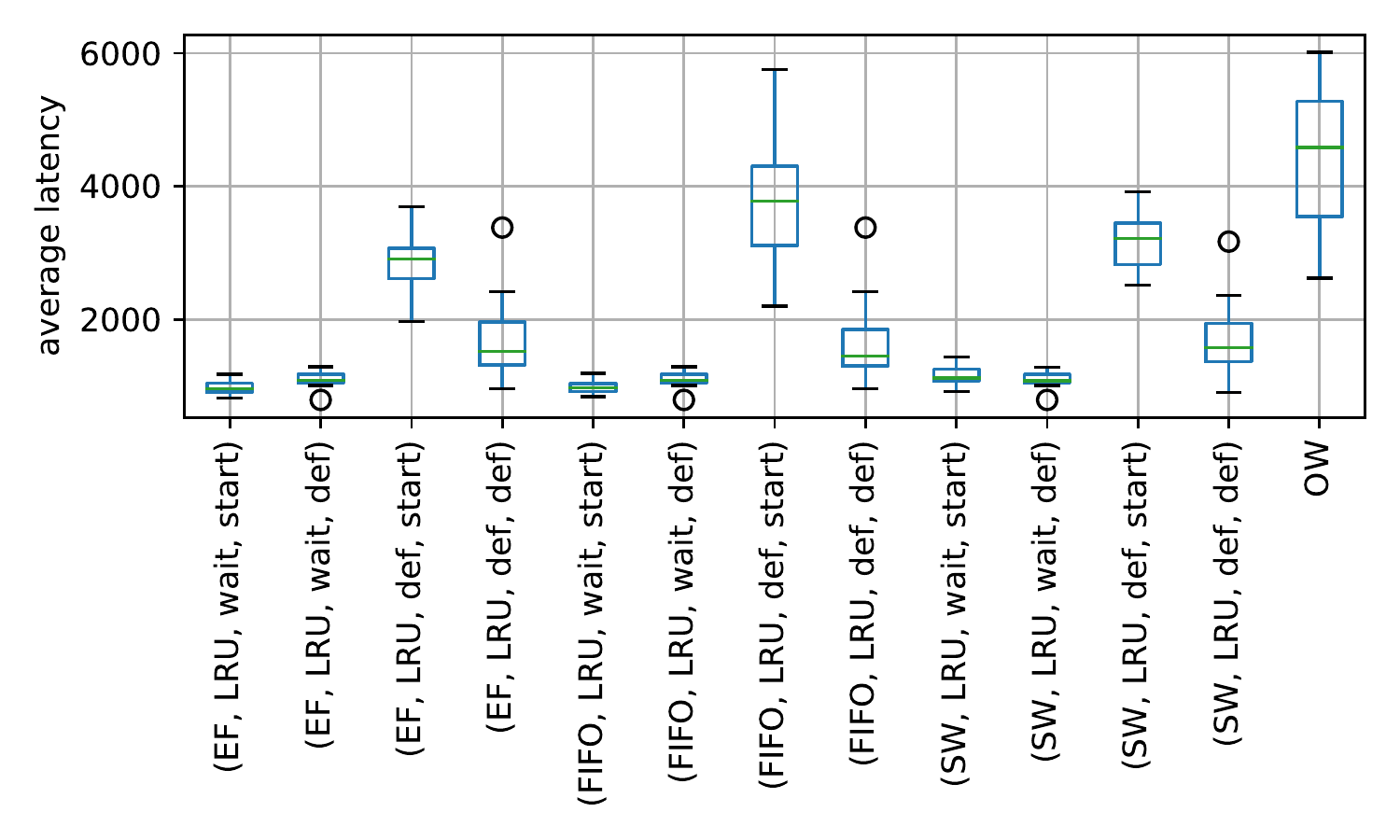}}}%
	\subfloat[50 families]{{\includegraphics[width=0.33\textwidth,trim=0 11 0 10,clip]{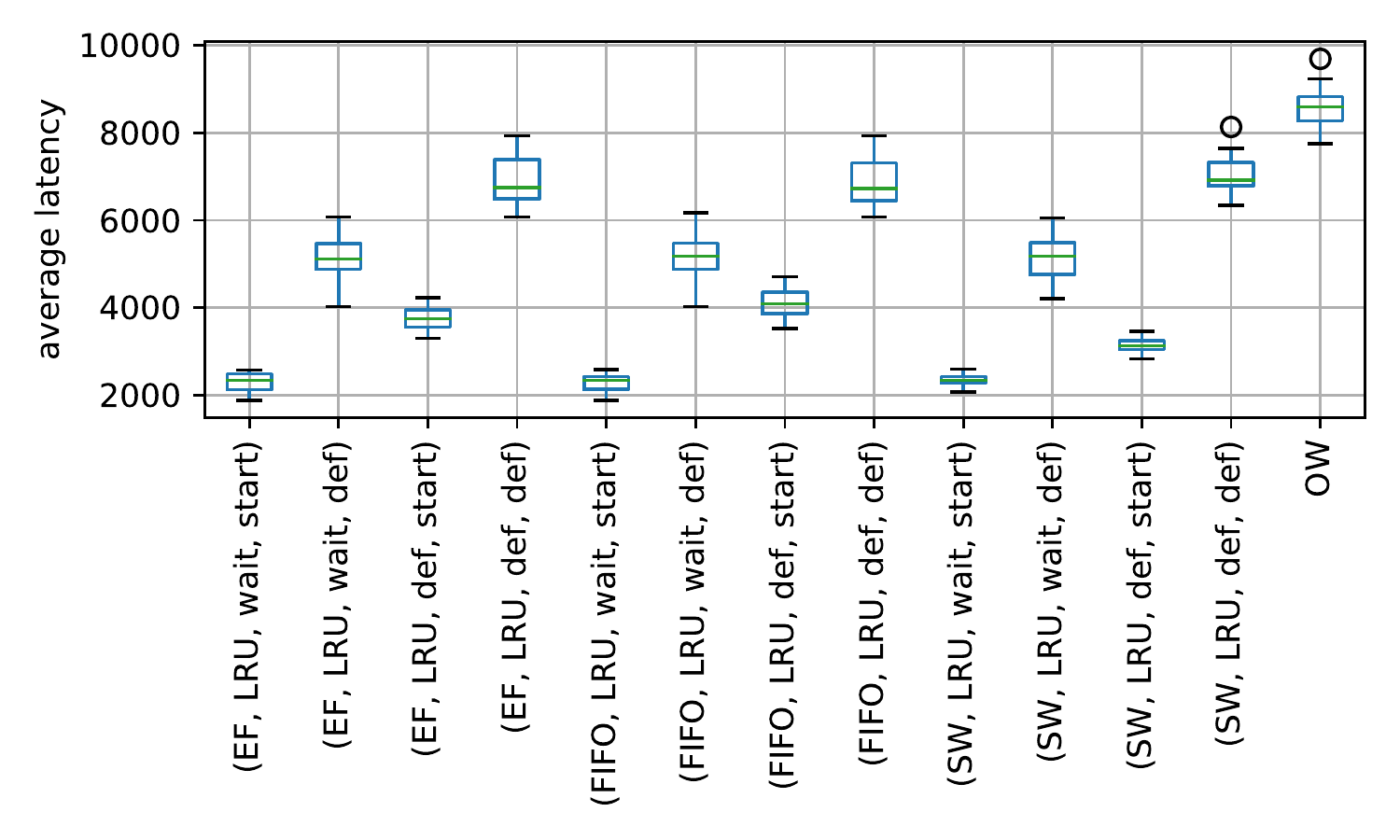}}}%
	\subfloat[200 families]{{\includegraphics[width=0.33\textwidth,trim=0 11 0 10,clip]{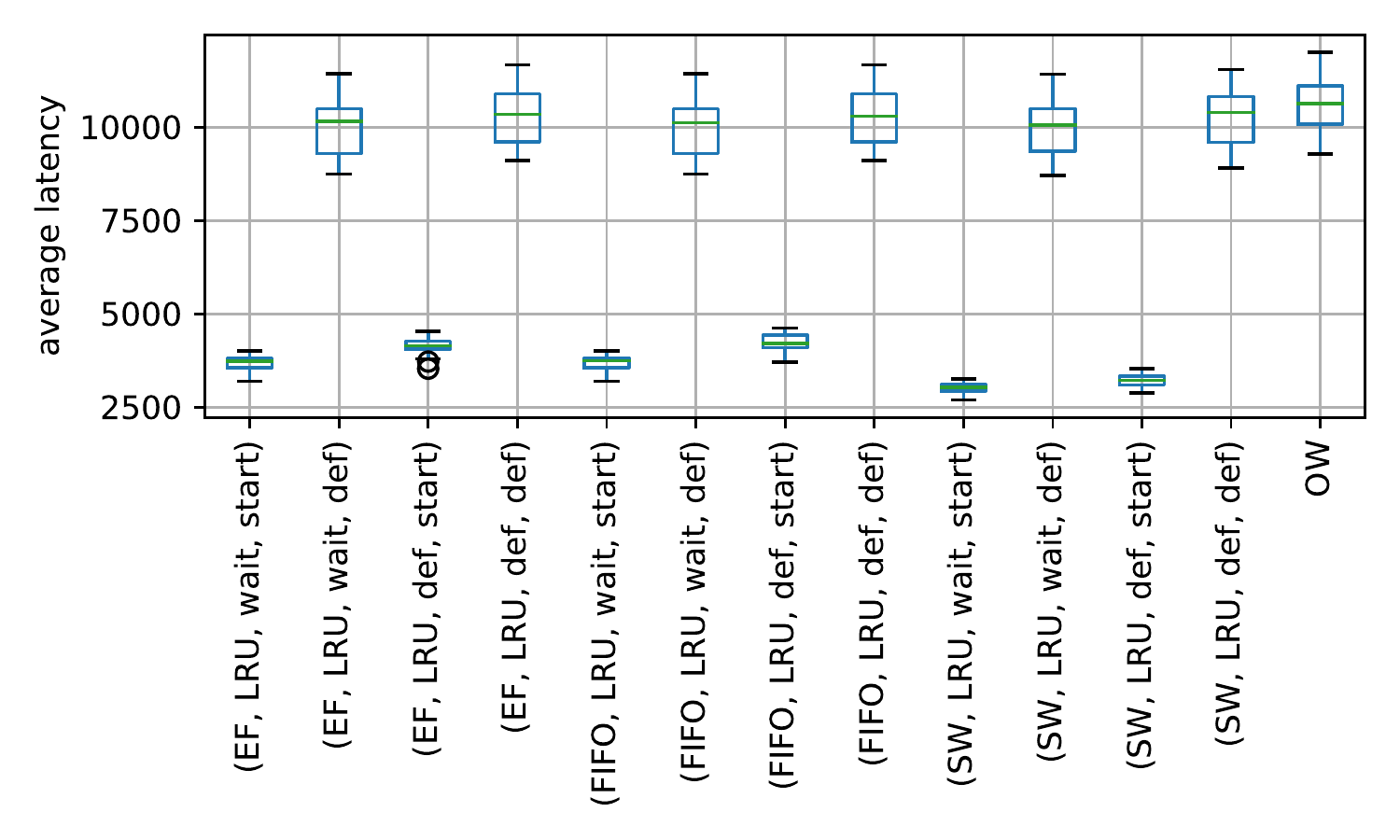}}}%
	\caption{Influence of the different number of families.
	To show general trend, we present results for 10, 50 and 200 families.
	For all instances $m=20$, $Q=10$, $s_f \in [100,200]$, $l \in [50, 100]$}
	\label{fig:compare_families}
\end{figure*}

Figure~\ref{fig:compare_families} compares results as a function of the number of task families in the system.
When the number of task families is small (up to 20), variants without dependency awareness (\emph{def}) and with \emph{wait} can give better results than dependency-aware variants.
In such cases, variants using EF method are slightly better than their equivalents using FIFO.
The same applies to the removal method: \emph{wait} variants give better results than their equivalents using plain LRU.
The higher the number of families, the higher the probability that the required type of environment is missing.
With at least $n_f=100$ families (Fig.~\ref{fig:compare_families}.c, similar results for $s_f \geq 100$, $l \geq 50$, $m \geq 20$, $Q \geq 10$ omitted due to space constraints), dependency awareness plays a crucial role -- variants using \emph{start} outperforms \emph{def} regardless of the used scheduling algorithm and removal policy.
Thus, in case of high variability of functions (i.e. requiring different environments), taking into account tasks' dependencies can significantly reduce the serving latency.

\subsection{Impact of the setup time}
\label{sec:setup_time_compare}

\begin{figure*}[tb]
	\centering
	\subfloat[setup time 0]{{\includegraphics[width=0.25\textwidth,trim=0 10 0 10,clip]{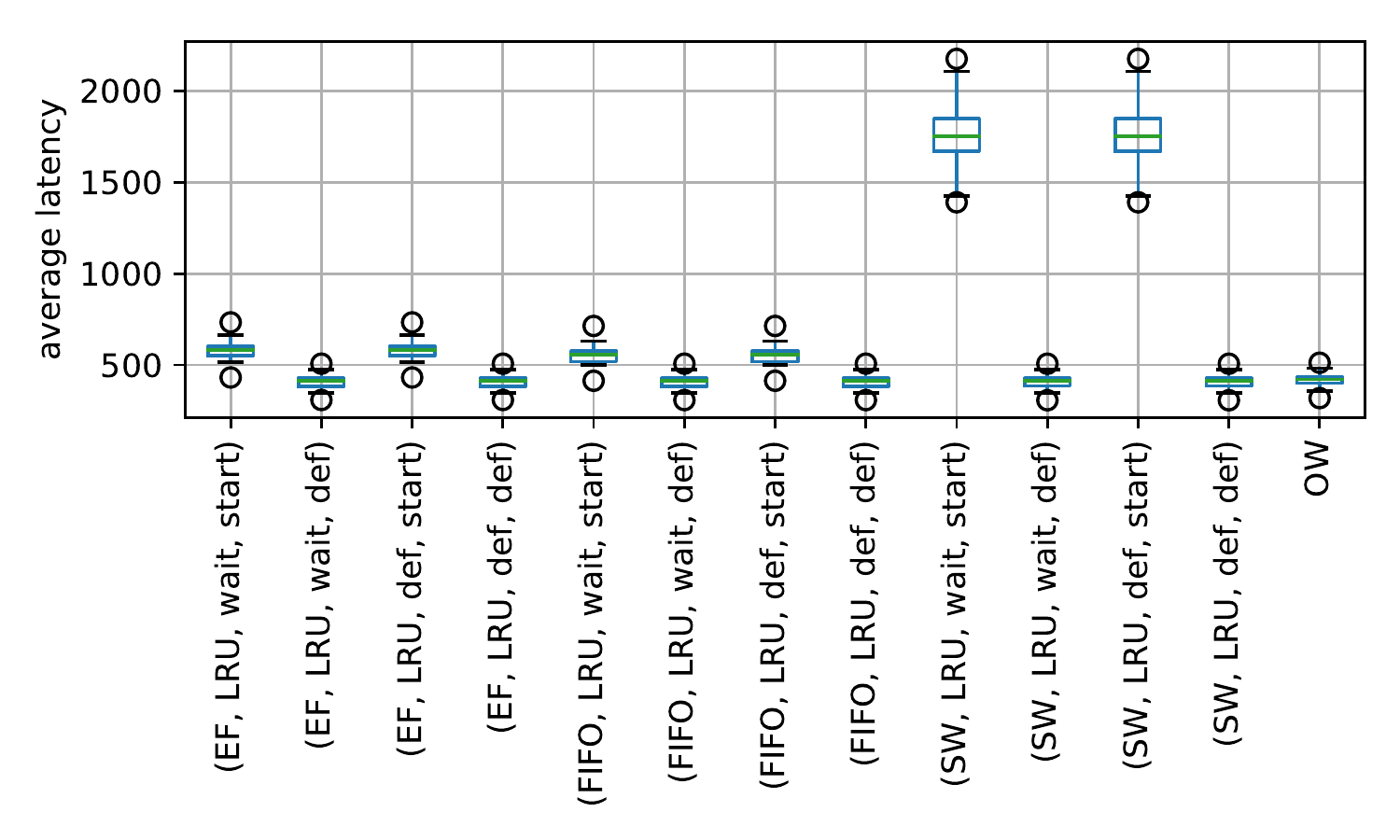}}}%
	\subfloat[setup time 10-20]{{\includegraphics[width=0.25\textwidth,trim=0 10 0 10,clip]{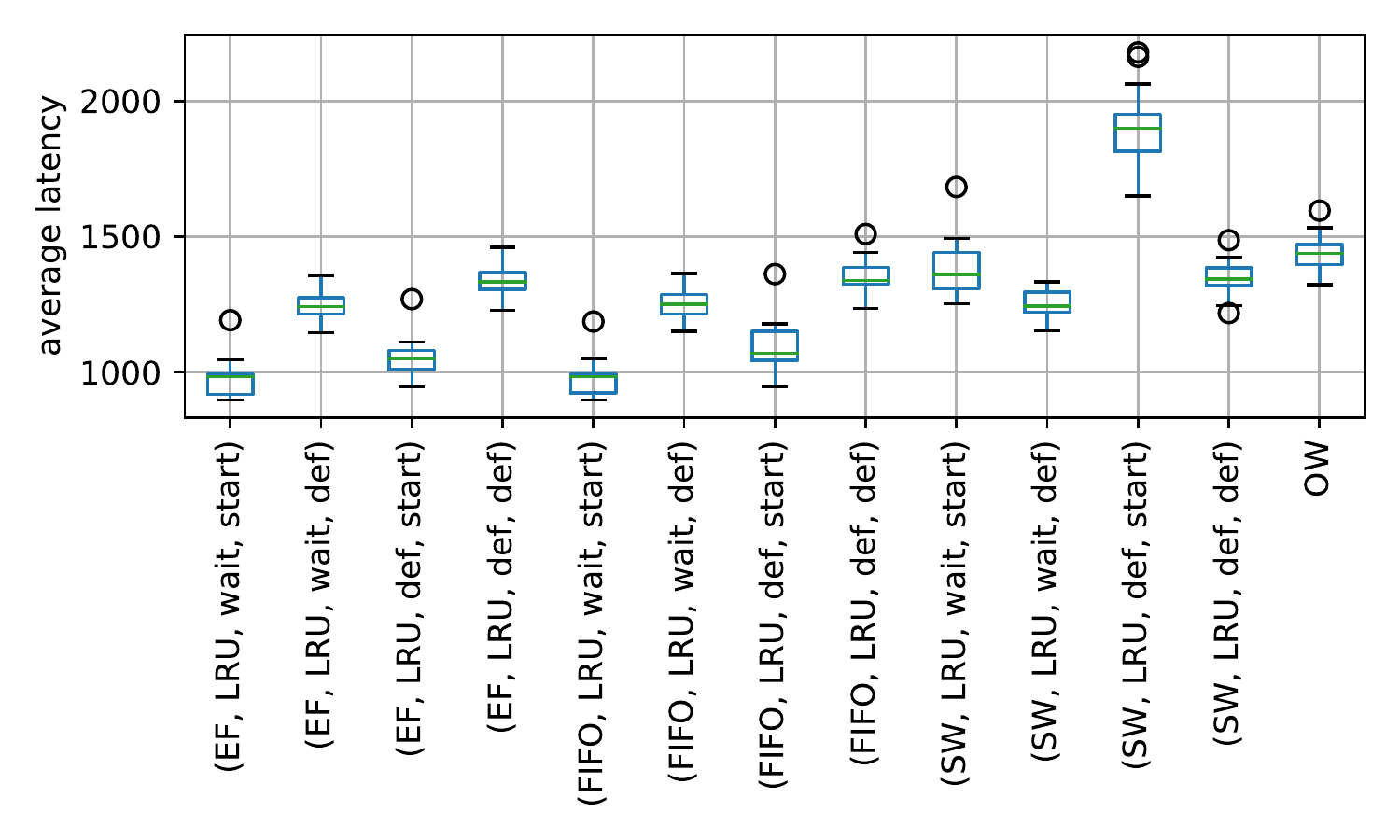}}}%
	\subfloat[setup time 100-200]{{\includegraphics[width=0.25\textwidth,trim=0 10 0 10,clip]{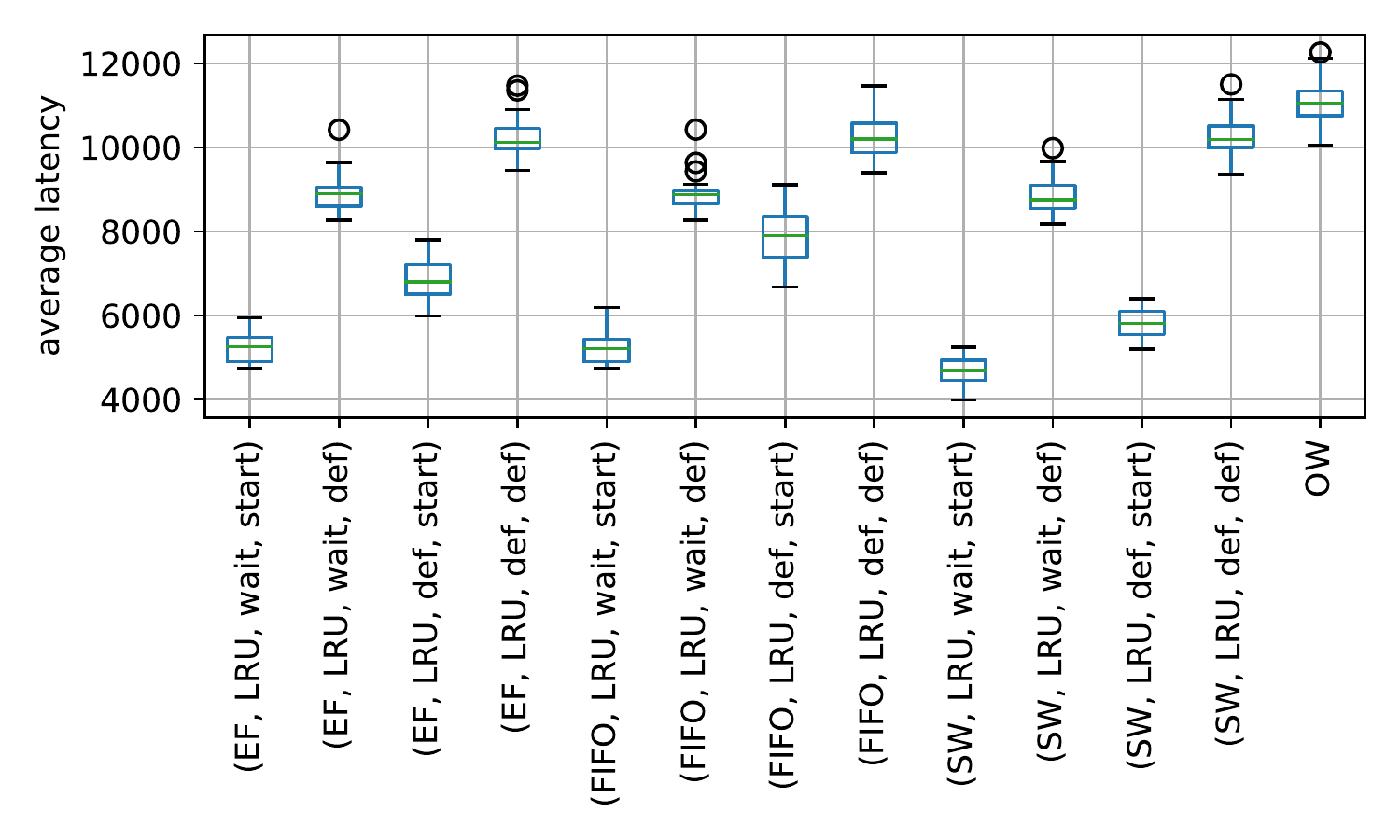}}}%
	\subfloat[setup time 1000-2000]{{\includegraphics[width=0.25\textwidth,trim=0 10 0 10,clip]{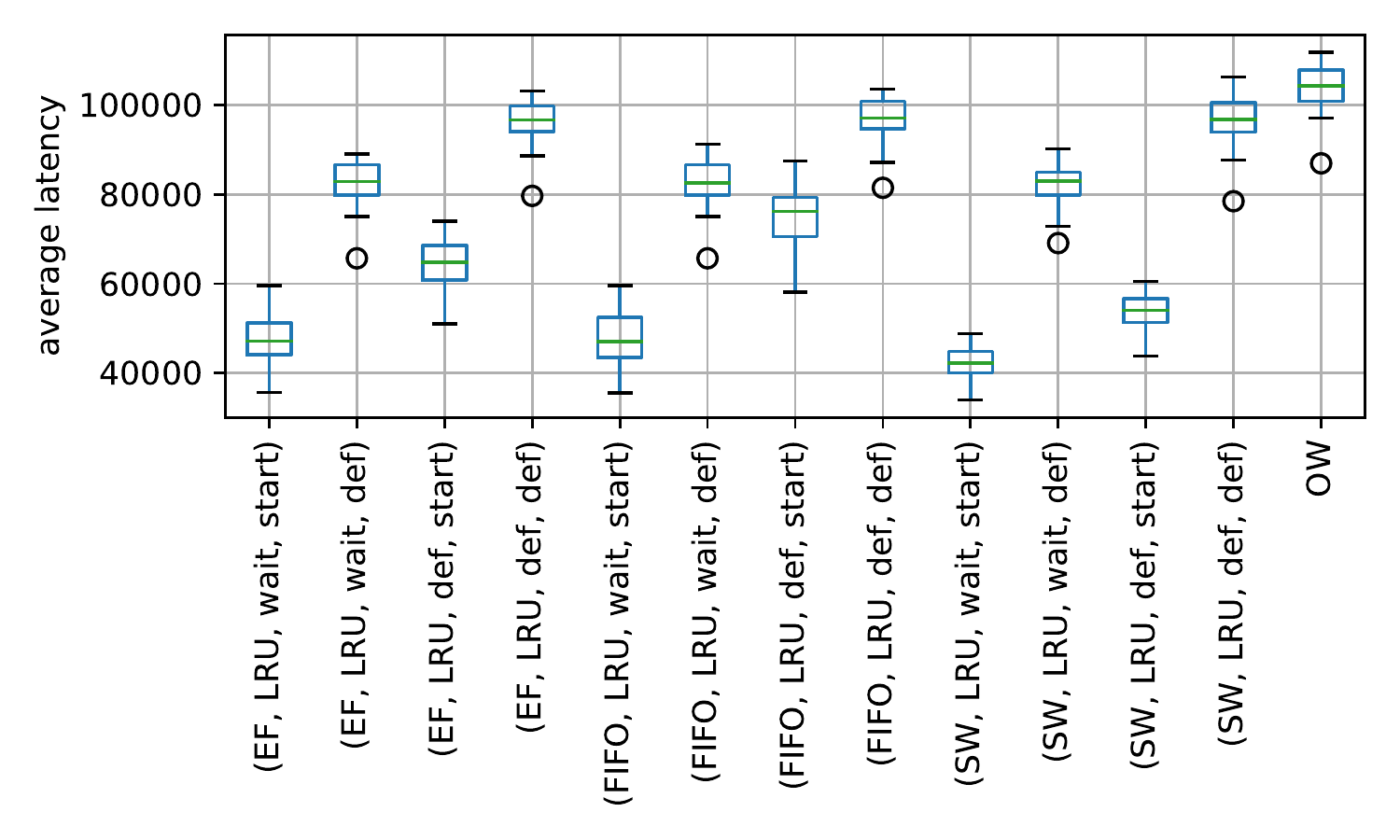}}}%
	\caption{Influence of the setup time.
	For all instances $n_f=50$, $m=10$, $Q=10$, $l \in [50, 100]$.}
	\label{fig:compare_setup_times}
\end{figure*}

Figure~\ref{fig:compare_setup_times} compares results as a function of different setup time ranges.
In the edge case with no setup times, $s_f=0$, we see no difference between the \emph{waiting} and the non-\emph{waiting} variants, 
as there is no additional penalty for inefficient environment re-creation.
Similarly, there are no differences between EF and FIFO. 
For non-zero setup times, dependency awareness (\emph{start}) reduces the latency.
However, with no setup time, \emph{start} latencies are \emph{longer}.
This behavior is caused by adding tasks with future release time to the queue (see Section~\ref{sec:awareness-dependencies}). 
Consider two jobs each of two tasks:
\begin{enumerate*}
	\item a \emph{long} job with task \emph{A} (duration 10) followed by task \emph{B} (duration 1);
	\item a \emph{short} job with task \emph{C} (duration 1), followed by task \emph{B} (same as in \emph{''long''}).
\end{enumerate*}
\emph{EF} and \emph{FIFO} using \emph{start} variants may assign the second task from the \emph{long} job to the environment of type B immediately after assigning the first task.
This might block the second task from the \emph{short} job until $t=11$; while the optimal schedule starts this task at $t=1$.
For the same reason, \emph{start} has worse results when there are more jobs (i.e. shorter chains) and the systems are smaller (less machines, smaller capacities).
%
%

We further investigate for which instance parameters the dependency-aware \emph{start} dominates the myopic \emph{def}, assuming non-negligible setup times $s_f \geq 100$.
We aggregate results by all simulation parameters (count of families $n_f$, machines $m$, machine sizes $Q$, range of chain lengths $l$, range of setup times $s_f$ and used algorithm variant) and compute the median average latency among 20 instances. 
Then we analyze in how many of resulting cases changing \emph{def} to \emph{start}  improves performance.
For long chains ($l \geq 50$), many task families ($n_F>100$), and many machines $(m \geq 10)$, changing the default (\emph{def}) variant to dependency-aware one improves performance in all cases. 

\subsection{Impact of machine capacity}
\label{sec:machine_cmp}
\begin{figure*}[tb]
	\centering
	\subfloat[5 machines, size 20]{{\includegraphics[width=0.5\columnwidth,trim=0 10 0 10,clip]{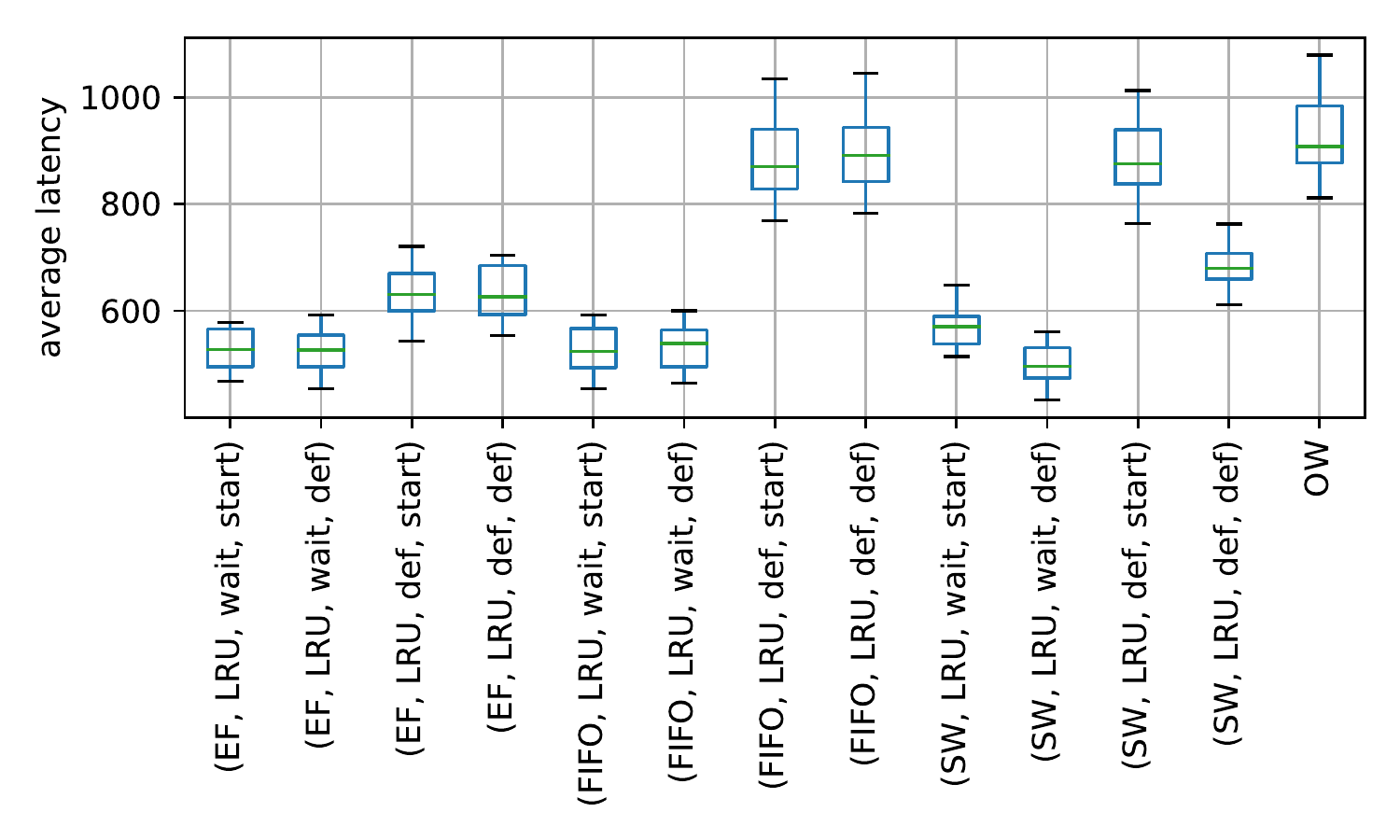}}}%
	\subfloat[20 machines, size 20]{{\includegraphics[width=0.5\columnwidth,trim=0 10 0 10,clip]{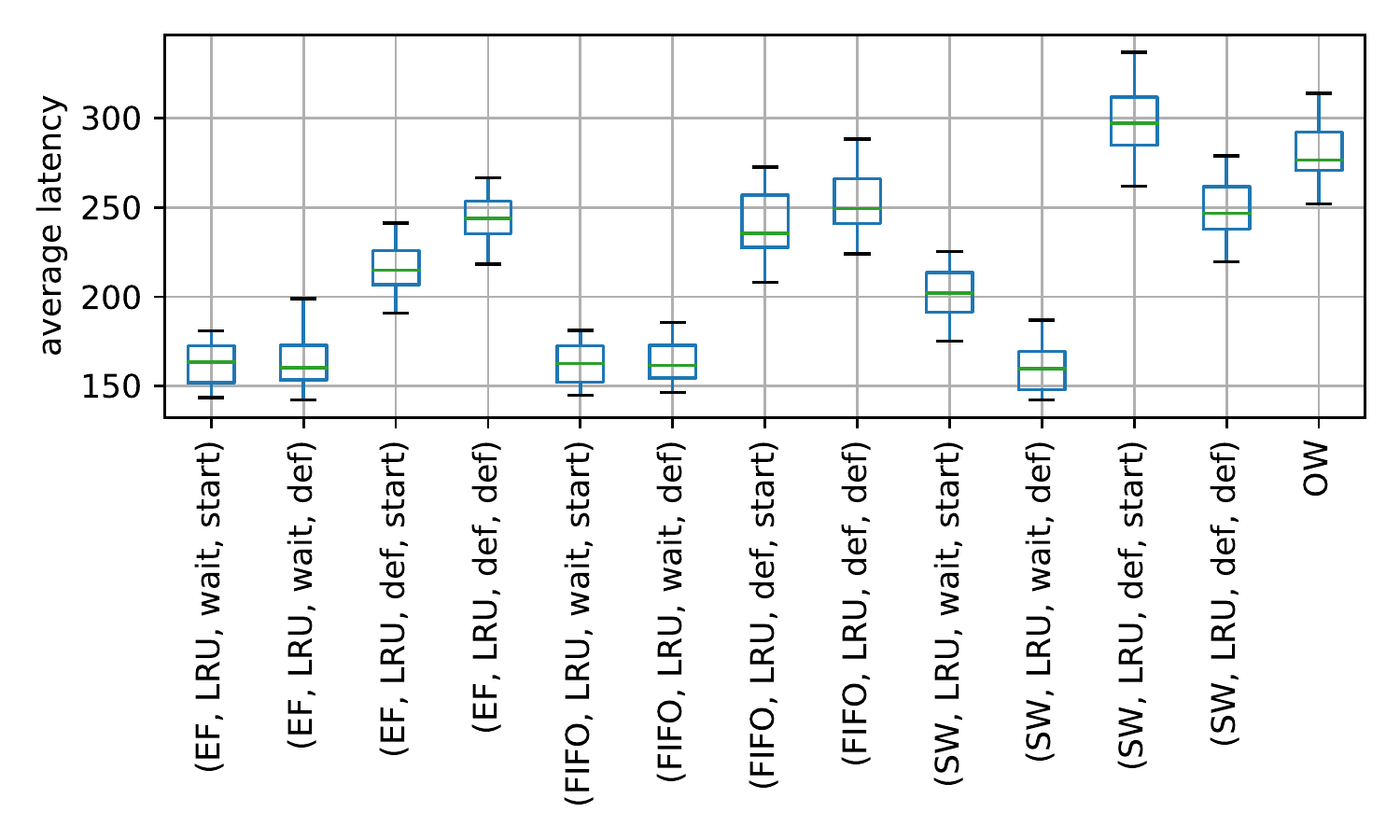}}}%
	\subfloat[50 machines, size 10]{{\includegraphics[width=0.5\columnwidth,trim=0 10 0 10,clip]{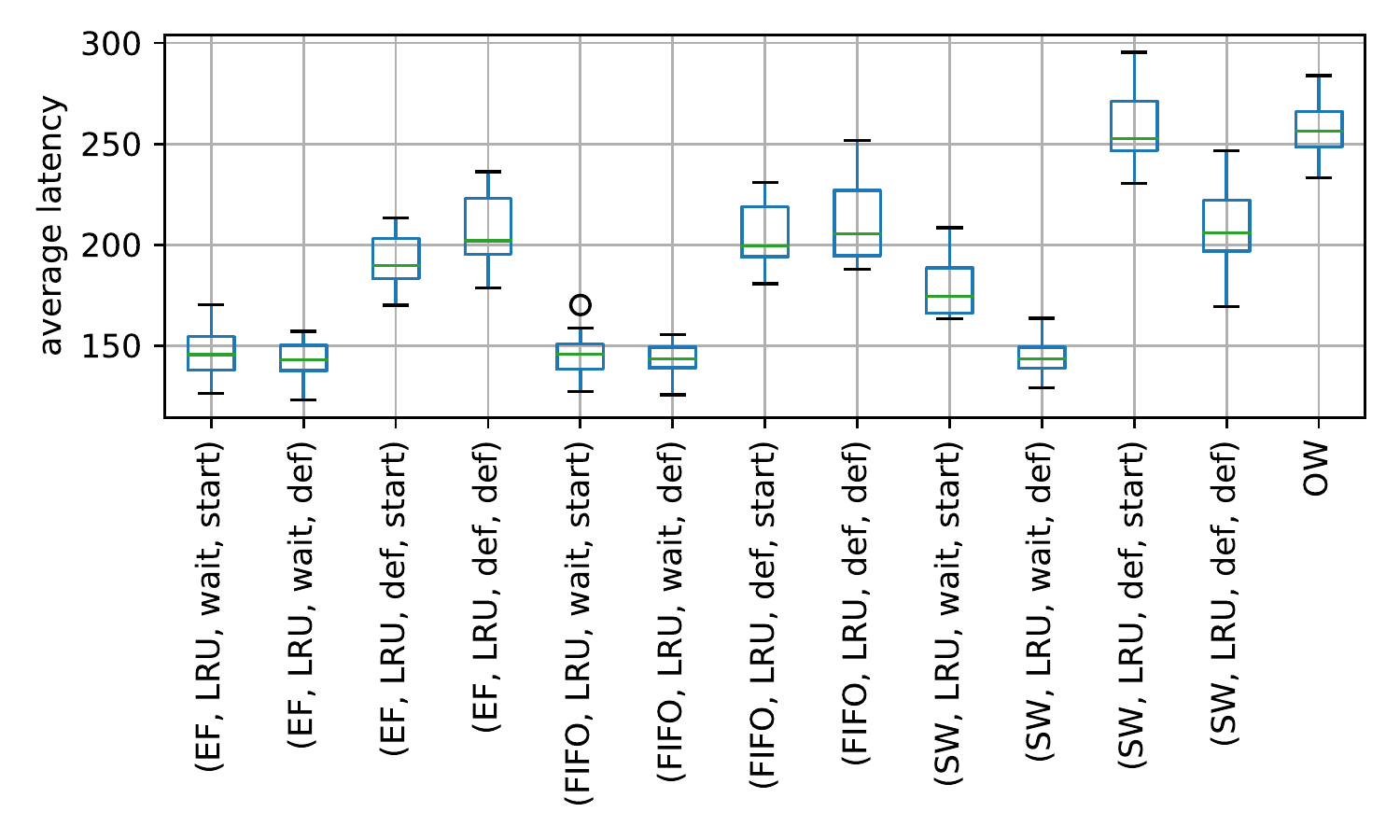}}}%
	\subfloat[50 machines, size 50]{{\includegraphics[width=0.5\columnwidth,trim=0 10 0 10,clip]{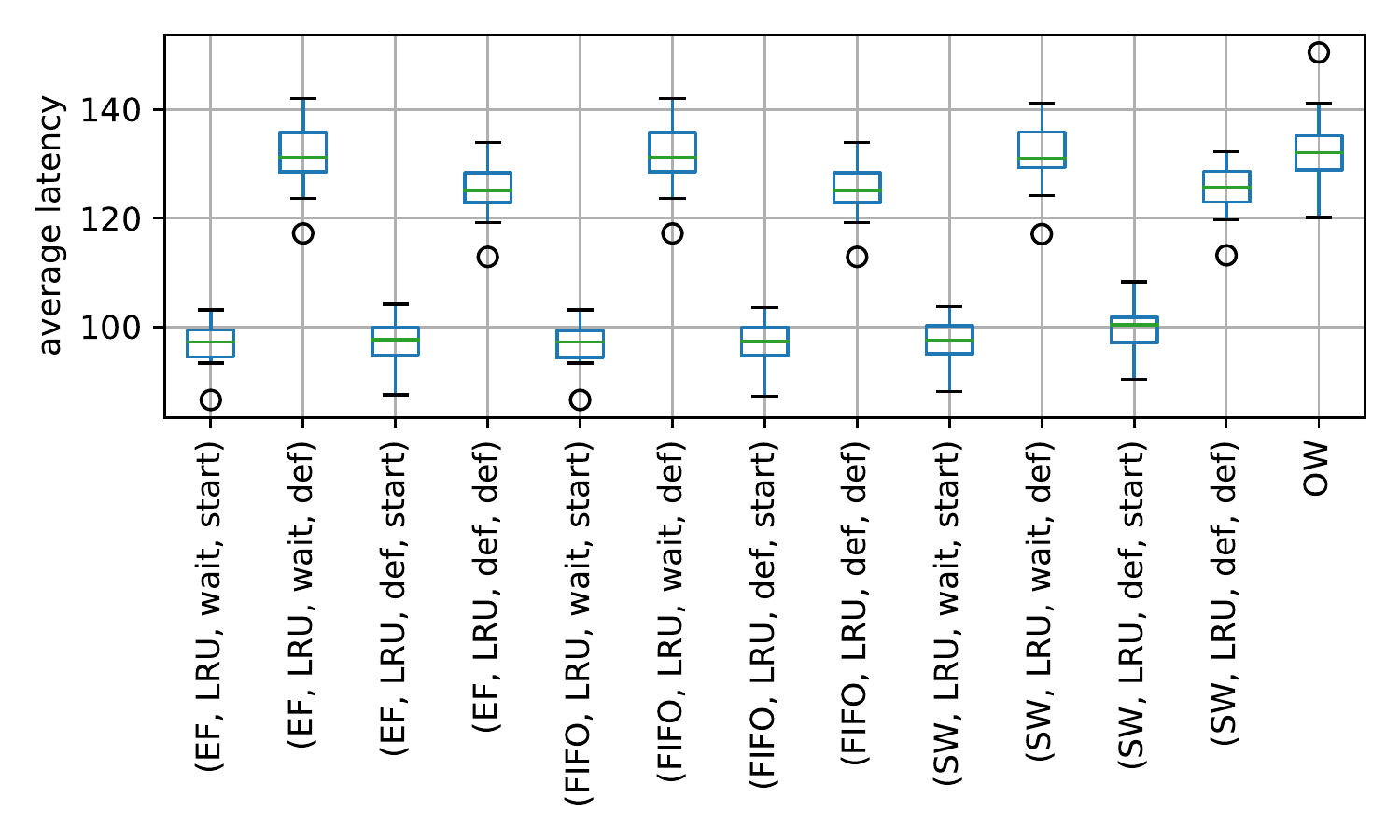}}}%
	\caption{Influence of the machine environment.
	For all instances $f_n=50$, $l\in[10,20]$, $s_f\in[10,20]$}
	\label{fig:compare_m_ms}
\end{figure*}

Figure~\ref{fig:compare_m_ms} compares results as a function of the number of machines and their size.
For all instances $n_f=50$, $l \in [10,20]$, $s_f\in[10, 20]$.
To show general trend and ensure clarity, out of 15 considered machine configurations we present results only for instances with $(m,Q)\in\{(5, 20), (20, 20), (50, 10), (50, 50)\}$.
For cases up to $(m, Q)=(5,20)$, the only observable differences between the plain and dependency-aware variants are for SW scheduling policy.
Due to large number of jobs (chain lengths are in range 10-20), when dependent tasks are added to the queue earlier, environments may get blocked as described in Section~\ref{sec:setup_time_compare}, therefore there is no additional benefit of dependency-awareness.
For capacities up to $(m, Q)=(50,10)$, using \emph{wait} variants outperform the default (def) variants using the same scheduling algorithm and with the same setting of dependency-awareness.
In all presented cases, for \emph{FIFO} and \emph{EF} scheduling policies, variants using \emph{wait} with \emph{start} have one of the lowest average latency.
The improvement on overall system performance is most visible in the case of highly-overloaded machines.
Therefore, our methods could be used to improve handling of situation when datacenter has to handle rapid increase (peak) of requests.

\section{Related work}
\label{sec:related}
Our model of FaaS resource management combines scheduling (with setup times and dependencies)~\cite{allahverdi2015third} with multiple knapsack (when environments of different sizes must fit into machines).
Our simulation results show that all these aspects have to be taken into account by the scheduler (the baseline \emph{OW} is consistently dominated by our policies).
Individually, these are classic problems in combinatorial optimization. Allahverdi~\cite{allahverdi2015third} performs a comprehensive review of about 500 papers on scheduling with setup times. Brucker~\cite{brucker2007scheduling} reviews scheduling results.
Below we describe only the applications in serverless and cloud computing.

\emph{Bin packing with setup times:}
With no dependencies, our problem reduces to bin packing with sequence independent setup times.
Weng et al.~\cite{weng2001unrelated} study similar problem of minimizing mean weighted completion time in case of tasks with sequence dependent setup times.
\cite{webster2001dynamic} presents dynamic algorithms addressing scheduling with setup times with objective of minimal weighted flow time.

\emph{Quadratic programming:}
We proposed heuristics, rather than using generic solvers or metaheuristics. Initially, we considered encoding our problem as an (integer) quadratic programming.
Nevertheless, Gurobi~\cite{gurobi} 
was unable to find an optimal schedule in 15 minutes (on a reasonable desktop machine) even for a small instance with $N=20$ jobs each of $n_l=20$ tasks. Schedulers in production systems need to respond in seconds, thus an approach based on a generic solver is probably not sufficient.

\emph{Workflow scheduling:}
With $s_f=0$ and task sizes equal to machine capacities $q_f=Q$, the problem reduces to workflow scheduling.
\cite{ilyushkin2018impact} measures how inaccurate runtime estimates influence the schedules --- which complements our study (as we assumed that estimates are known). 
\cite{10.1145/3337821.3337872} analyzes possible performance benefits of resource interleaving across the parallel stages.
\cite{sakellariou2004hybrid} proposes Balanced Minimum Completion Time, an algorithm for scheduling tasks with dependencies (and without setup times) on heterogeneous systems.
\cite{gacias2010parallel} schedules workflows with setup times using branch-and-bound. 
 While they considered small instances (up to $N*n_l=100$ task and $m=4$ machines); their method required 100s time limit for execution. Such long running times makes this method unusable in data-center schedulers.
A comprehensive survey on workflow scheduling in the cloud is presented in \cite{wu2015workflow}.
\cite{afzalirad2016resource} analyzes scheduling tasks with sequence-dependent setup times, precedence constraints, release dates on unrelated machines with resource constraints and machine eligibility.
The authors present two solutions: based on genetic algorithm and based on an artificial immune system.
Their largest instances had 60 tasks and 8 machines and needed 25 minutes (on the average) to solve, again rendering these methods unusable for FaaS.

\section{Conclusions}
\label{sec:summary}

%
Our experimental results clearly show that the performance of FaaS can be improved by considering the composition of functions and installing environments in advance. 
The \emph{EF} ordering prioritizes tasks that can be started using already prepared environments. 
The \emph{waiting} variant binds the task to the existing, used environment if such environment will be ready to process task earlier than a newly-created one. The \emph{start} variant adds successors of the task to the queue as soon as their release times could be determined.
For non-negligible setup times ($s_f \geq 100$, or at least 20 times longer than the average task duration), larger systems ($m \geq 10$ machines with $Q \geq 10$, or hosting at least two average-sized environments) for 80.7\% of cases, the average response latencies are reduced \emph{by the factor of two} when the scheduler is dependency- and startup-times aware  $(\cdot, \cdot, wait, start)$, compared to the baseline (OW).

Compared with the baseline, dependency- and startup-times aware scheduling is more efficient when the load of the system is high. Our methods can be used to mitigate the impact of the increased demand in the short term. If the demand increase is longer-term, the underlying infrastructure will be eventually scaled out by, e.g., adding new VMs. However, such scale-out takes considerably longer time (minutes); meanwhile, the load has to be handled.

Although our experiments were offline, the \emph{waiting} variant and the \emph{start} variant can be easily implemented in the existing FaaS schedulers (controllers). Changing the invocation order (as in SJF and EF variants) is less straightforward, as when new jobs arrive on-line,
existing jobs might be starved: these policies would additionally need to consider fairness. 
Alternatively, as our results show, \emph{waiting} and \emph{start} variants are beneficial even with the standard FIFO ordering.

Finally, while FaaS was the main motivation of this work, these ideas can be applied also in other systems executing workflows on shared machines (a machine executing multiple tasks in parallel), such as Apache Beam.

\section*{Acknowledgments}
This research is supported by a Polish National Science Center grant Opus (UMO-2017/25/B/ST6/00116).

\bibliographystyle{IEEEtran}
\bibliography{bibliography}

\ifdefined\arxivpreprint
\appendix
\ifdefined\arxivpreprint
\newcommand{\sectionhead}[1]{\section*{#1}}
\else
\documentclass[conference]{IEEEtran}
\IEEEoverridecommandlockouts
\usepackage{cite}
\usepackage{amsmath,amssymb,amsfonts}
\usepackage{algorithm}
\usepackage{algorithmicx}
\usepackage[noend]{algpseudocode}
\usepackage{float}
\usepackage[inline]{enumitem}
\usepackage{graphicx}
\usepackage{textcomp}
\usepackage{xcolor}

\usepackage[utf8]{inputenc}
\usepackage{url}
\usepackage{subfig}

\MakeRobust{\Call}
\newcommand\fixme[1]{
	\textcolor{blue}{{{\em\bf{[{\sc{{Fixme:}}} #1]}}}}
	}

\begin{document}
\title{Scheduling Methods to Reduce Response Latency of Function as a Service - Appendix}

\maketitle
\newcommand{\sectionhead}[1]{\section{#1}}
\fi
\sectionhead{Validation of algorithms performance on DAGs}
In this paper we focus on chains, as they are simplest function compositions which allows us to analyze impact of proposed optimizations.
Moreover chains are particularly interesting as they are directly supported by OpenWhisk, therefor our research can be applied to the real system. 

Nevertheless, a single function is able to spawn arbitrary number of other functions by connecting directly to the platform API.
While spawning new function using API is straightforward, defining function that has more than one predecessor without direct platform support is more sophisticated, as it requires e.g. to store information which of the predecessors completed their execution.
Therefore, we are particularly interested in out-trees.

In general, executing DAGs by appending to each function code invoking successors using platform's API, hides structure of the DAG from scheduler.
In this section we assume that scheduler has information about defined DAGs.
We validate if results obtained for chains are applicable to more generic DAGs.

We generate out-trees using our existing dataset using following procedure: for each task in job (chain), we select randomly new parent from preceding tasks in the same chain.
Therefore a single task may have more than one successor, while each of the tasks still have at most one predecessor.

\begin{figure*}[!tb]
	\centering
	\subfloat[2-10 tasks]{{\includegraphics[width=0.33\textwidth]{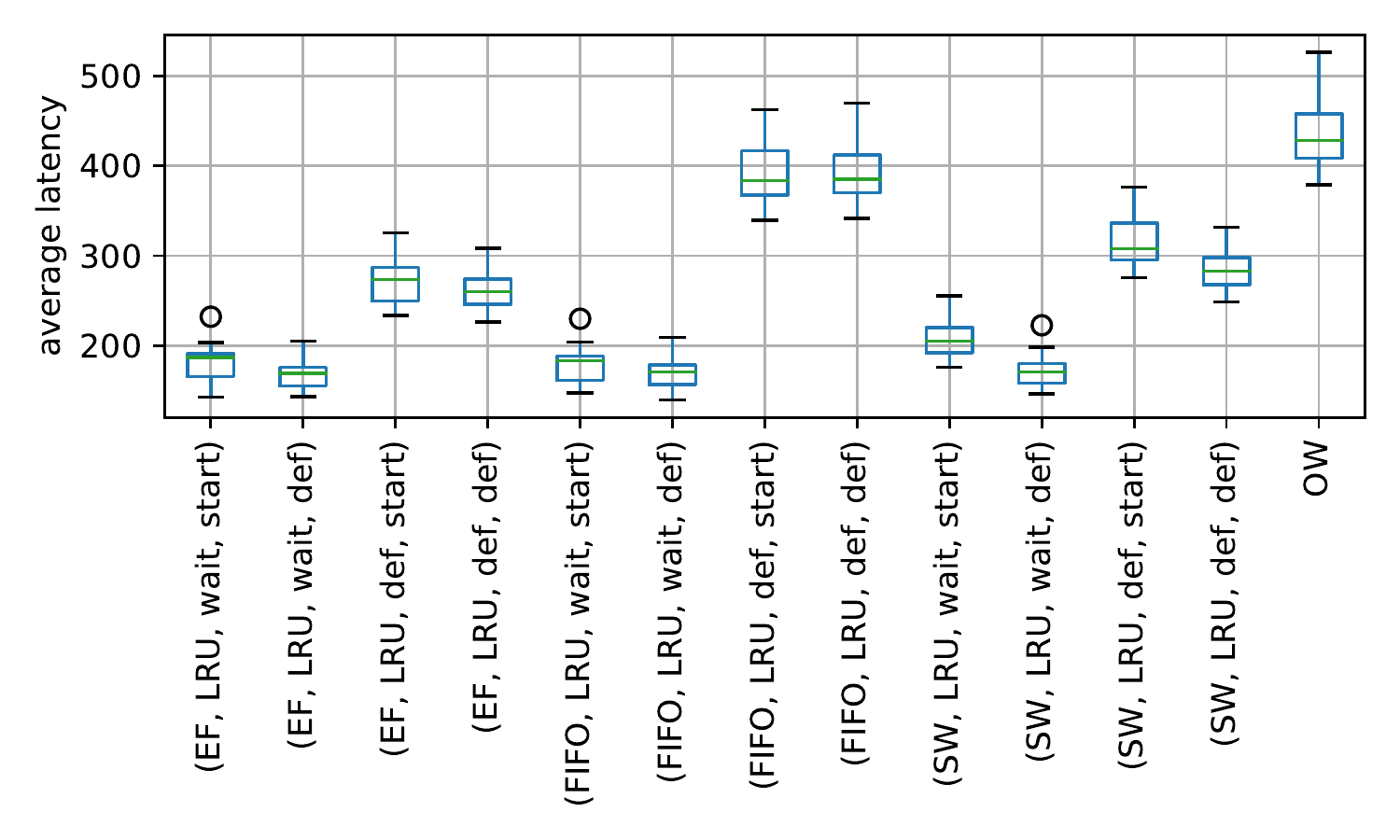}}}%
	\subfloat[10-20 tasks]{{\includegraphics[width=0.33\textwidth]{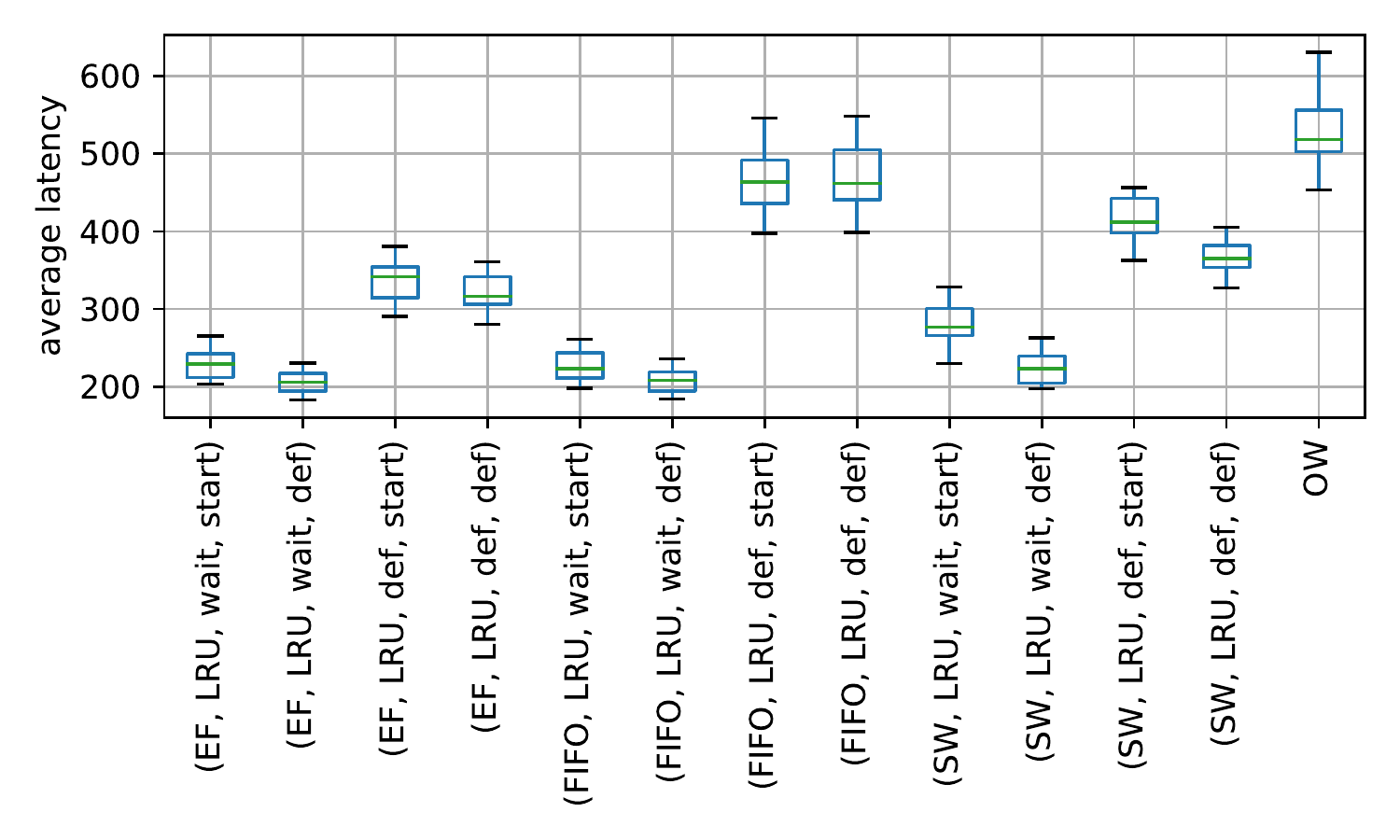}}}%
	\subfloat[50-100 tasks]{{\includegraphics[width=0.33\textwidth]{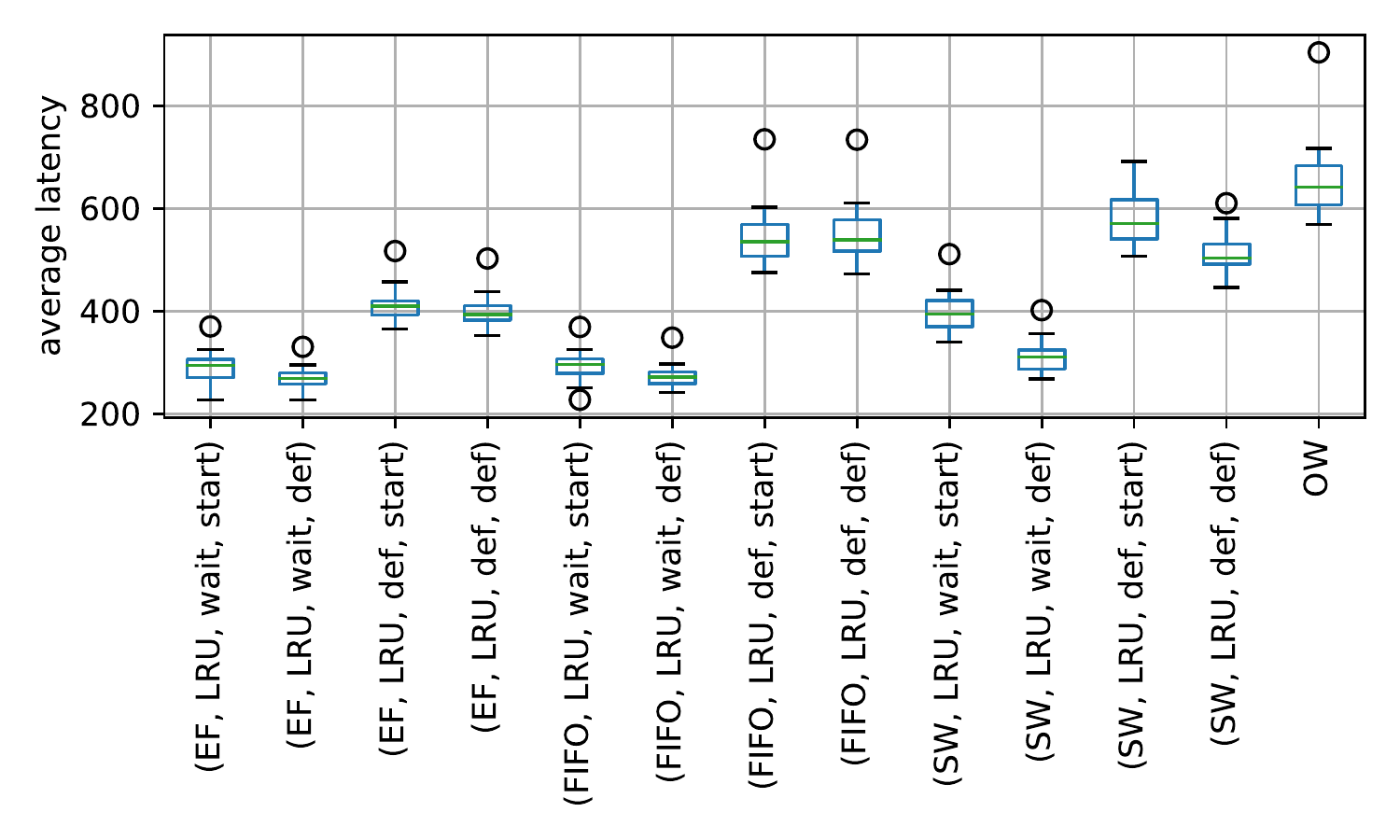}}}%
	\caption{Comparison of different job sizes in DAG.
	In each case dataset contains 50 families.  
	Setup times 10-20, 20 machines of size 10.}
	\label{fig:dag_compare_chain_len}
\end{figure*}

\begin{figure*}[!tb]
	\centering
	\subfloat[setup time 0]{{\includegraphics[width=0.25\textwidth,trim=0 10 0 10,clip]{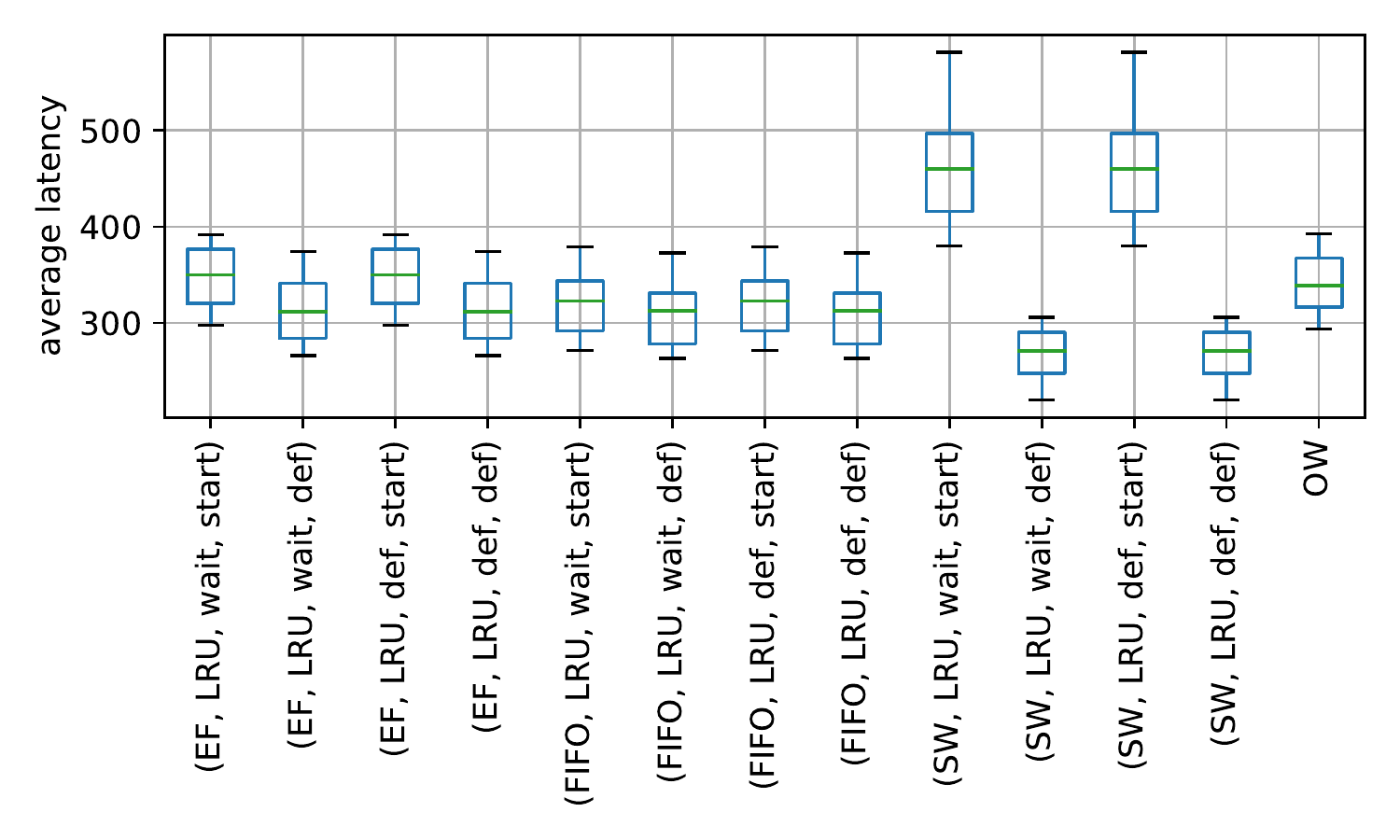}}}%
	\subfloat[setup time 10-20]{{\includegraphics[width=0.25\textwidth,trim=0 10 0 10,clip]{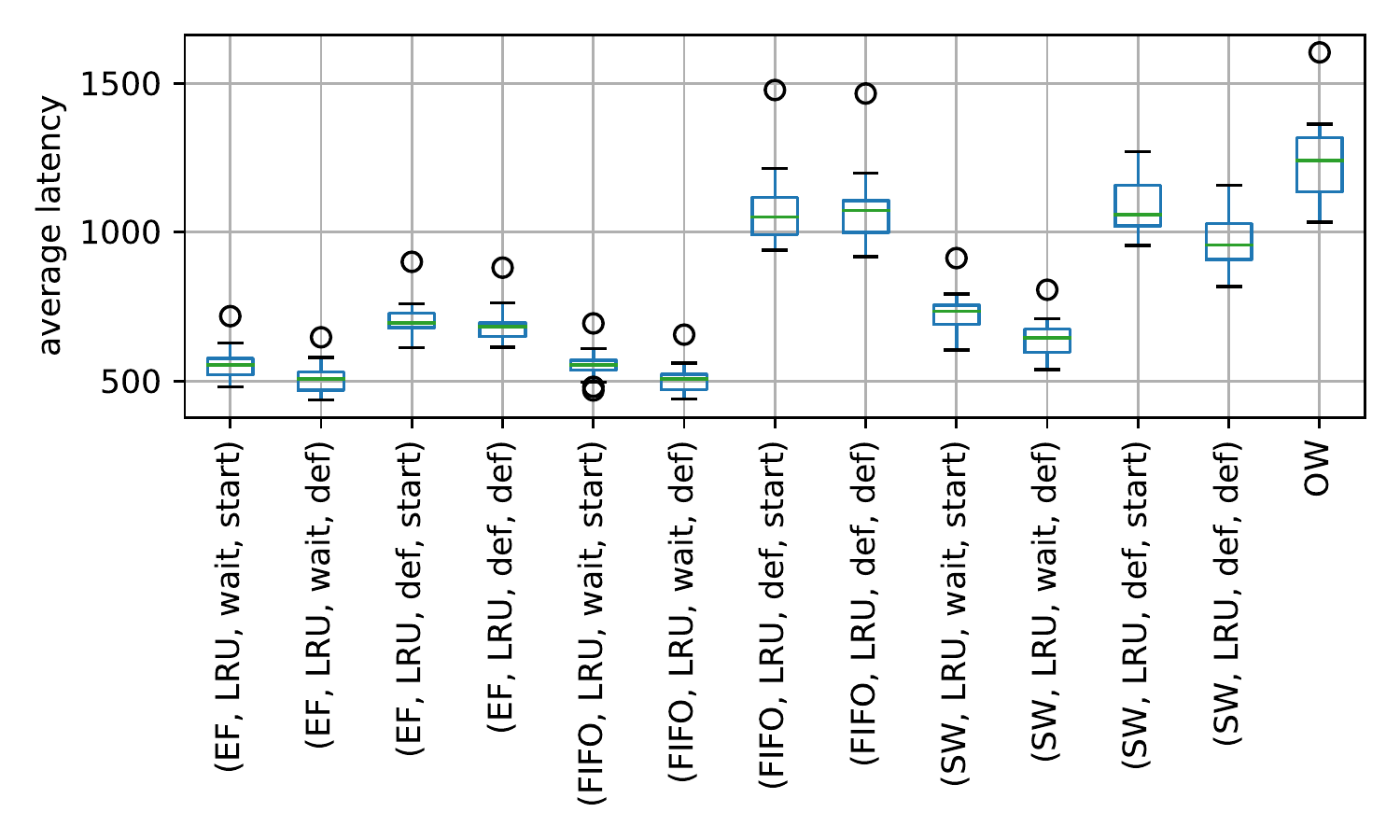}}}%
	\subfloat[setup time 100-200]{{\includegraphics[width=0.25\textwidth,trim=0 10 0 10,clip]{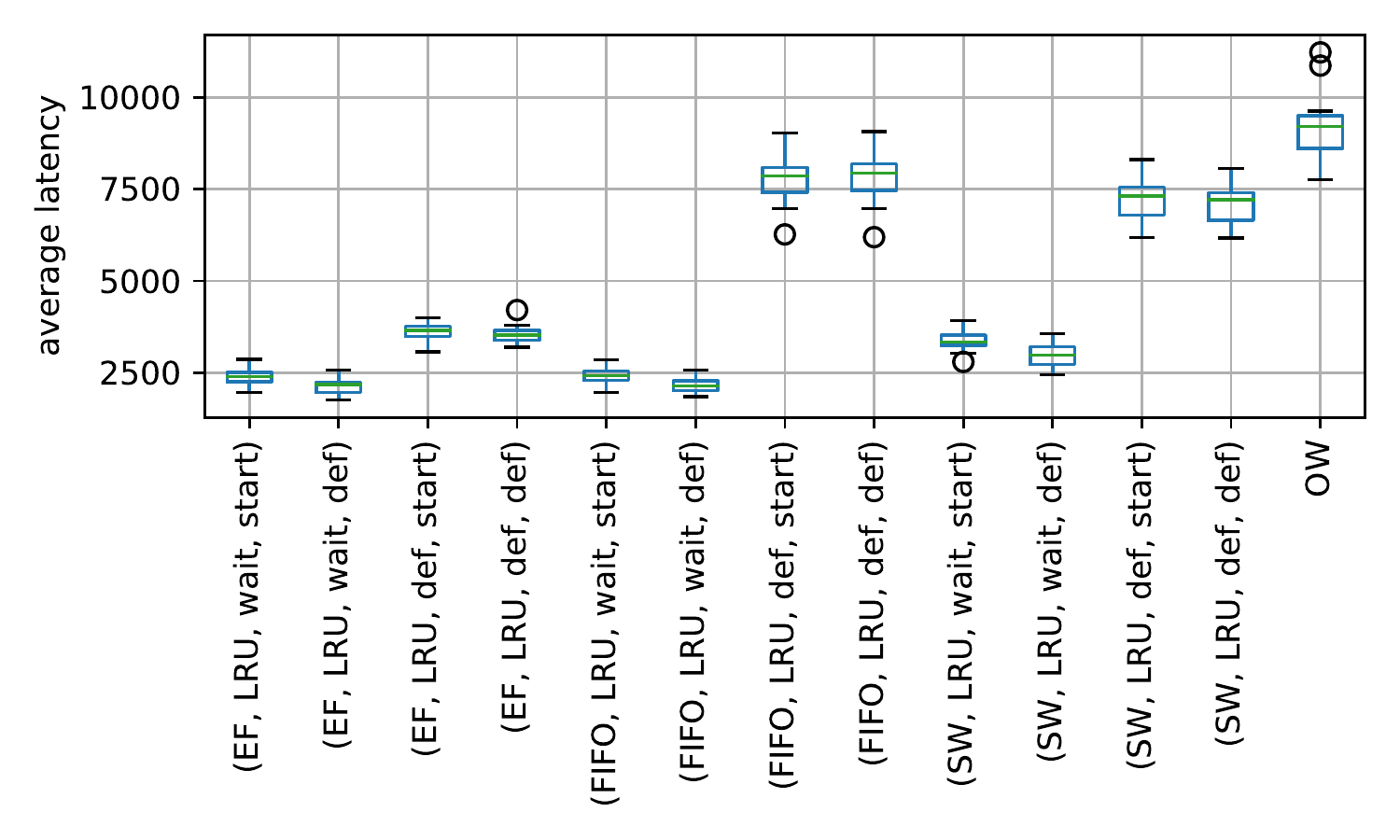}}}%
	\subfloat[setup time 1000-2000]{{\includegraphics[width=0.25\textwidth,trim=0 10 0 10,clip]{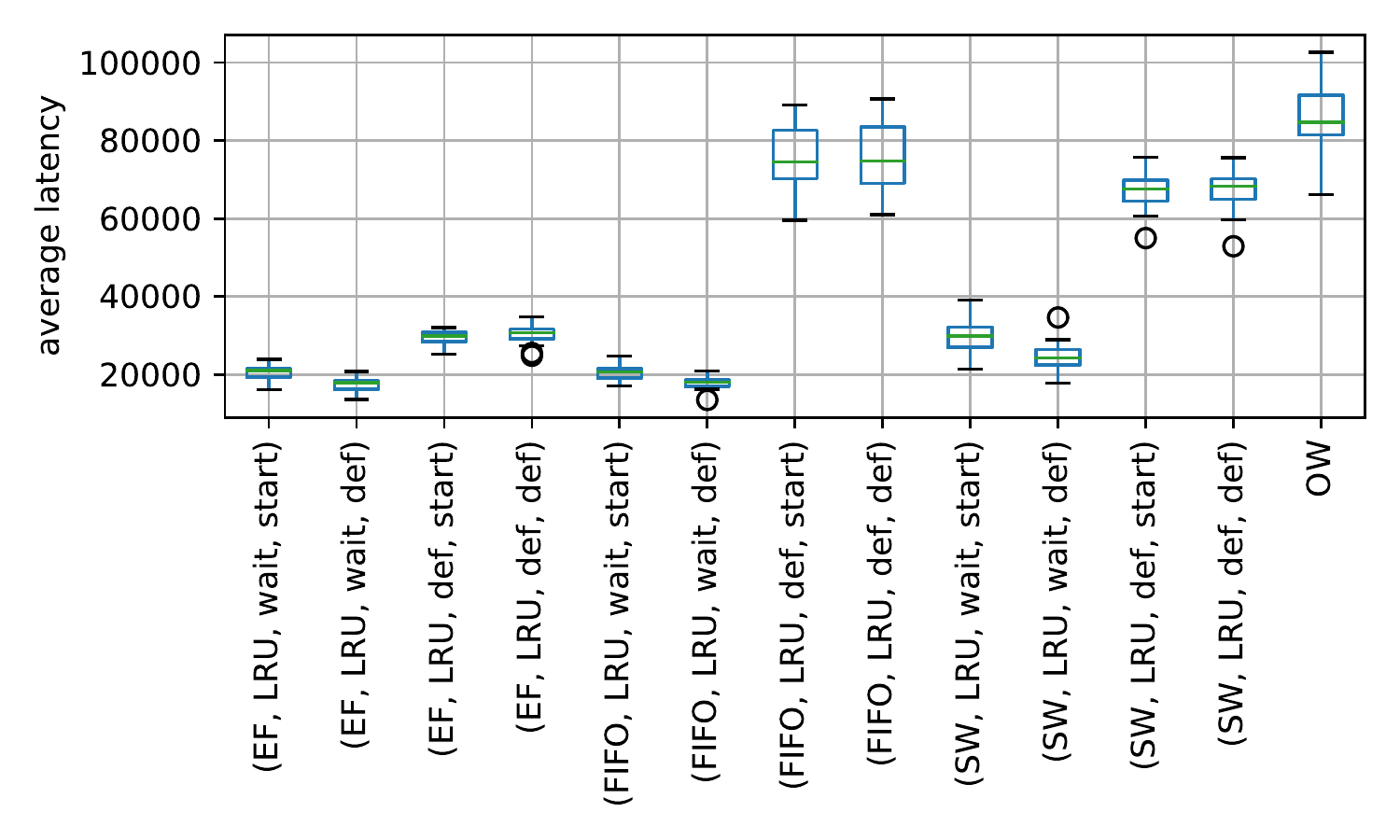}}}%
	\caption{Comparison of impact of setup times.
	All datasets contains 50 families and 1000 tasks in jobs of 50-100 tasks. Experiments were run on 10 machines of size 10}
	\label{fig:dag_compare_setup_times}
\end{figure*}

\begin{figure*}[!tb]
	\centering
	\subfloat[2 machines, size 10]{{\includegraphics[width=0.33\textwidth]{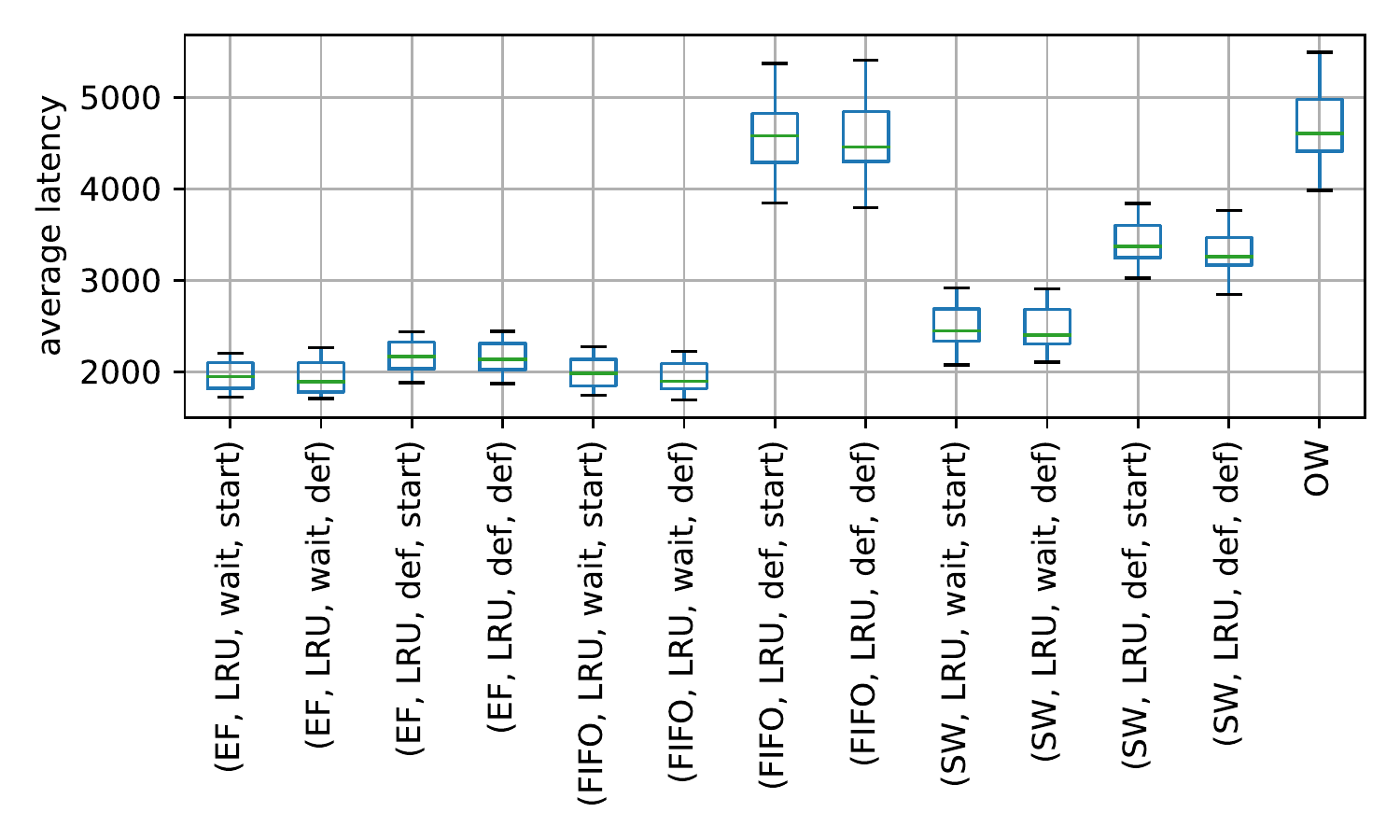}}}%
	\subfloat[2 machines, size 20]{{\includegraphics[width=0.33\textwidth]{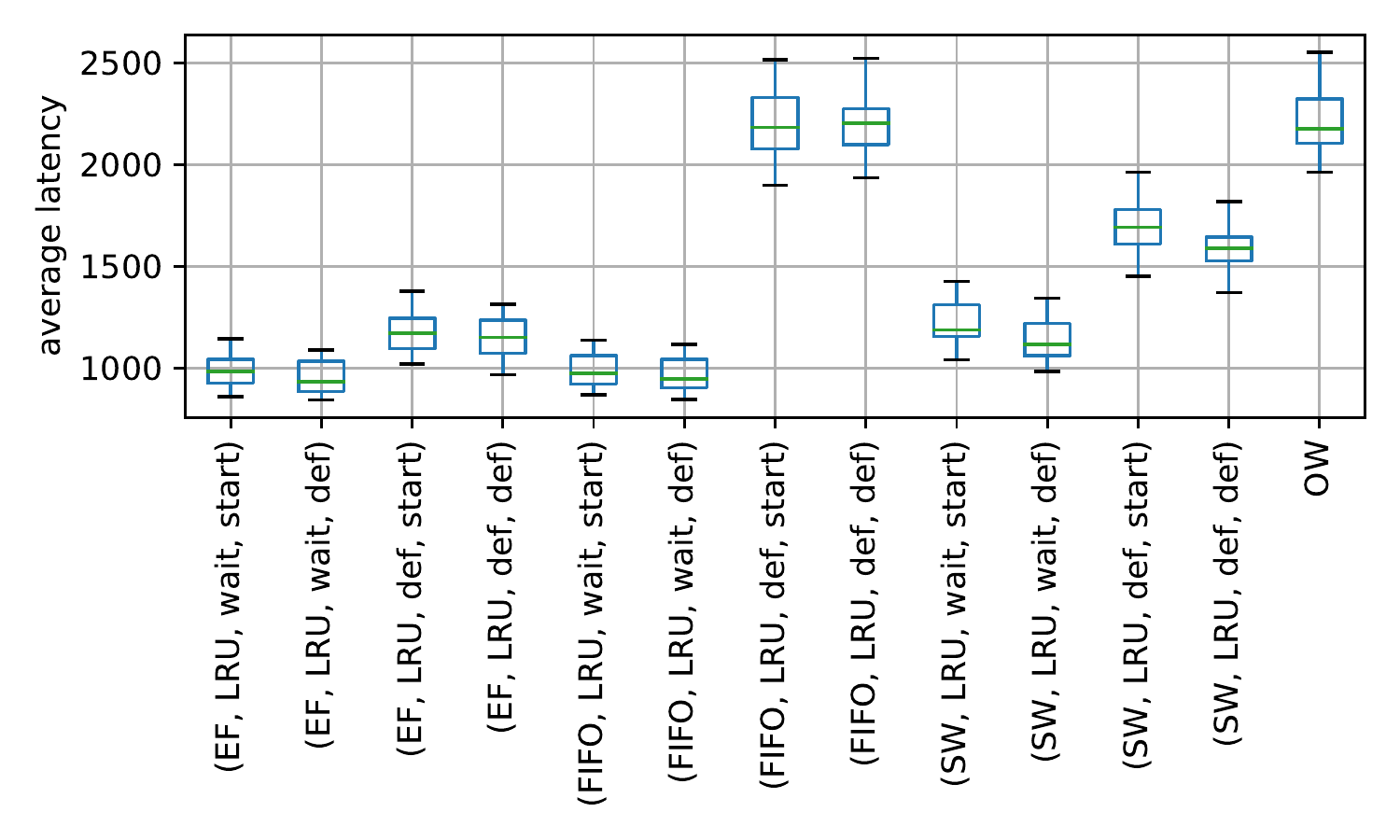}}}%
	\subfloat[2 machines, size 50]{{\includegraphics[width=0.33\textwidth]{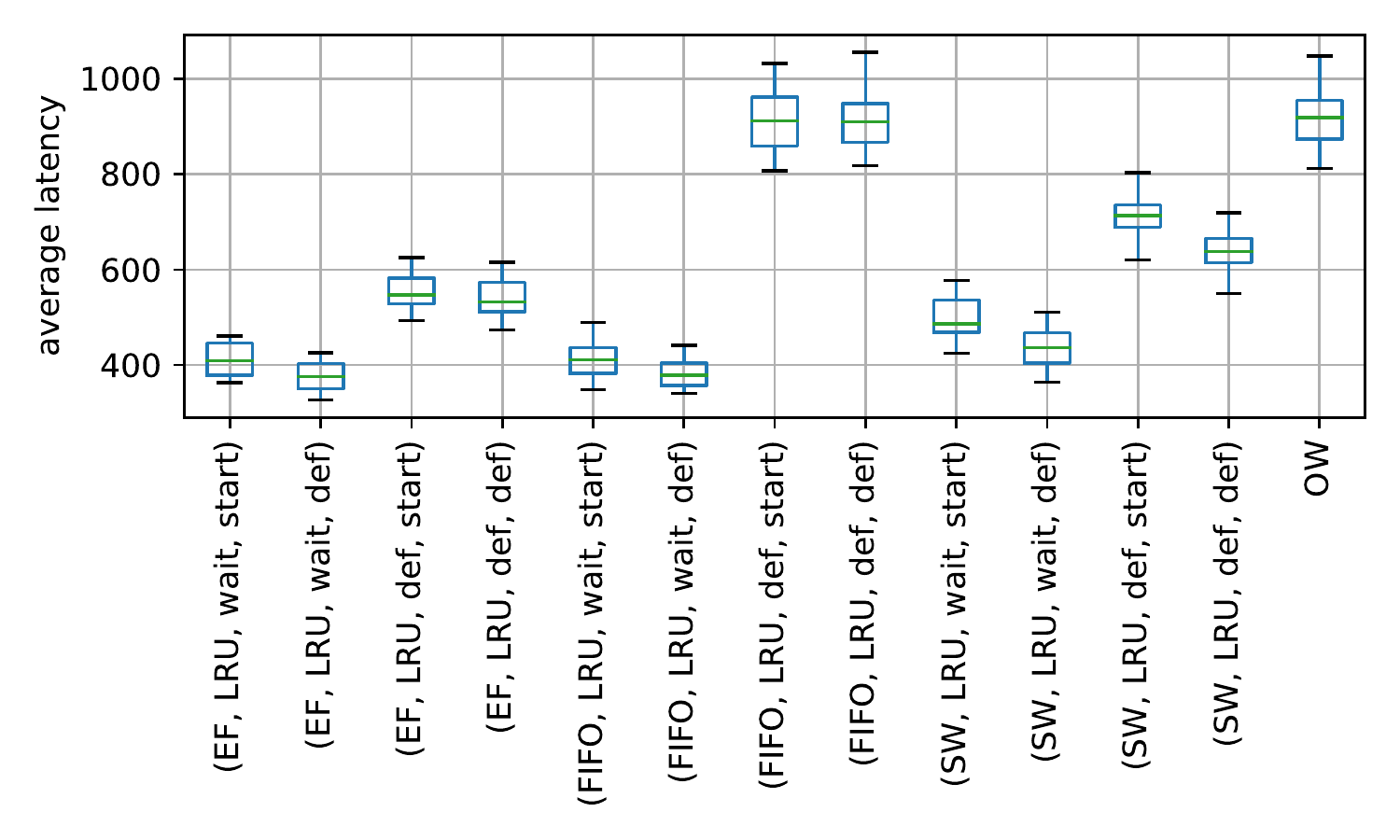}}}%
	\\
	\subfloat[5 machines, size 10]{{\includegraphics[width=0.33\textwidth]{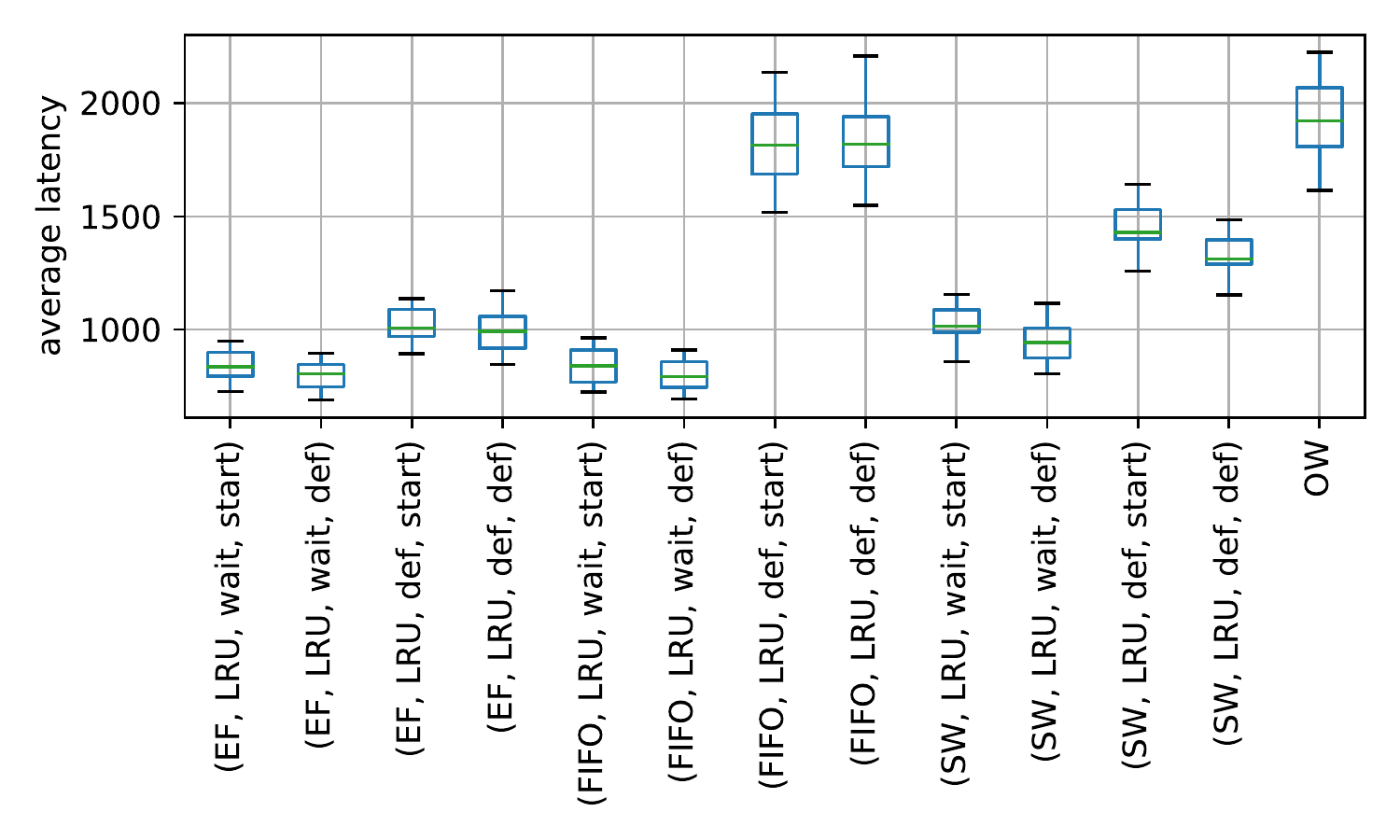}}}%
	\subfloat[5 machines, size 20]{{\includegraphics[width=0.33\textwidth]{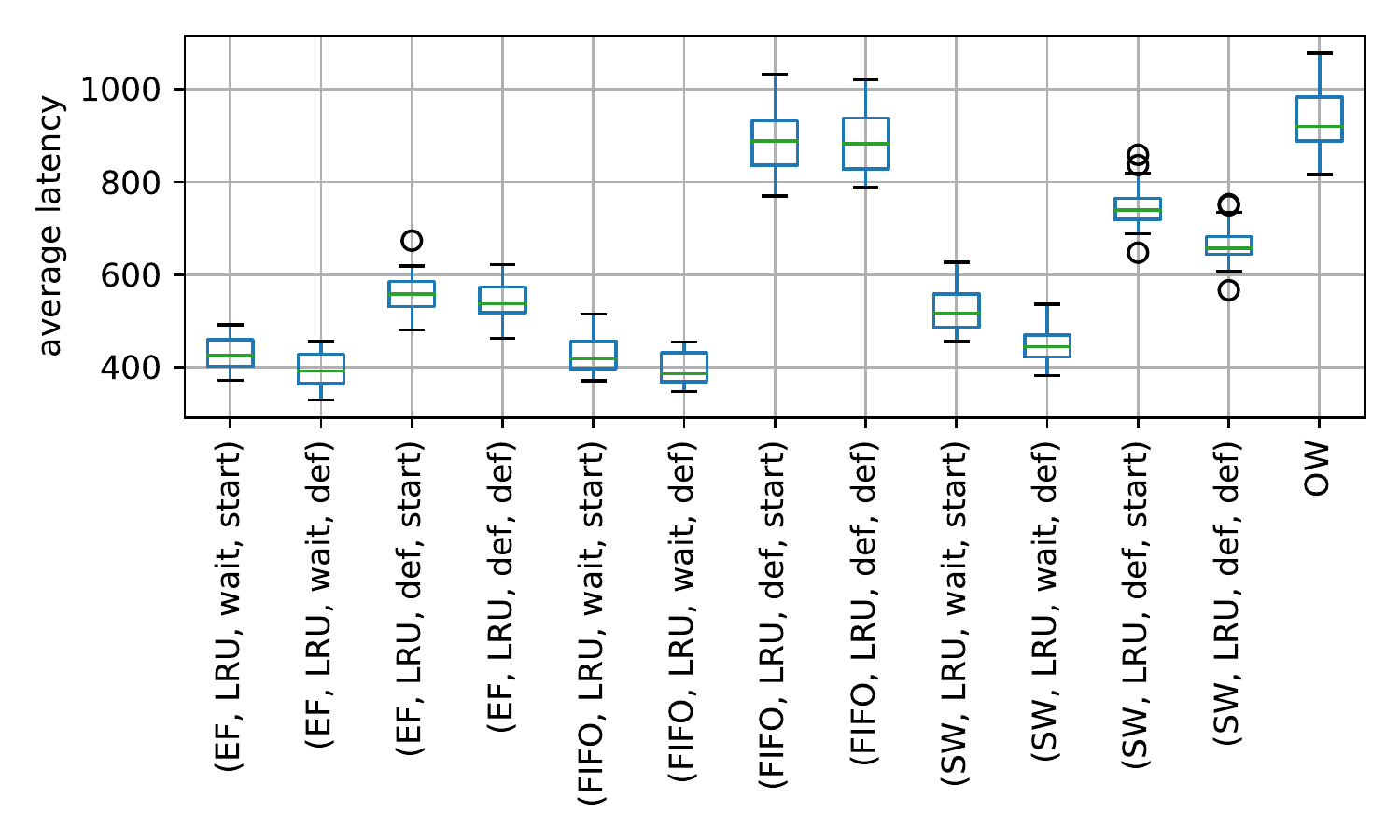}}}%
	\subfloat[5 machines, size 50]{{\includegraphics[width=0.33\textwidth]{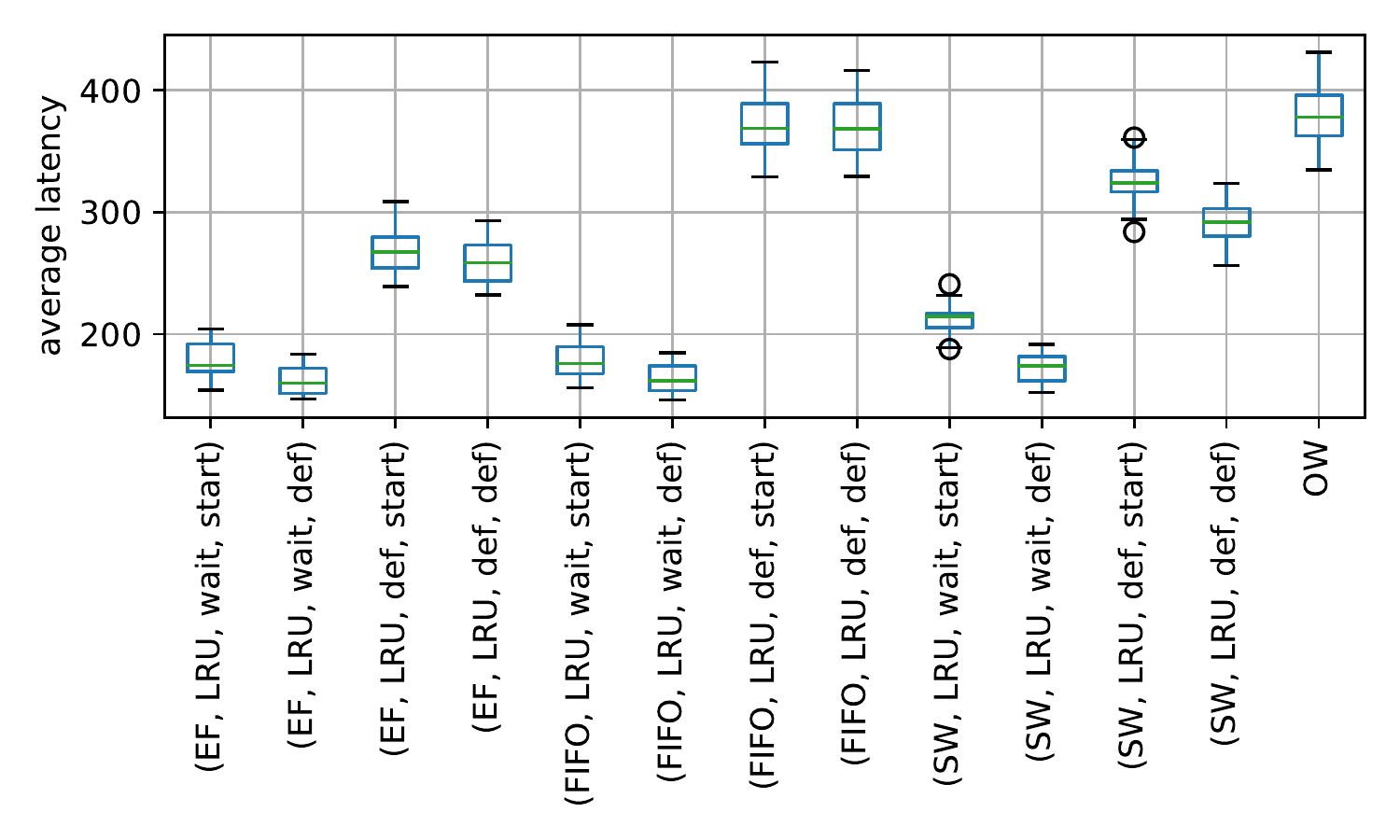}}}%
	\\
	\subfloat[10 machines, size 10]{{\includegraphics[width=0.33\textwidth]{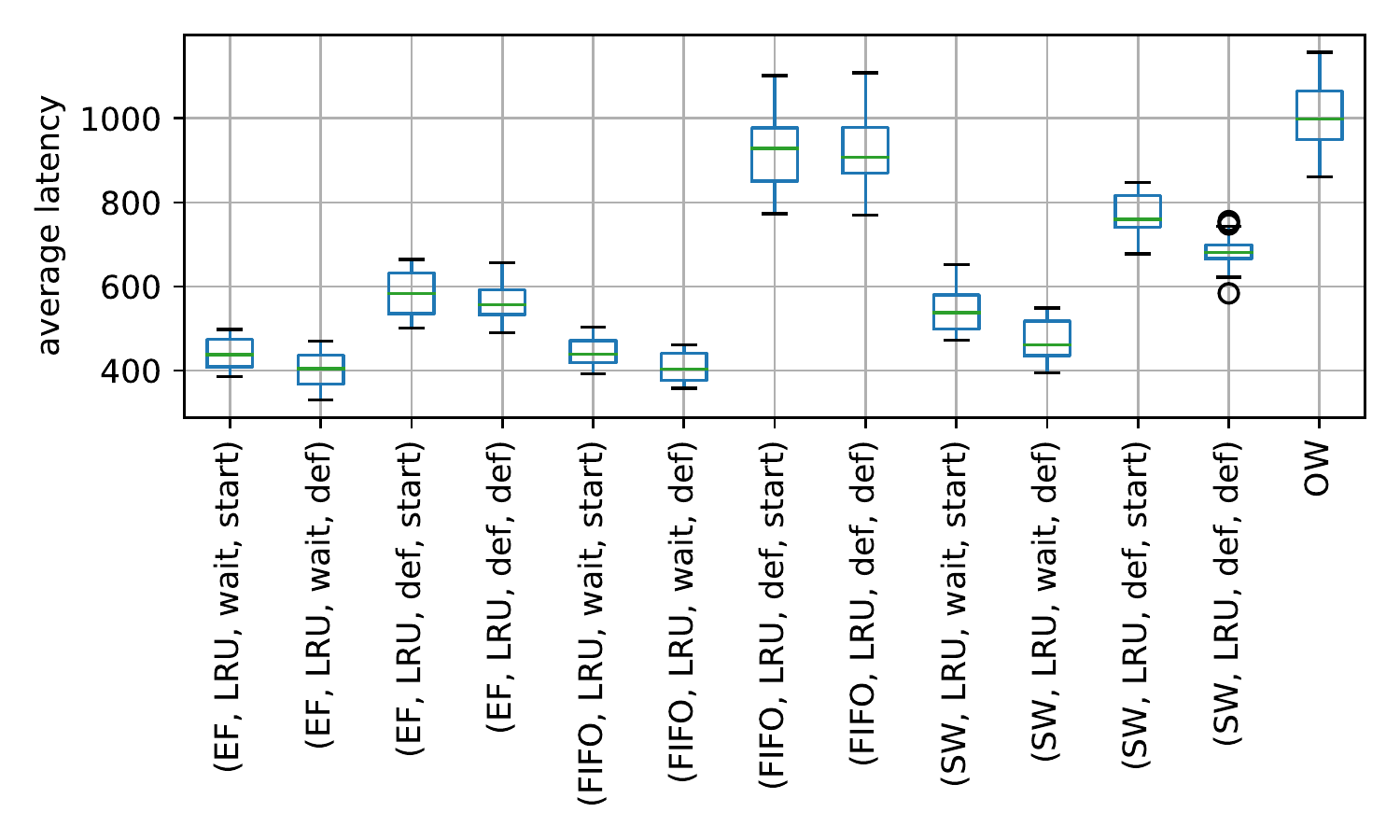}}}%
	\subfloat[10 machines, size 20]{{\includegraphics[width=0.33\textwidth]{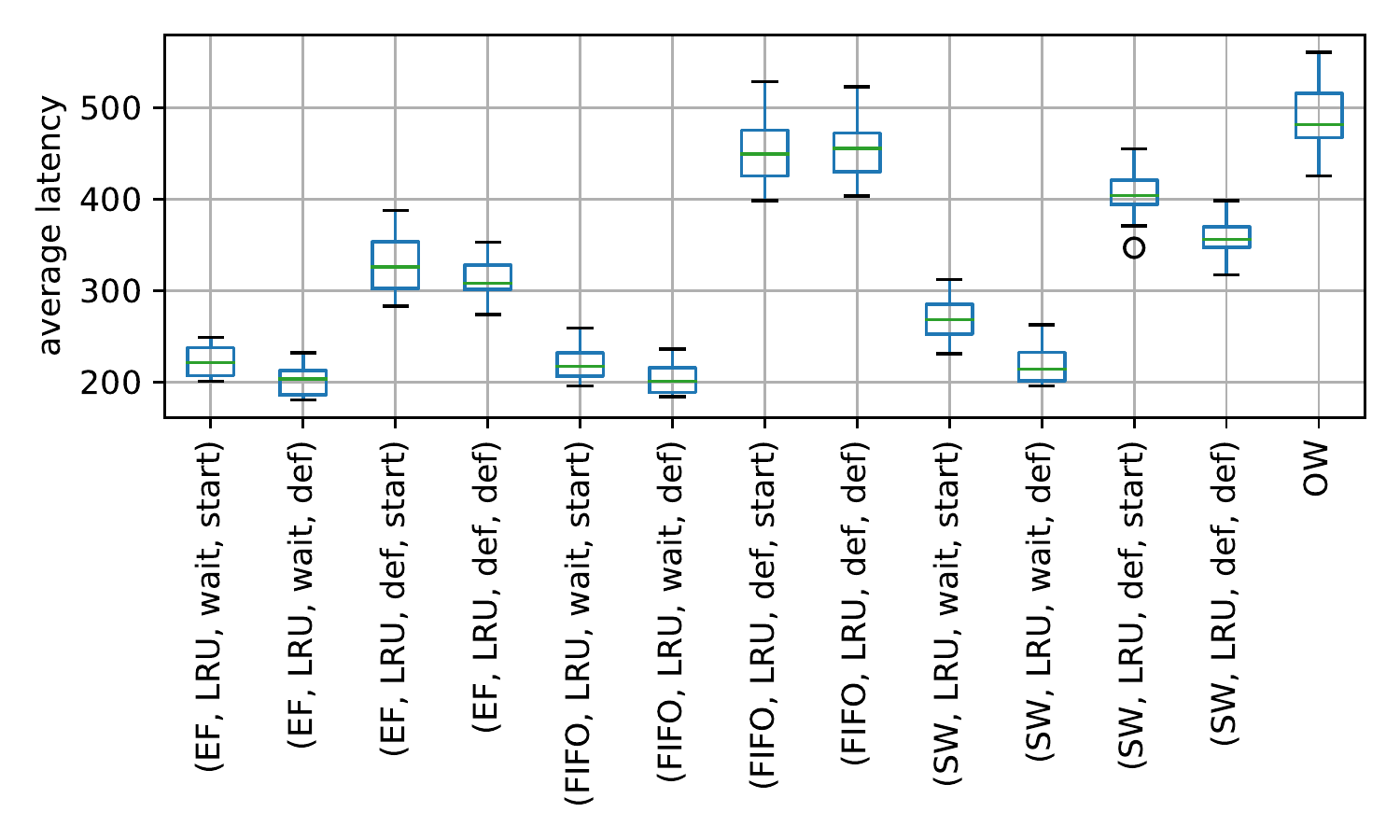}}}%
	\subfloat[10 machines, size 50]{{\includegraphics[width=0.33\textwidth]{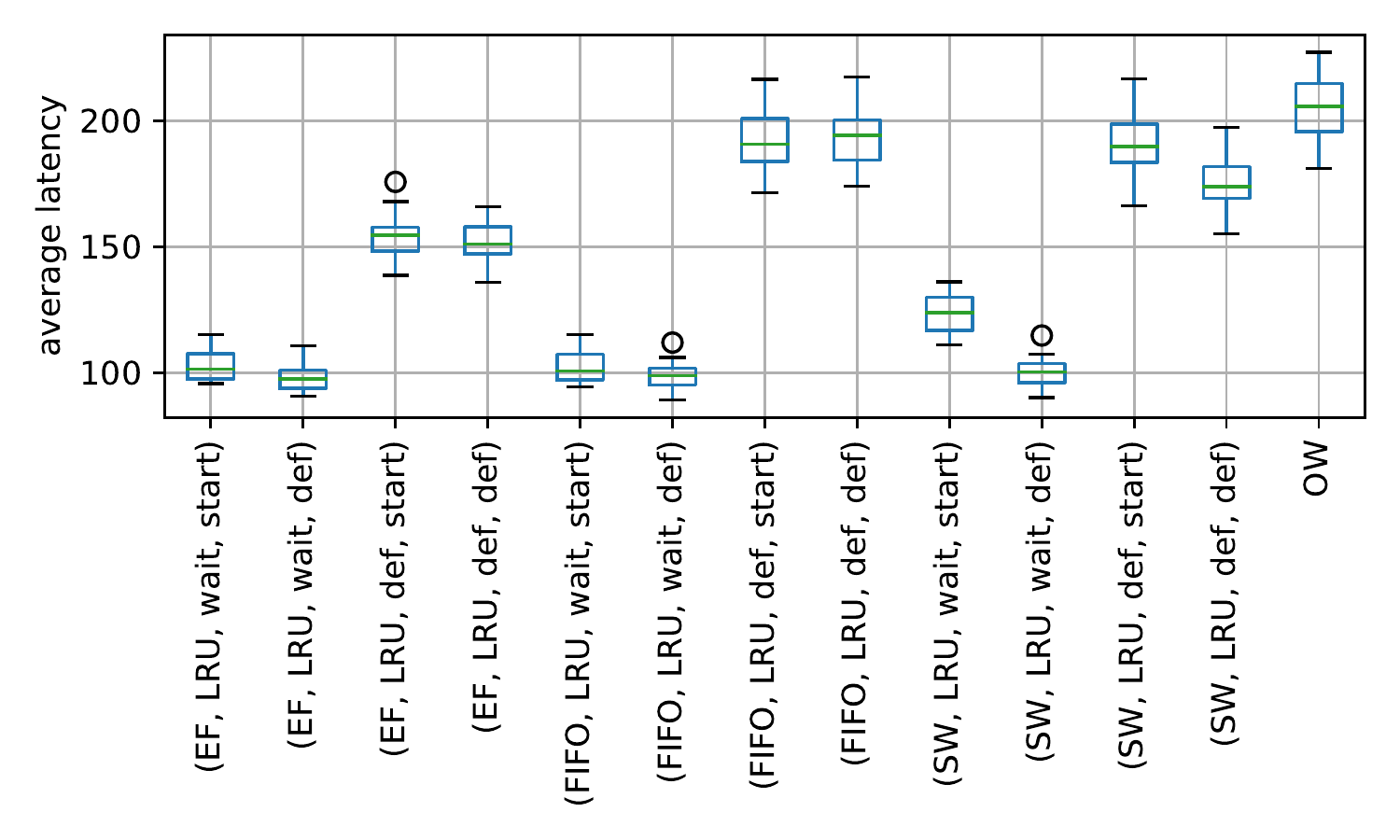}}}%
	\\
	\subfloat[20 machines, size 10]{{\includegraphics[width=0.33\textwidth]{fig/dag_fc_50_st_10_len_10_m_20_ms_10_avg.pdf}}}%
	\subfloat[20 machines, size 20]{{\includegraphics[width=0.33\textwidth]{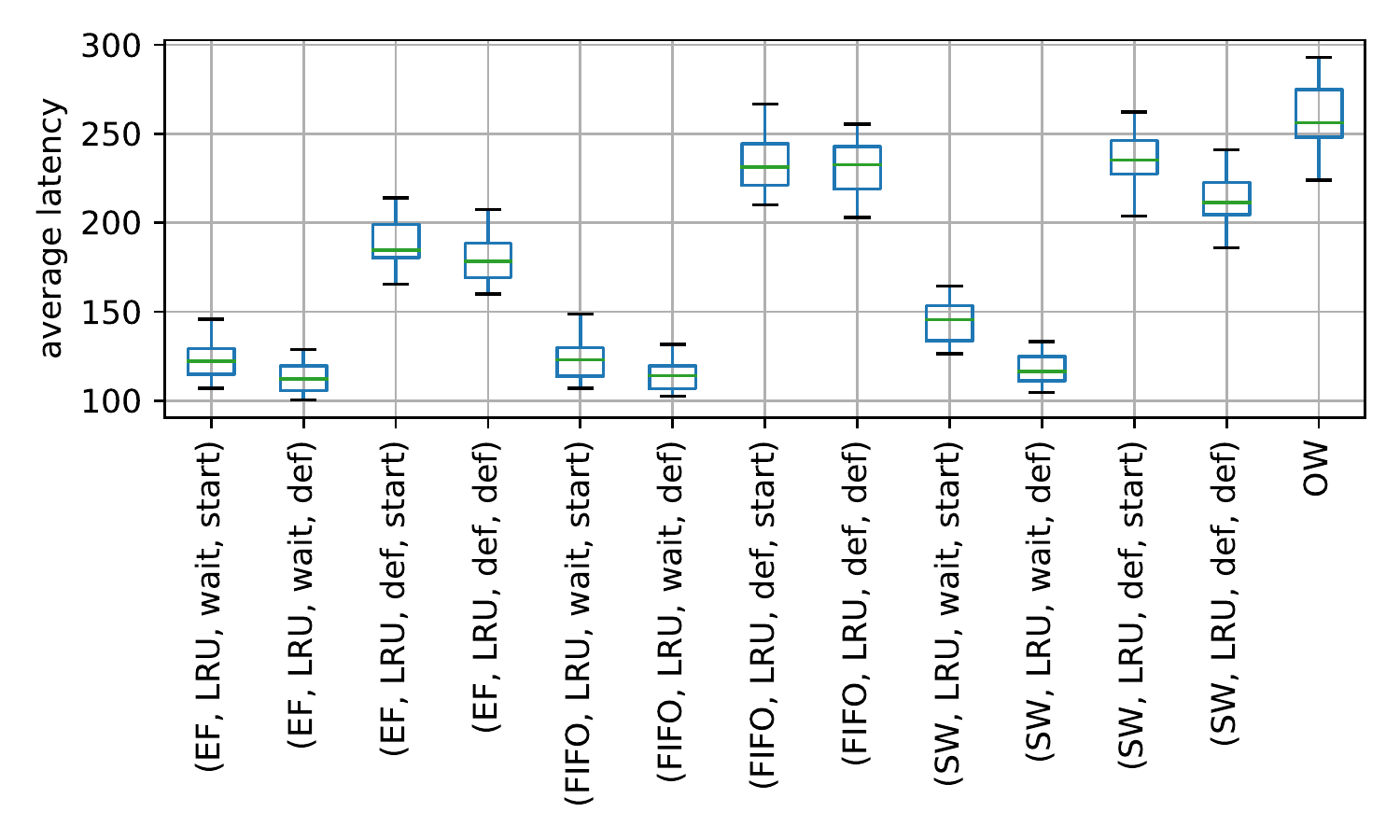}}}%
	\subfloat[20 machines, size 50]{{\includegraphics[width=0.33\textwidth]{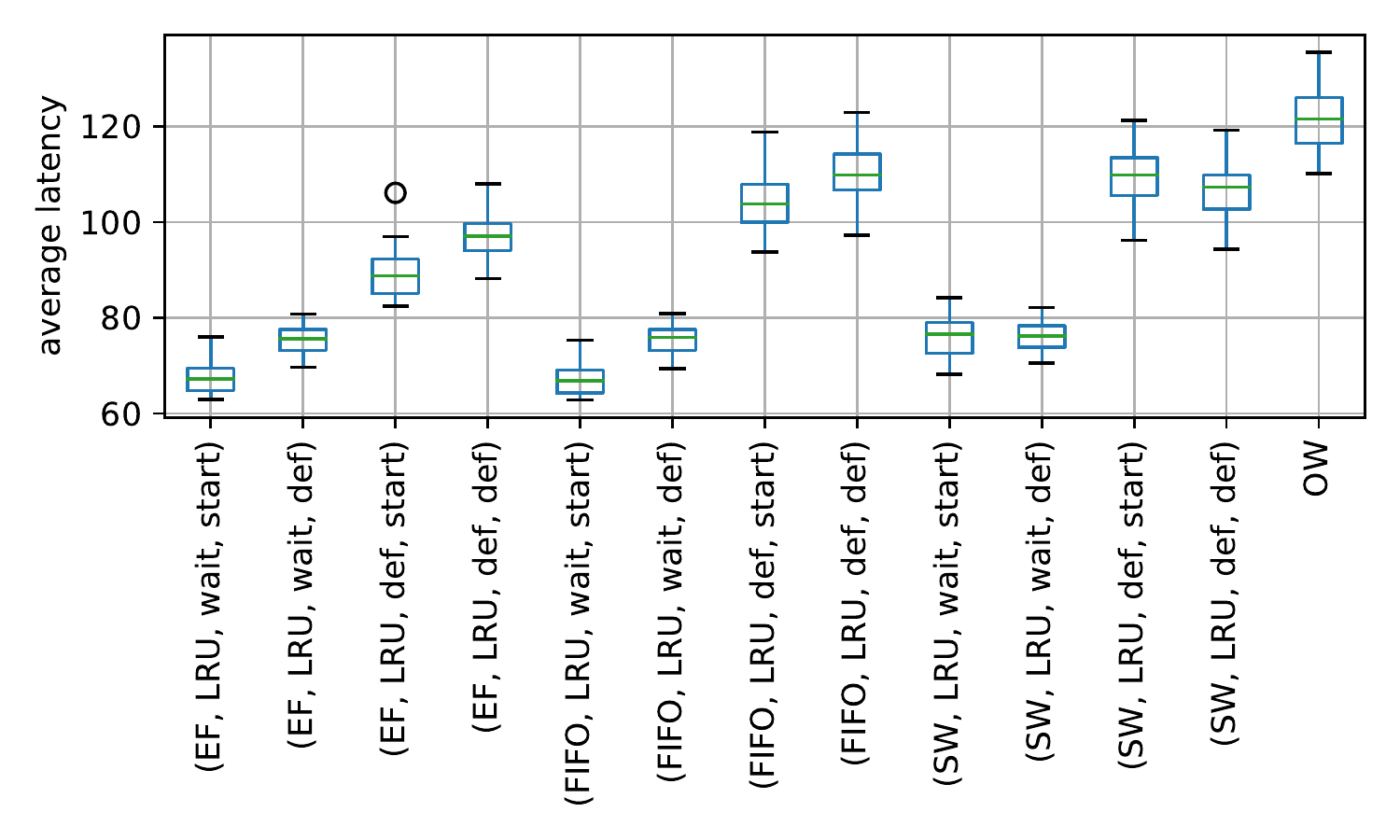}}}%
	\\
	\subfloat[50 machines, size 10]{{\includegraphics[width=0.33\textwidth]{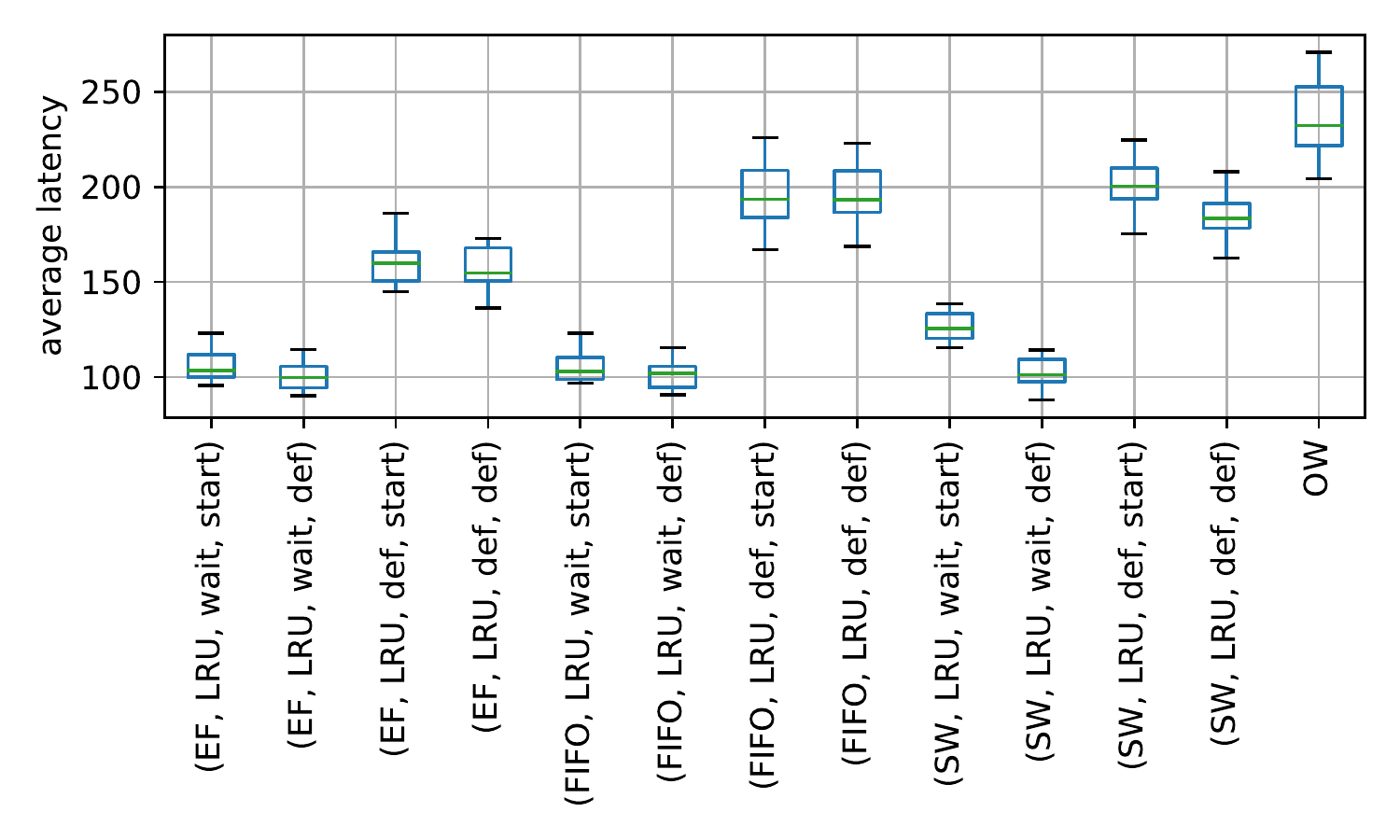}}}%
	\subfloat[50 machines, size 20]{{\includegraphics[width=0.33\textwidth]{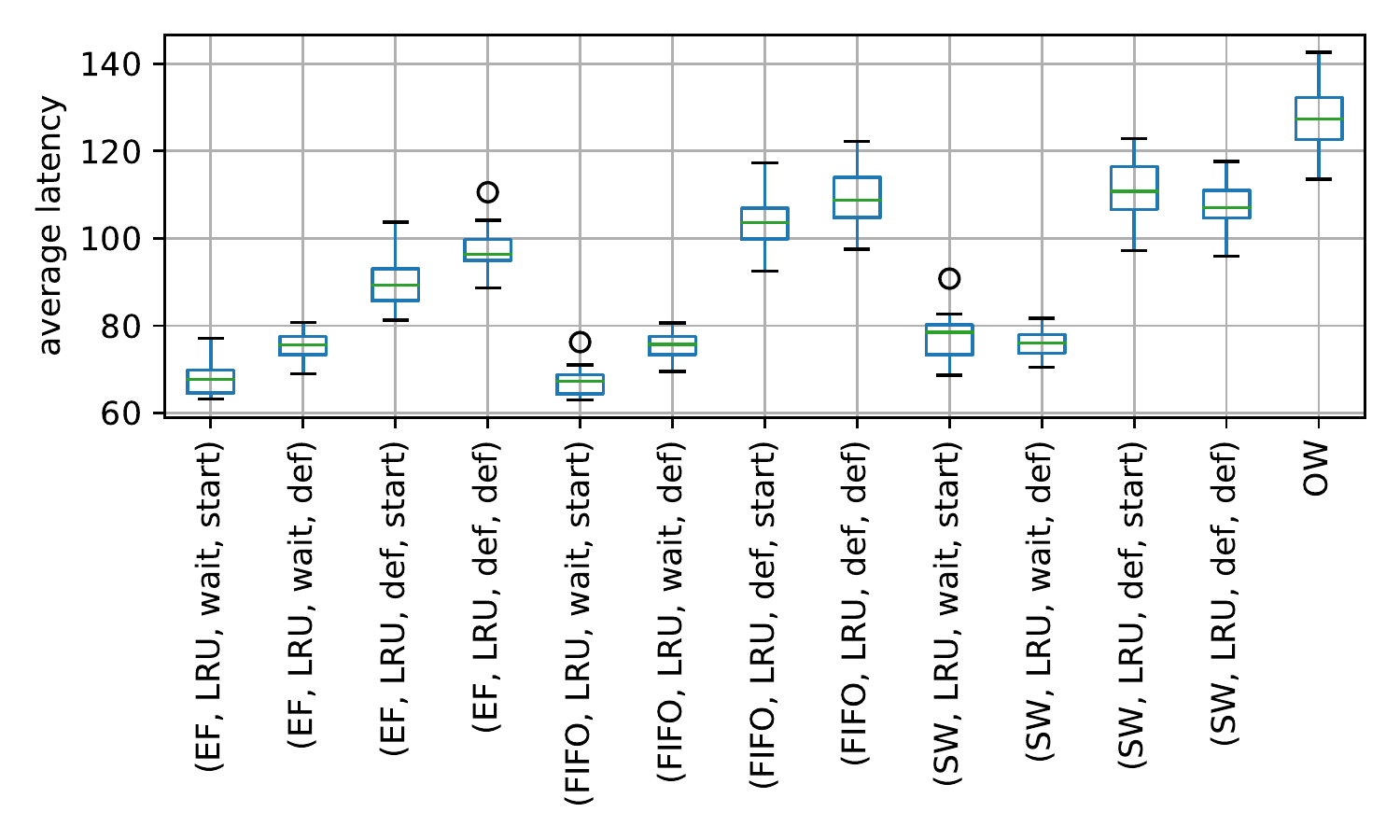}}}%
	\subfloat[50 machines, size 50]{{\includegraphics[width=0.33\textwidth]{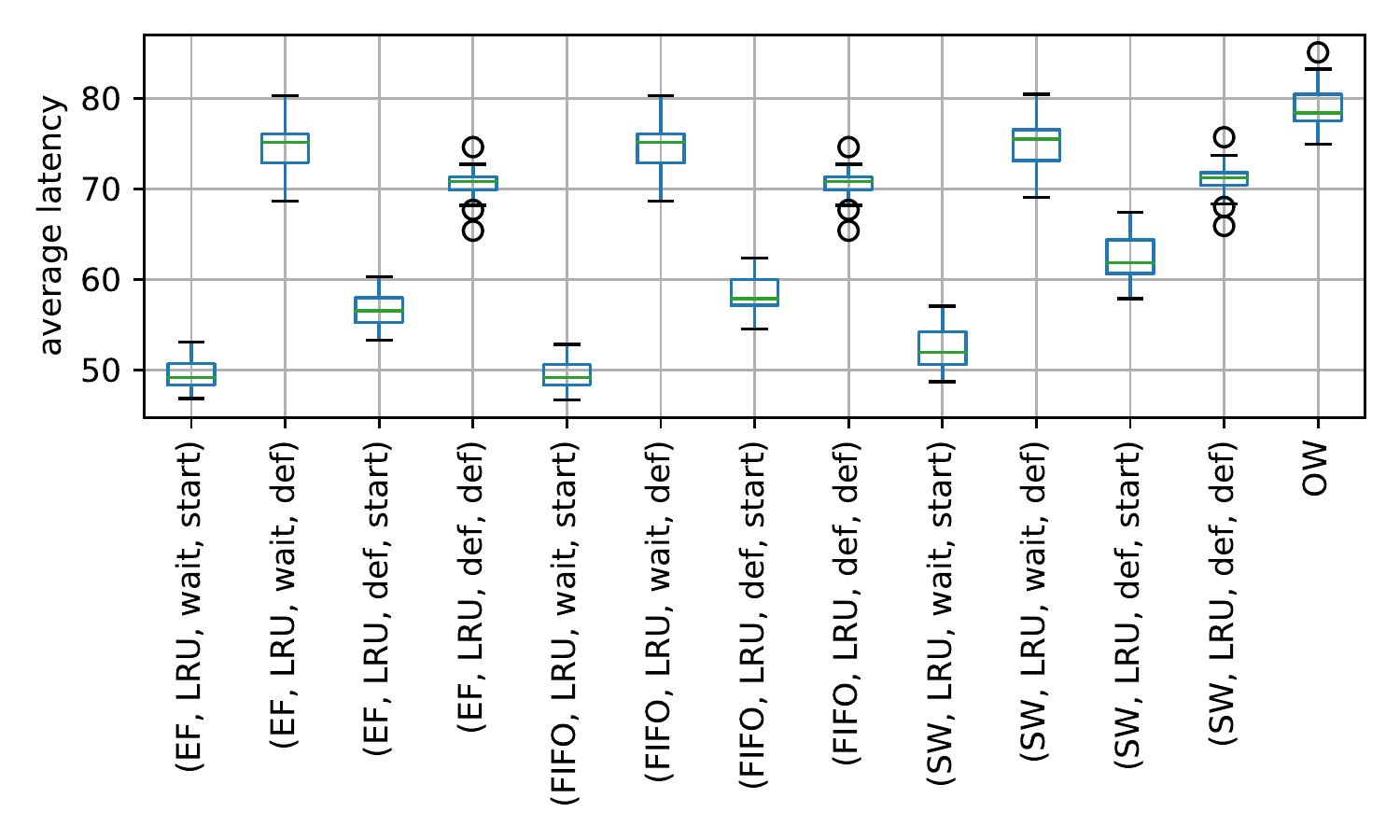}}}%
	\caption{Comparison of impact of machines count and capacities.
	All experiments were run on the same data, containing 1000 tasks in 50 families.
	Tasks are grouped in jobs containing 10-20 tasks.
	Family setup times are in range 10-20.}
	\label{fig:dag_compare_m_ms}
\end{figure*}

We performed analogous analysis to presented in the
Sections~4.E-4.G. 
We present how behavior of the algorithms changes for different job sizes (Figure~\ref{fig:dag_compare_chain_len}), setup times (Figure~\ref{fig:dag_compare_setup_times}), machine count and sizes (Figure~\ref{fig:dag_compare_m_ms}).

In Figure~\ref{fig:dag_compare_chain_len} we observe that increasing job size has lower impact on observed average latency than in chains.
This behavior is connected with the fact that datasets have the same number of tasks and datasets with longer chains have less tasks executed in parallel which limits concurrency.
Contrary, for datasets with DAGs there could be more than one task within a job that could be executed simultaneously.
Similar to results obtained for chains, EF scheduling policy along with LRU and waiting outperforms OW in all cases.

If we compare behavior for different setup times (Figure~\ref{fig:dag_compare_setup_times}), we can observe that for non-zero setup times our algorithms perform better than baseline (OW).
However, contrary to results for chains, there is very little difference between variants without (def) and with dependency awareness (\emph{start}).

\begin{figure*}[!tb]
	\centering
	\subfloat[10 families]{{\includegraphics[width=0.33\textwidth]{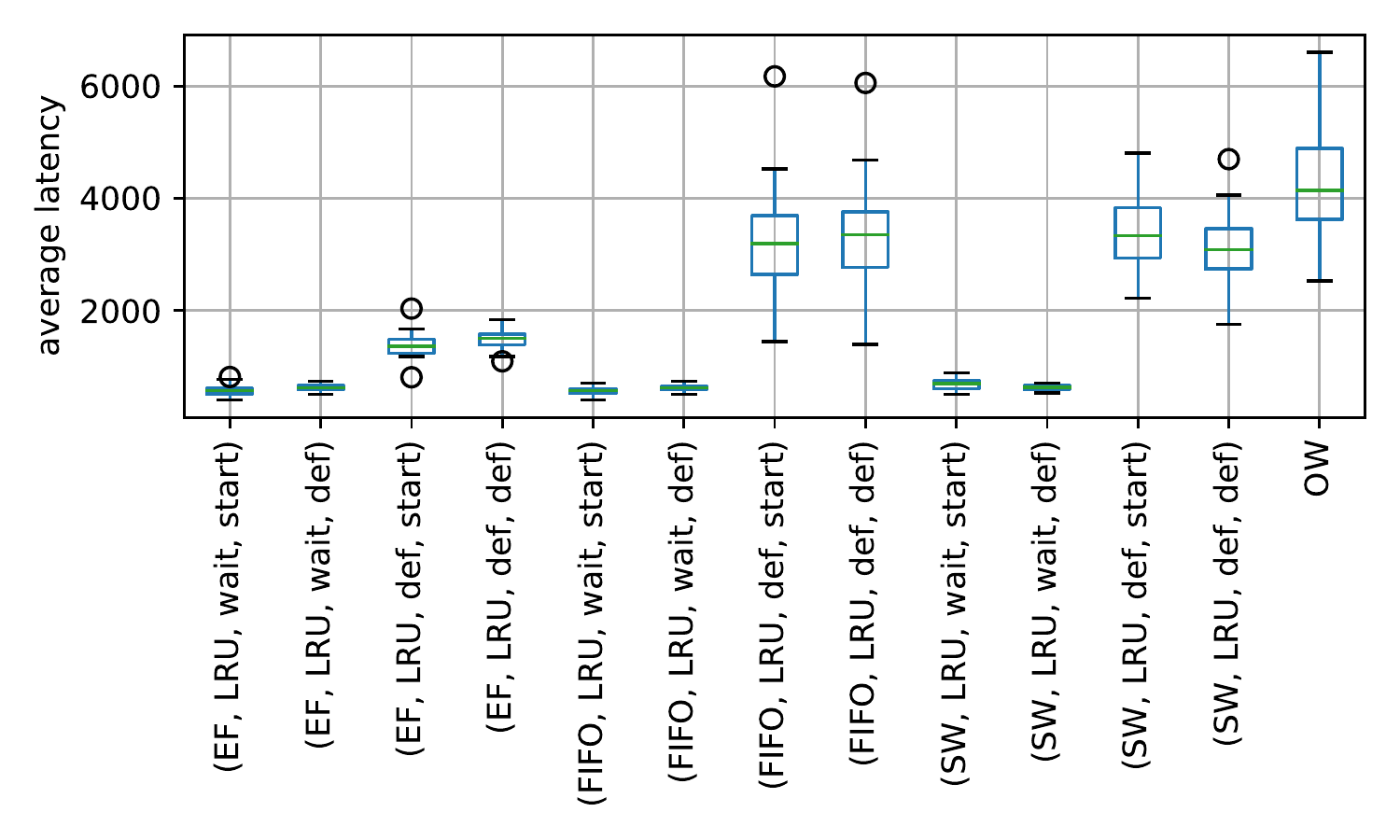}}}%
	\subfloat[20 families]{{\includegraphics[width=0.33\textwidth]{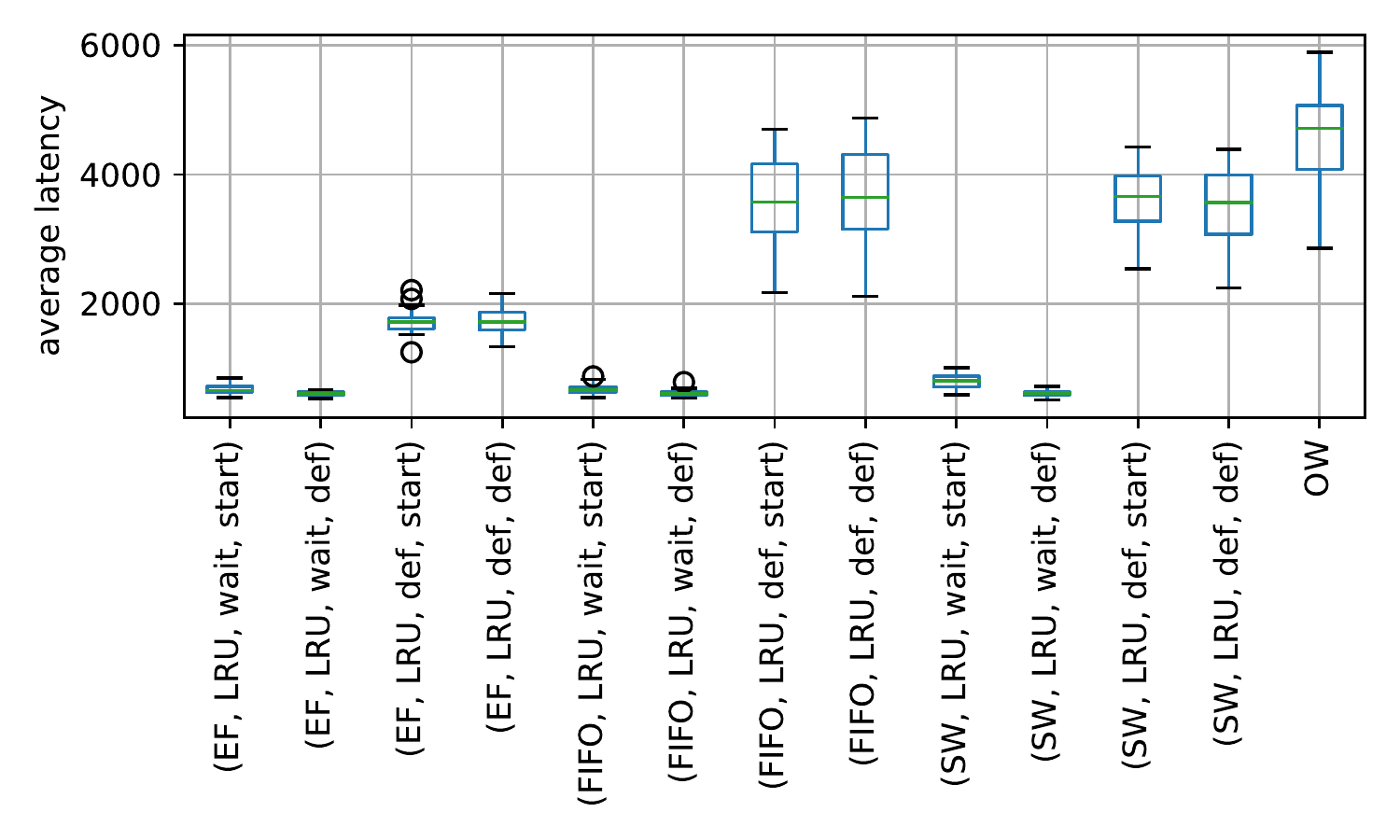}}}%
	\subfloat[50 families]{{\includegraphics[width=0.33\textwidth]{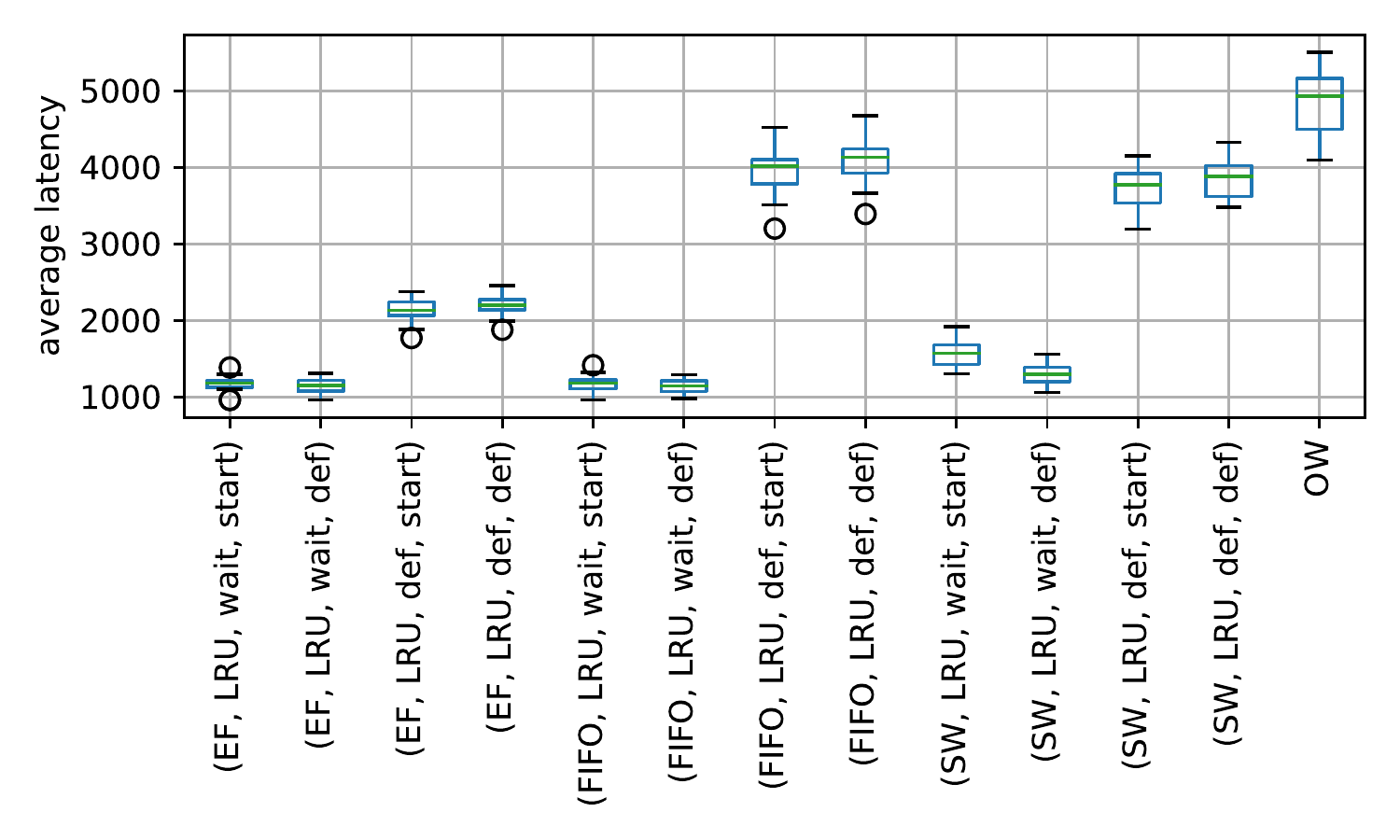}}}%
	\\
	\subfloat[100 families]{{\includegraphics[width=0.33\textwidth]{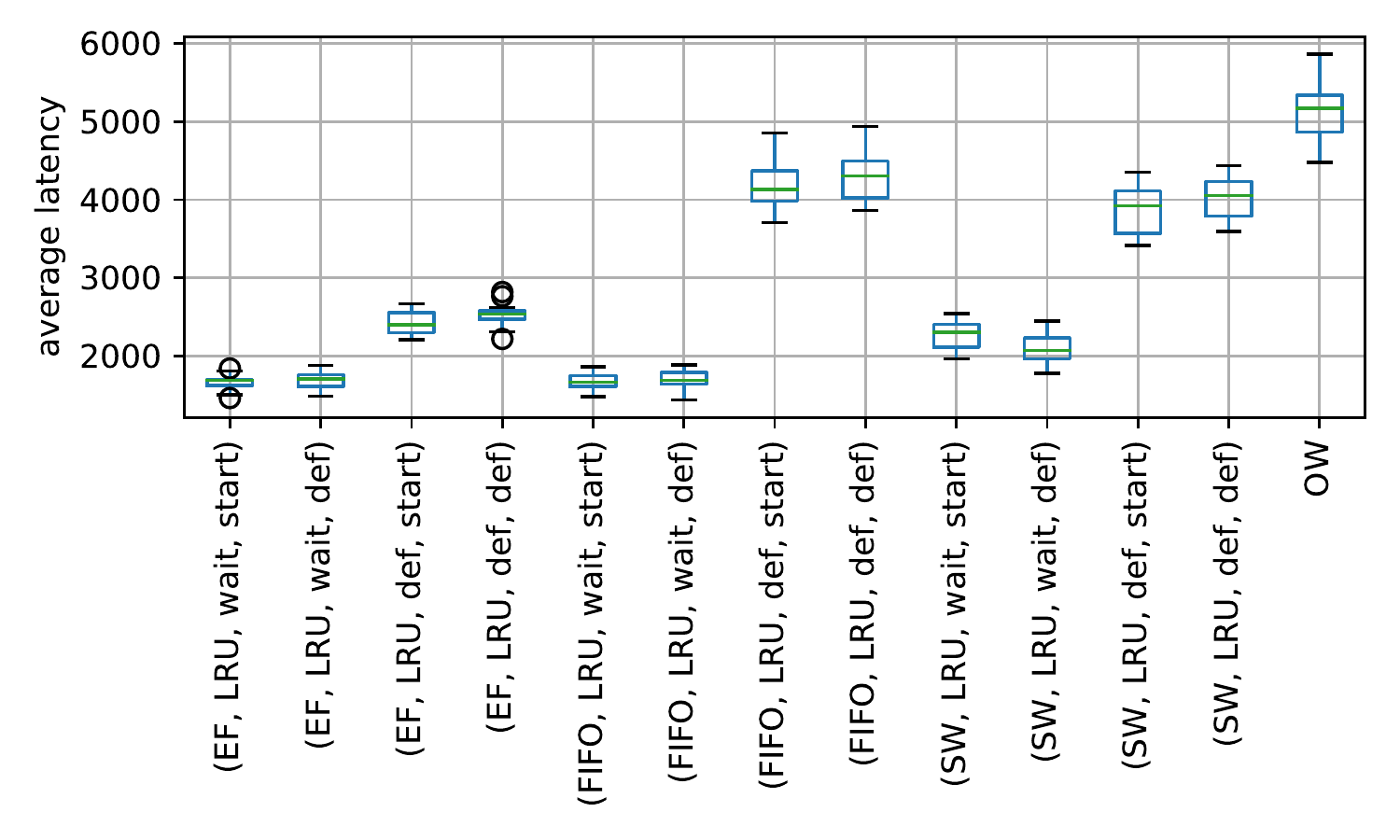}}}%
	\subfloat[200 families]{{\includegraphics[width=0.33\textwidth]{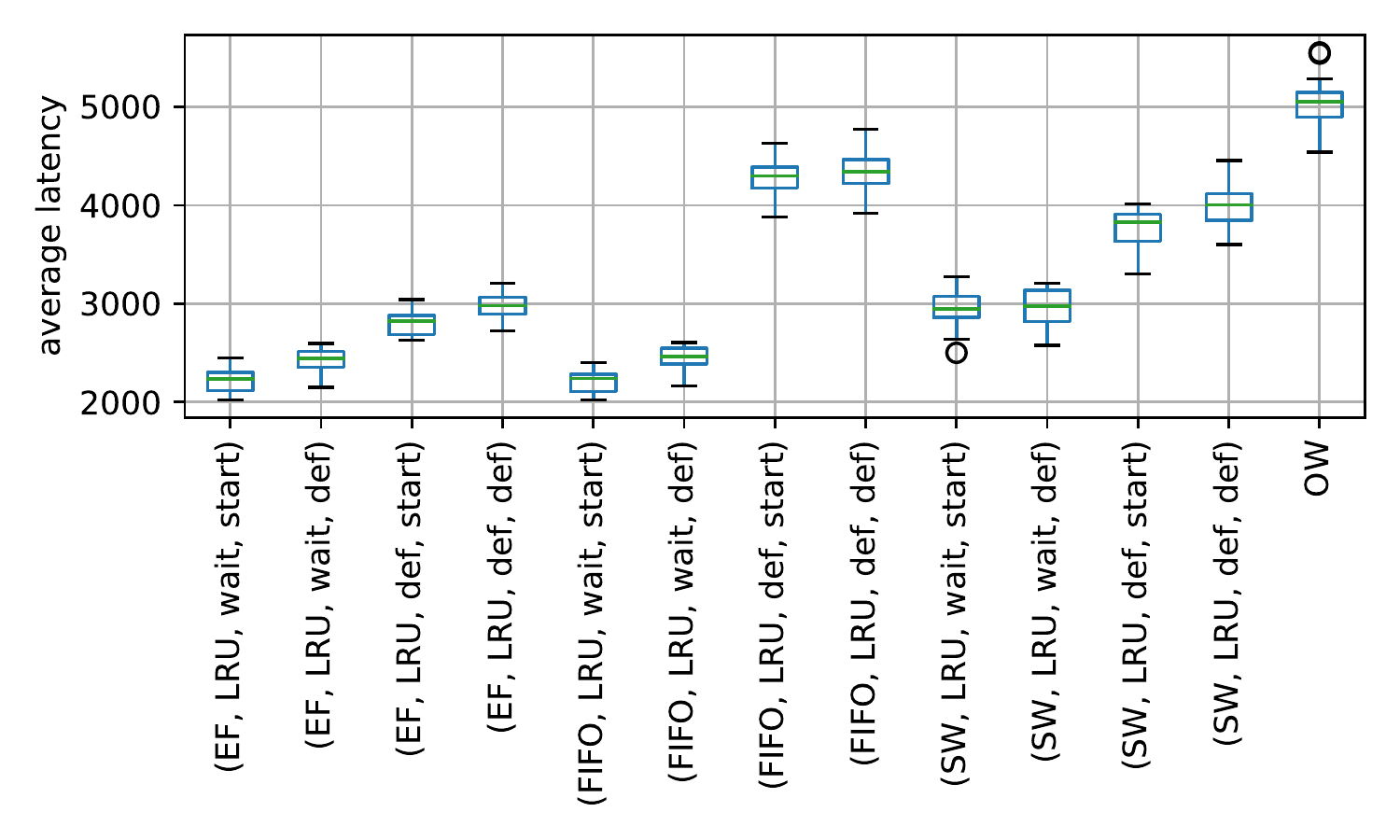}}}%
	\subfloat[500 families]{{\includegraphics[width=0.33\textwidth]{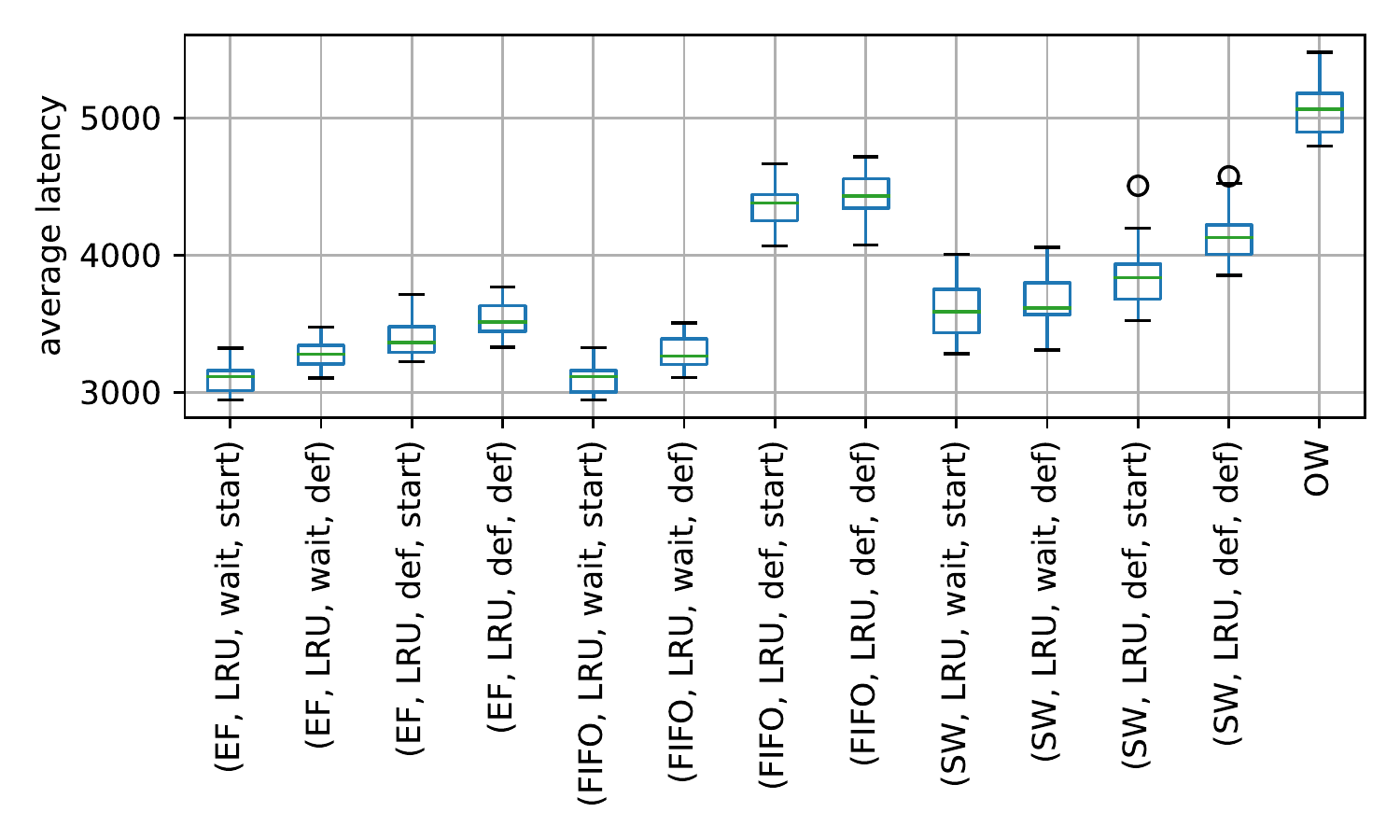}}}%
	\caption{Comparison of different family count. In all cases setup times are in range 100-200. We present results for 20 machines of size 10, jobs containing 50-100 tasks (note: except last generated one)}
	\label{fig:dag_compare_families}
\end{figure*}

Similar behavior can be observed for datasets with different family count (Figure~\ref{fig:dag_compare_families}).
Moreover, the best improvement over baseline can be observed for datasets with up to 50 families.
For datasets with more families, our algorithms still provide more optimal schedules than OW, but difference between the best schedule and the baseline is noticeably smaller.

Figure~\ref{fig:dag_compare_m_ms} compares behavior of the algorithms under different machine sizes and the number of machines.
All task durations are in range $[1, 10]$ and tasks are grouped in jobs containing $[10,20]$ tasks (apart from the last one which may be smaller).
In all presented cases we evaluate the same data with 50 task families.
Only for the largest processing capacity (50 machines of size 50) there is observable fundamental improvement of dependency-aware (\emph{start}) variants over \emph{def}.
For all machine configurations except the largest one (50 machines of size 50), \emph{wait} variants performed better than the default (\emph{non-waiting}) variants using the same scheduling method and with the same setting of dependency-awareness.
For large capacities (over 20 machines of size 50 or over 50 machines of size 20), for \emph{FIFO} and \emph{EF} scheduling policies, variants using \emph{wait} with \emph{start} have one of the lowest average latency.

As we can see in all analyzed cases with non-zero setup times, proposed algorithms behave better than OW baseline.
Therefore our modifications should improve average latency in more general scenarios.

\sectionhead{Experiments with cluster workloads}
To our best knowledge currently there is no publicly available cluster trace containing information about tasks with dependencies, setup times and introducing tasks families.
However, we can generate dateset resembling the trace, making some rational assumptions about missing data in existing cluster trace.
Thus we can verify how such dataset would behave if executed in analyzed model.

Recently published Google Cluster Trace 2019~\cite{clusterdata:Wilkes2020a} contains information about dependencies between executed computations.
The trace defines \emph{jobs} as a set of \emph{tasks} (processes) which may be executed simultaneously.
Jobs and tasks belongs to exactly one \emph{collection}.
The collections may form a DAG  -- i.e. none of tasks in the collection can be started until all computations in predecessors are completed.
Moreover, the collections specify information about collection purpose in field \emph{collection\_logical\_name} -- e.g. if multiple collections execute the same program, all of them should have the same \emph{collection\_logical\_name}. 
We create test workload preserving the dependency structure and information about computation type -- thus, the cluster collection is equivalent of the function invocation in our model and collections with the same \emph{collection\_logical\_name} belong to the same family.


The full trace contains information from 8 datacenters from different locations.
Each of them is independent source of data - in our analysis we use data obtained from New York cluster (2019-05-a).

We generate input data as follows.
We extract all DAGs by using data in \emph{start\_after\_collection\_ids} field of \emph{collection\_events} table.
Each DAG is given unique job id and we use \emph{collection\_id} as task index within a job
 We obtained 8740 different jobs in this procedures.

Next, we skip all DAGs which match any of following criterion:
\begin{enumerate}
	\item contain task without start time (i.e. started before trace period),
	\item contain task without end time (i.e. still running at end of trace period),
	\item are not chains,
	\item have collection depending on collection not existing in trace data
\end{enumerate}
In such procedure we removed 3894 jobs.

We use \emph{collection\_logical\_name} to indicate which tasks belongs to the same family.
Resulting collections belong to 424 different families.
For each family we estimate size, duration and setup time as follows.

Let $F$ be set of collections belonging the same family and $C(j)$ be set of all tasks belonging to collection $j$.
For each task $i \in C(j)$ we obtain maximum value of used memory $c_{task}(i)$ from field \emph{assigned\_memory} in \emph{instance\_usage} table.
Then we compute an average memory required to run any tasks belonging collection $j$: $c_{avg}(j) = \frac{1}{|C(j|}\sum_{i \in C(j)} c_{task}(i)$.
We set family size as $c_{family}(F) = \max_{j \in F} c_{avg}(j)$.

By obtaining time of SCHEDULE, FAIL, FINISH, KILL, LOST events, we compute start (first SCHEDULE event) and end (last all of rest events) time of all collections.
We set family duration time as average value of duration of collections with the same \emph{collection\_logical\_name}.
We estimate family setup time multiplying its duration by random factor obtained from discrete uniform distribution over range $[10, 100)$.

\begin{figure}[!tb]
	\centering
	\includegraphics[width=\columnwidth]{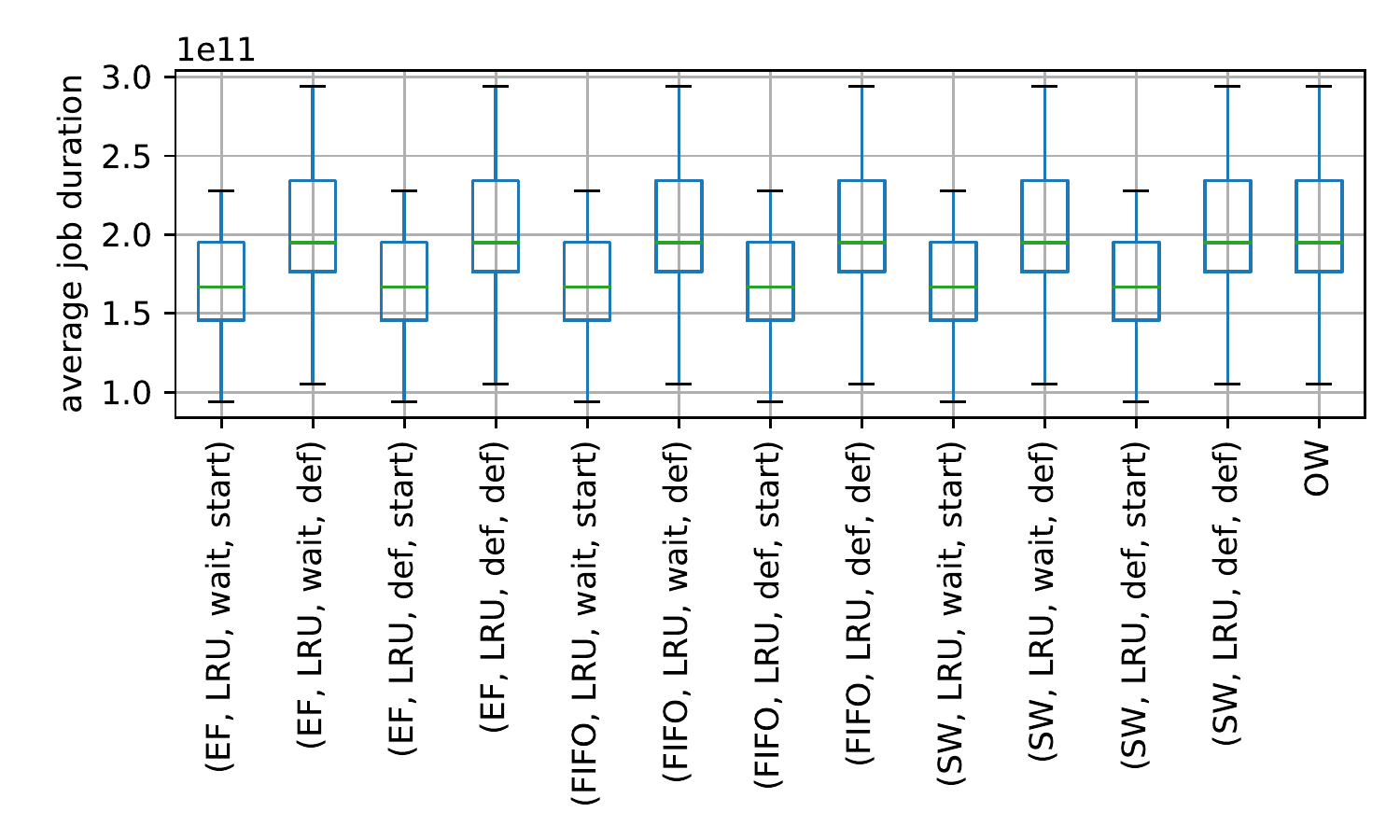}
	
	\caption{Comparison of different scheduling policies for dataset created from Google Cluster Trace. Each dataset contains 100 different chains of tasks.
	Each task corresponds to one collection in Google Trace.
	For presented figure $m=20$, $Q=10$ - to preserve clarity, we omit results for other machine counts and sizes as changing those parameters give no difference in observed results.
	}
	\label{fig:gcp_compare}
\end{figure}

We generate 20 samples each containing 100 randomly chosen (without repetition) jobs.
Figure~\ref{fig:gcp_compare} presents difference between behavior of the algorithms for 20 machines of size 10.
To preserve clarity,  we omit results for other machine counts and sizes as changing those parameters give no difference in observed results.

This result confirms our results for experiments with generated data -- we observed similar behavior for datasets with larger (200, 500) number of families.
While we observe no difference between different scheduling and eviction methods, enabling dependency awareness reduces observed average latency.

\ifdefined\arxivpreprint
\else
\bibliographystyle{IEEEtran}
\bibliography{bibliography}

\end{document}
\fi

\fi

\end{document}